\documentclass{article}
\usepackage{authblk}

\usepackage[nomain,acronym]{glossaries}
\usepackage{url}
\usepackage{bbm}
\usepackage{adjustbox}

\usepackage{amssymb}
\usepackage[english]{babel}
\usepackage[utf8]{inputenc}
\usepackage{amsmath}
\allowdisplaybreaks

\DeclareMathOperator*{\argmin}{arg\,min}
\usepackage{graphicx}
\usepackage{imakeidx}
\usepackage{subfig}
\usepackage{float}
\usepackage{rotating}
\usepackage{enumerate}
\usepackage{comment}

\usepackage[numbers]{natbib}
\usepackage{proof-at-the-end}
\usepackage{soul}

\usepackage{stmaryrd}
\usepackage{amsthm}
\usepackage{mathtools}
\usepackage[ruled,vlined]{algorithm2e}
\usepackage{mleftright}
\graphicspath{ {./images/} }

\newcommand{\norm}[1]{\left\lVert#1\right\rVert}

\newcommand{\Var}{\mathrm{Var}}

\makeatletter
\def\moverlay{\mathpalette\mov@rlay}
\def\mov@rlay#1#2{\leavevmode\vtop{%
   \baselineskip\z@skip \lineskiplimit-\maxdimen
   \ialign{\hfil$\m@th#1##$\hfil\cr#2\crcr}}}
\newcommand{\charfusion}[3][\mathord]{
    #1{\ifx#1\mathop\vphantom{#2}\fi
        \mathpalette\mov@rlay{#2\cr#3}
      }
    \ifx#1\mathop\expandafter\displaylimits\fi}
\makeatother

\newcommand{\bigcupdot}{\charfusion[\mathop]{\bigcup}{\cdot}}

\usepackage{array}

\usepackage[a4paper,top=2.5cm, bottom=4.5cm, left=2.5cm, right=2.5cm]{geometry}

\usepackage{nomencl}
\makenomenclature

\usepackage{etoolbox}
\renewcommand\nomgroup[1]{%
  \item[\bfseries
  \ifstrequal{#1}{F}{Functions}{%
  \ifstrequal{#1}{A}{Acronyms}{%
  \ifstrequal{#1}{K}{Kernels}{%
  \ifstrequal{#1}{O}{Other Symbols}{}}}}%
]}
\usepackage{hyperref}
\usepackage[capitalise,nameinlink
    ,noabbrev]{cleveref}

\newtheorem{definition}{Definition}[section]
\newtheorem{prop}{Proposition}[section]
\newtheorem{corollary}{Corollary}[section]
\newtheorem{lemma}{Lemma}[section]

\newtheorem{rmk}{Remark}[section]

\newcommand{\Keywords}[1]{\par\noindent
{\textbf{Keywords\/}: #1}}
\newcommand{\Subjclass}[1]{\par\noindent
{\textbf{MSC\/}: #1}}

\title{Gradient-based estimation of linear Hawkes processes with general kernels}

\author[1,2]{\'Alvaro Cartea}
\author[1,3]{Samuel N. Cohen}
\author[1]{Saad Labyad}
\affil[1]{Mathematical Institute, University of Oxford}
\affil[2]{Oxford--Man Institute of Quantitative Finance}
\affil[3]{The Alan Turing Institute}

\date{\today}
\begin{document}

\maketitle
\begin{abstract}
Linear multivariate Hawkes processes (MHP) are a fundamental class of point processes with self-excitation. When estimating parameters for these processes, a difficulty is that the two main error functionals, the log-likelihood and the least squares error (LSE), as well as the evaluation of their gradients, have a quadratic complexity in the number of observed events. In practice, this prohibits the use of exact gradient-based algorithms for parameter estimation. We construct an adaptive stratified sampling estimator of the gradient of the LSE. This results in a fast parametric estimation method for MHP with general kernels, applicable to large datasets, which compares favourably with existing methods.
\Keywords{Hawkes processes; stochastic gradient descent; point processes; Monte Carlo methods; adaptive stratified sampling.}  
\Subjclass{60G55,62M09,90C52,93E10.}  
\end{abstract}
\setcounter{secnumdepth}{3}
\setcounter{tocdepth}{1}

\newenvironment{acknowledgements} {\renewcommand\abstractname{Acknowledgements}\begin{abstract}} {\end{abstract}}
\begin{acknowledgements}
\'Alvaro Cartea and Samuel Cohen acknowledge the support of the Oxford--Man Institute for Quantitative Finance. Samuel Cohen also acknowledges the Alan Turing Institute under the Engineering and Physical Sciences Research Council grant EP/N510129/1.
Saad Labyad acknowledges the support of St John's College, University of Oxford, under the Ioan and Rosemary Jones scholarship.
\end{acknowledgements}

\section{Introduction}
Temporal point processes are widely applied as models of asynchronous streams of events. One way to specify these models is through their conditional intensity, that is, the expected infinitesimal rate of events per unit of time, conditioned on the history of the process. Linear multivariate Hawkes processes (MHP) are a specific class of point processes in which the conditional intensity takes a linear auto-regressive form parameterized by a matrix of kernel functions and a vector of constant background rates; see \citet*{hawkes1971spectra}. The widespread use of MHP is mainly due to their explainability: their matrix of kernel functions accounts for self-excitation and cross-excitation between different types of events, and their cluster representation can be a proxy for causality between events. 

MHP have applications in a variety of domains including \emph{finance}, particularly in market microstructure (see \citet*{cartea2014buy}, \citet*{bacry2015hawkes}, and \citet*{hawkes2018hawkes} for an extensive review); \emph{social networks}, with an emphasis on modeling information cascades such as retweets (see \citet{zhao2015seismic}, \citet{kobayashi2016tideh} and \citet{chen2018marked}); \emph{seismology}, to study the occurrence of earthquakes and their aftershocks (see \citet*{veen2008estimation}); and \emph{criminology}, to examine criminal contagion mechanisms, notably in burglaries and gang violence (see \citet{mohler2011self}, \citet{lewis2012self}, and \citet*{mohler2014marked}). 

Despite their popularity, estimation of the kernel matrix and background rates of MHP remains difficult. In practice, the absence of a fast parametric estimation method prohibits the use of MHP with significant amounts of data (i.e., of order higher than $10^6$--$10^7$ jumps in an observed sample path), and with  arbitrary kernels, in particular non-Markovian kernels. These limitations arise because the evaluation of the conditional intensity at each time $t$ has linear complexity in the number of jumps up to time $t$, leading to quadratic complexity overall. Therefore, objective functions based on the conditional intensity of the MHP are expensive to evaluate and to minimize (see \cref{subsubsec:minimize_lse} for a detailed analysis). A notable exception is the MHP with exponential kernels (see \cref{subsubsec:Some kernels}), where the conditional intensity can be evaluated recursively, which explains the predominance of exponential MHP in the literature. 

The goal of this paper is to overcome these limitations. We develop a stochastic optimization algorithm for parametric MHP estimation that does not directly evaluate the conditional intensity of the MHP; we call this the \texttt{ASLSD} algorithm (Adaptively Stratified Least Squares Descent). As an accompaniment to our derivation and analysis, an implementation in python is available at \url{https://github.com/saadlabyad/aslsd}. This algorithm is computationally efficient, accurate for a wide range of sample sizes, and flexible enough to allow for regularization and sparsity terms to be easily included.

\subsection{MHP Estimation methods}
We briefly review the state of the art for MHP estimation.  The time complexity, assumptions, objective function, and regularization type of the algorithms discussed here are summarized in \cref{table:recap_methods}. Estimation procedures fall into three main categories:
\begin{itemize}
    \item \emph{Method of moments}. These procedures are typically based on spectral properties of the MHP, and usually aim to convert the estimation problem into solving a system of equations. Most of these methods are non-parametric, and require mild assumptions beyond stationarity of the MHP as they use second order properties of the process. 
    \item \emph{Maximum likelihood estimation (MLE)}. As in other statistical problems, MLE benefits from sound theoretical guarantees. However, for general MHP, the evaluation of the log-likelihood of the MHP has a quadratic time complexity in the number of jumps observed. The application of the expectation-maximization (EM) algorithm to MHP estimation usually improves the convergence of optimization algorithms, but the high computational cost of the E step in EM methods does not allow for efficient algorithms. 
    \item \emph{Least squares estimation}. This class of methods is rarely used in the context of MHP. The cost of evaluating the least squares objective function is roughly as expensive as that of evaluating the log-likelihood. Nonetheless, in this paper we show that, unlike the log-likelihood, the least squares error (LSE) has an additive decomposition that is particularly suitable for efficient stochastic approximation.
\end{itemize}

\paragraph{Method of moments} \citet*{hawkes1971point,hawkes1971spectra} applies methods developed for the general analysis of the spectra of point processes by \citet*{bartlett1967spectral}. Hawkes shows a link between the Laplace transform of the autocovariance function $\nu$ of the increments of stationary MHP and the mean conditional intensity and kernels of the MHP. \citet*{bacry2012non} use this link to propose a non-parametric estimation method in the specific case of stationary MHP with symmetric kernels and Laplace transforms diagonalizable in the same orthogonal basis. All these assumptions (except stationarity) are relaxed by \citet*{bacry2016first}, who show that the MHP parameters solve a system of Wiener\textendash Hopf equations; we use this algorithm as a nonparametric baseline in our numerical examples. \citet{achab2017uncovering} use the first three cumulants of the MHP to propose a non-parametric estimation method for the adjacency matrix of the MHP (the matrix of $L_1$ norms of the kernels). This algorithm is fast, as it depends linearly on the number of jumps of the MHP; however, this method is not meant for the estimation of the kernels themselves. Finally, the work of \citet*{gao2018transform} relies on the spectrum of the cumulative number of jumps of different types instead of the autocovariance property.

While algorithms based on the method of moments apply to a wide range of models, they are particularly inefficient when the number of observations is small. These moment-based methods are also particularly prone to the curse of dimensionality (with respect to the number of dimensions $d$ of the MHP), and regularization for the sake of dimensionality reduction seems difficult for these models.

\paragraph{Maximum likelihood estimation} A different paradigm consists in maximising the log-likelihood of the sample path, see \citet*{daley2007introduction}. To the best of our knowledge, the fastest parametric approach, in the case where the kernels are a sum of exponentials with fixed decay rates, is that in \citet*{bompaire2018dual}; we use this algorithm as a parametric baseline in our numerical examples. \citet*{lemonnier2014nonparametric} use Bernstein polynomials to give a density argument to justify the choice of a linear combination of exponential decays. In the case of linear combinations of non-exponential kernels, \citet{bacry2016mean} propose a mean field approximation of the log-likelihood to speed up standard parametric estimation. Despite the speed of this method, it is difficult to generalize it due to the mean-field and linearity assumptions.

Another limitation of log-likelihood methods for MHP estimation is the flatness of the log-likelihood. A classic approach to solve this issue is the EM procedure introduced by \citet{veen2008estimation} and \citet{lewis2011nonparametric}, which is based on \citeauthor{hawkes1974cluster}'s \citep{hawkes1974cluster}  cluster representation of the MHP. In the general case, the complexity of an EM iteration remains quadratic, but significantly smoothes the objective. The ADM4 algorithm of \citet*{zhou2013adm4} also builds on the EM approach, with the assumption that the kernels of the MHP are of a fixed form with a single scale coefficient. This method uses sparsity and low rank penalties to estimate high-dimensional MHP. Finally, \citet*{zhou2013learning} show that the kernels satisfy an Euler\textendash Lagrange equation and use a Seidel method to solve it numerically. Again, these methods are not applicable to general kernels without a significant computational burden.

\paragraph{Least squares estimation}
Among M-estimation methods for MHP, the log-likelihood is significantly more popular than the least squares functional. To the best of our knowledge, the work of \citet*{reynaud2010adaptive} is the first to introduce this objective for MHP. Their estimation method is meant for piece-wise constant kernels with finite support, with a view towards applications to genomics. \citet{bacry2020sparse} develop an approach for more conventional kernels; namely, linear combinations of exponential kernels with fixed decays. They are interested in dimensionality reduction via sparsity inducing penalties, as the number of kernels in the MHP is quadratic in the number of event types. However, their method is not applicable to general kernels. We review least-square estimation methods in \cref{subsec:least_squares_estimation}.

\subsection{Work outline}
\cref{sec: Background} recalls the definitions of MHP and the least squares estimation framework. In \cref{sec:The Method}, \cref{theorem:lse} provides a decomposition of the LSE as a sum of functions. The rationale behind this decomposition is that, if these functions and their partial derivatives can be evaluated quickly, then a Monte Carlo estimator, of the gradient of the LSE is inexpensive to evaluate. We construct this Monte Carlo estimator using adaptive stratified sampling for variance reduction purposes, allowing for general kernels for the MHP. We combine this estimator with numerical schemes from the stochastic gradient descent literature to propose a new fast estimation method for MHP with general kernels and large datasets. We evaluate our procedure on synthetic data in \cref{sec:Numerics}, and benchmark it against state of the art algorithms. Finally, we give example applications of our method in \cref{sec:applications} using the MemeTracker dataset, and a dataset of malaria infections in China.

\section{Background on MHP estimation}\label{sec: Background}
Before defining MHP, we recall the definition of the larger class of point processes and their conditional intensity. The objective function we use to estimate MHP in this work is the LSE; we recall its definition and discuss the challenges of LSE minimization with standard exact or stochastic first order methods in \cref{subsec:least_squares_estimation}.

\subsection{Hawkes processes}\label{subsec:hawkes_processes}
For any positive integer $n$, we denote by  $[n]$ the set of integers from $1$ to $n$. We fix a dimension $d \in \mathbb{N}^*$ for our process, where $\mathbb{N}^*$ denotes the set of strictly positive integers. 
\subsubsection{Model definition}
Following the notation and terminology in \citet{daley2003introduction}, we briefly introduce point processes. 
\begin{definition}[Point process]\label{def:pointprocess}
A $d$-dimensional orderly point process is a random sequence of times $\mathcal{T}=\{t^i_m: m\in \mathbb{N}^*, i\in [ d ], t^i_m<t^i_{m+1}	\}$. The associated counting process $\boldsymbol{N}$ is defined for $t\geq 0$ by
\begin{equation*}
   \boldsymbol{N_t}:=(N^i_t)_{i\in [ d ]}, \quad \textrm{where} \quad N^i_{t}:=\sum_{i=1}^{+\infty} \mathbbm{1}_{\left\{t \geq t^{j}_{i}\right\}}\,.
\end{equation*}
We denote the total number of jumps up to time $t$ by $N_t:=\sum_{i=1}^d N^i_t$.
\end{definition}
Consider a $d$-dimensional point process $\mathcal{T}$ and the associated counting process $\boldsymbol{N}$. Let $\mathcal{F}=(\mathcal{F}_t)_{t\geq 0}$ be the natural filtration of $\boldsymbol{N}$.
The counting process $\boldsymbol{N}$ is characterized by its conditional intensity, the infinitesimal rate of events per unit of time given the history of the process.
\begin{definition}[Conditional Intensity]
For $i\in [ d ]$, the conditional intensity of $N^i$ is defined by
\begin{equation*}
\lambda_i\left(t \right):=\lim _{h \rightarrow 0^+} \frac{ \mathbb{P}( N^i_{t+h}-N^i_{t}=1 | \mathcal{F}_{t^{-}})}{h} \, ,    
\end{equation*}
and we write $\boldsymbol{\lambda}:=(\lambda_1,\dots,\lambda_d)^\intercal$.
\end{definition}
\begin{definition}[Compensator]
For $t\geq 0$, the compensator of the counting process $N^i$ is defined by
\begin{equation*}
    \Lambda_i(t):=\int_{0}^{t}\lambda_i(s)\mathrm{d}s\,.
\end{equation*}
The Doob--Meyer decomposition of $N^i$ is $N^i=\Lambda_i+M^i$, that is, 
$   M^i_t:=N^i_t-\Lambda_i(t)
$
defines a local martingale.
\end{definition}
We define the MHP model as in \citet{liniger2009multivariate}.
\begin{definition}[MHP]\label{def:linear_mhp}
Let $\boldsymbol{N}$ be a $d$-dimensional counting process with conditional intensity $\boldsymbol{\lambda}$. We say that $\boldsymbol{N}$ is a (linear) MHP if, for all $i \in [ d ]$ and for all $t \geq 0$, we have
\begin{equation}\label{eq:linear_mhp}
    \lambda_i(t)=\mu_i+\sum_{j=1}^d \sum_{\{ m:t^j_m<t \}	}\phi_{ij}(t-t^j_m),
\end{equation}
where 
\begin{itemize}
    \item $\forall i,j\in [ d ]$, $\phi_{ij}:[0,+\infty)\to [0,+\infty)$ is in $L_1$. The functions $\phi_{ij}$ are called the kernels of the MHP, and we write in matrix notation $\Phi(\cdot)=\big(\phi_{ij}(\cdot)\big)_{ij}$.
    \item $\forall i\in [ d ]$, $\mu_i>0$. The scalars $\mu_i$ are called baseline intensities, and we write in vector notation $\boldsymbol{\mu}=(\mu_1, \dots , \mu_d)^\intercal$. 
\end{itemize}
We refer to such a process as a $(\boldsymbol{\mu},\Phi)$-MHP.
\end{definition}

\subsubsection{Examples of kernels}\label{subsubsec:Some kernels}
\begin{figure}%
    \centering
\resizebox{1.\textwidth}{!}{
\includegraphics{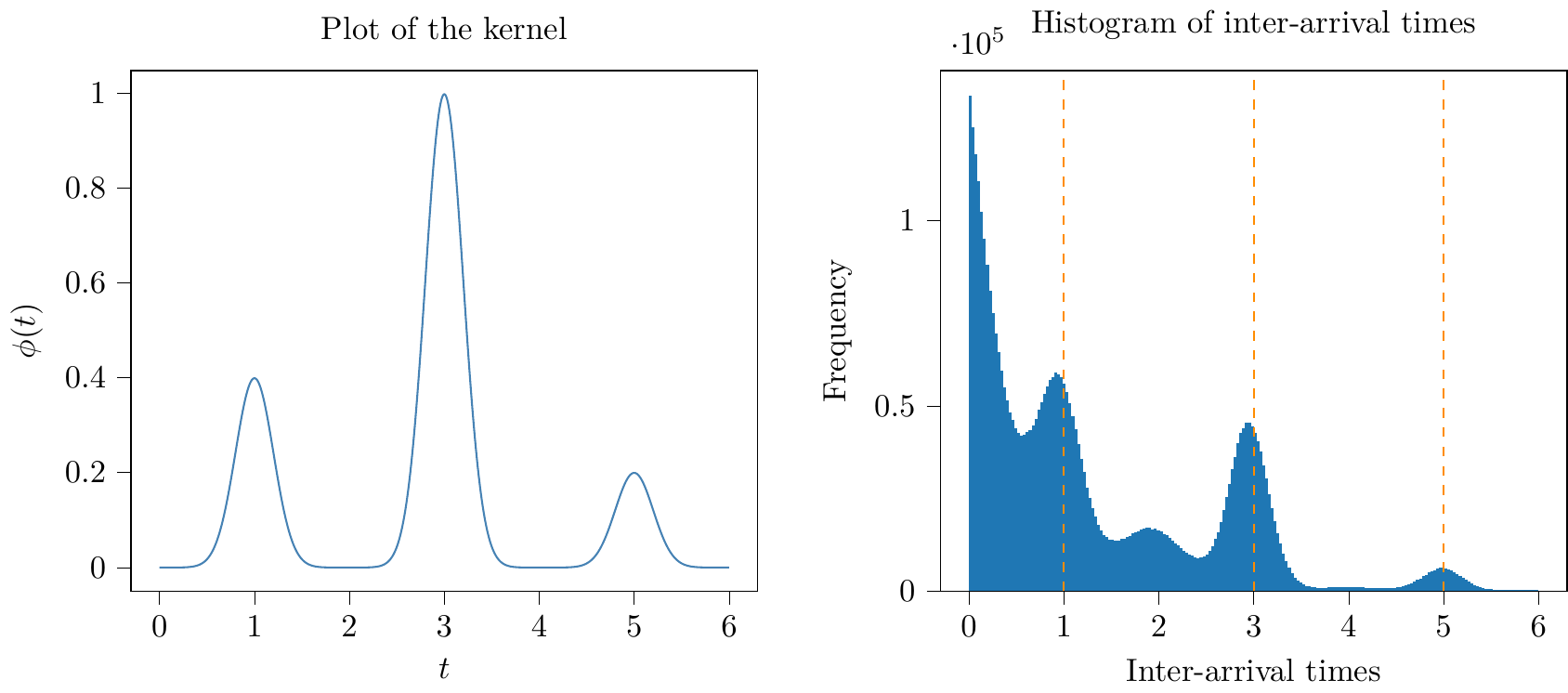}
}
    \caption{Simulated path of a Gaussian MHP}%

    \medskip
\begin{minipage}{0.65\textwidth} 
{\footnotesize Kernel plot and histogram of inter-arrival times of a univariate Hawkes process with a Gaussian kernel, $r=3$. Orange vertical dashed lines correspond to the means of the Gaussian mixture. The apparent mode near $t=2$ corresponds to the difference $3-1$.  \par}
\end{minipage}

    \label{fig:hist_iatimes_delayed_gaussian}%
\end{figure}

This paper is aimed at general parametric classes of kernel models. In both theoretical and numerical applications of MHP, monotonically decaying kernels predominate. In particular, exponential kernels are commonly used in the literature (see the discussion in \cref{subsec:least_squares_estimation}). We now provide some examples which we use to test our method. 
\begin{definition}[Exponential kernel]
Let $r\in \mathbb{N}^*$. For $x\geq 0$, the exponential kernel $\phi^{\mathcal{E}}_{(\boldsymbol{\omega},\boldsymbol{\beta})}$ is
\begin{equation*}
    \phi^{\mathcal{E}}_{(\boldsymbol{\omega},\boldsymbol{\beta})}(x):=\sum_{l=1}^r \omega_l \beta_l e^{-\beta_l x},
\end{equation*}
where the parameters are the vector of weights $\boldsymbol{\omega}:=(\omega_l)_{l\in \llbracket 1, r \rrbracket}\in [0,+\infty)^r$ and the vector of decays $\boldsymbol{\beta}:=(\beta_l)_{l\in \llbracket 1, r \rrbracket}\in (0,+\infty)^r$. 
We have $\| \phi^{\mathcal{E}}_{(\boldsymbol{\omega},\boldsymbol{\beta})} \|_1=\sum_{l=1}^r\omega_l$.
\end{definition}
In nature, there are systems generating streams of data with non-monotonic kernels; for example, when events might trigger each other with some delay. A famous example in the seismology literature is the \emph{linlin} model in \citet{hawkes1974cluster}, which uses Gamma mixture kernels. For computational reasons, we will consider a model built from truncated Gaussian mixture distributions. While the method we propose is designed as a parametric estimation algorithm, non-parametric variations of our algorithm with large Gaussian kernel bases (which allows for general kernels through a standard density argument) are, in principle, possible.
\begin{definition}[Gaussian kernel]
Let $r\in \mathbb{N}^*$. For $x\geq 0$, the Gaussian kernel $\phi^{\mathcal{N}}_{(\boldsymbol{\omega},\boldsymbol{\beta},\boldsymbol{\delta})}$ is
\begin{equation*}
    \phi^{\mathcal{N}}_{(\boldsymbol{\omega},\boldsymbol{\beta},\boldsymbol{\delta})}(x)=\sum_{l=1}^r \omega_l \frac{1}{\beta_l \sqrt{2 \pi}} \exp \left(-\frac{(x-\delta_l)^{2}}{2 \beta_l^{2}}\right),
\end{equation*}
where the parameters are the vector of weights $\boldsymbol{\omega}:=(\omega_l)_{l\in \llbracket 1, r \rrbracket}\in [0,+\infty)^r$, the vector of means $\boldsymbol{\delta}:=(\delta_l)_{l\in \llbracket 1, r \rrbracket}\in \mathbb{R}^r$, and the vector of standard deviations $\boldsymbol{\beta}:=(\beta_l)_{l\in \llbracket 1, r \rrbracket}\in (0,+\infty)^r$.
We have $\| \phi^{\mathcal{N}}_{(\boldsymbol{\omega},\boldsymbol{\beta},\boldsymbol{\delta})} \|_1=\sum_{l=1}^r\omega_l \big(1-F_{\mathcal{N}}(\delta_l / \beta_l ) \big)$, with $F_{\mathcal{N}}$ the standard normal distribution function.
\end{definition}
To demonstrate the effect of non-monotonicity of the kernel on inter-arrival times, we simulate a univariate Hawkes process with a Gaussian kernel. \cref{fig:hist_iatimes_delayed_gaussian} shows a multi-modal empirical distribution of inter-arrival times, with modes near the component means and their harmonics. 
 
 Further examples of kernels suitable for our method can be found in \cref{appendix: kernel specific}.
 
\subsection{Least squares estimation}\label{subsec:least_squares_estimation}
We are interested in estimation of the parameters of a $(\boldsymbol{\mu^{\diamond}},\Phi^{\diamond})-$MHP. In what follows, we fix a horizon $T>0$. We observe a single sample path of jump times $\mathcal{T}_T$ of $\boldsymbol{N}$ on the interval $[0,T] \,$ and we would like to estimate $(\boldsymbol{\mu^{\diamond}},\Phi^{\diamond})$ given these observations. For all $i \in [ d ]$ and $t \leq T$, we consider the intensity model for $\lambda_i$ as in \eqref{eq:linear_mhp}, with candidate baseline and kernel model $(\boldsymbol{\mu},\Phi)$. The idea behind least squares estimation is to chose $(\boldsymbol{\mu},\Phi)$ such that the conditional intensity model $\boldsymbol{\lambda}$ is close to the ground truth $\boldsymbol{\lambda^\diamond}$ in the $L_2$ sense.
\subsubsection{The LSE}
\begin{definition}[Least squares error]
The LSE of the linear MHP model $(\boldsymbol{\mu},\Phi)$ is 
\begin{equation}\label{eq:lse_def_eq}
    \mathcal{R}_T(\boldsymbol{\mu},\Phi):=\frac{1}{T}\sum_{k=1}^d \int_0^T \lambda_k(t)^2\mathrm{d}t-\frac{2}{T}\sum_{k=1}^d\sum_{m=1}^{N^k_T}\lambda_k(t^k_m).
\end{equation}
\end{definition}
Using the Doob\textendash Meyer decomposition of $\boldsymbol{N}$, one can show that 
\begin{equation}\label{eq:rationale lse}
 \mathbb{E}\left[\mathcal{R}_{T}(\boldsymbol{\mu},\Phi)\right]=\frac{1}{T}\sum_{k=1}^d \int_0^T\mathbb{E}\left[(\lambda_k(t)-\lambda^{\diamond}_k(t) )^{2}\right]\mathrm{d}t-\frac{1}{T}\sum_{k=1}^d \int_0^T\mathbb{E}\left[\lambda^{\diamond}_k(t)^{2} \right]\mathrm{d}t,
\end{equation}
where $\boldsymbol{\lambda^{\diamond}}$ is the intensity of the MHP that generates the observations. As highlighted in \citet[Lemma 3]{reynaud2010adaptive}, the main rationale behind the definition of the LSE is that $\mathcal{R}_T$ is a statistical contrast functional. From \eqref{eq:rationale lse} one can see that $ \mathbb{E}\left[\mathcal{R}_{T}(\boldsymbol{\mu},\Phi)\right]$ is minimized if and only if
\begin{equation}\label{eq:intensity_equality}
    \lambda_k(t)=\lambda^{\diamond}_k(t) \quad \textrm{a.s.},\quad  \text{for almost all} \quad t\leq T, k \in 	[d] .
\end{equation}
In this work, we consider parametric models $(\boldsymbol{\mu^{\theta}},\Phi^{\theta})$ and denote by $\Theta$ the parameter space. To simplify notation, we drop the superscript $\theta$ when the dependence on $\theta$ is clear and write $\boldsymbol{\mu},\Phi,\boldsymbol{\lambda}$. By abuse of notation, we see the LSE in \eqref{eq:lse_def_eq} as a function on $\Theta$, and so write $\mathcal{R}_{T}(\theta)$ and consider the estimator 
\begin{equation*}
    \theta^*_T \in \argmin_{\theta\in \Theta}\mathcal{R}_T(\theta).
\end{equation*}
It is not clear whether the estimator $\theta^*_T$ is unbiased or consistent and, if so, at which rate $\theta^*_T$ converges. Nevertheless, this estimator gives satisfactory results in practice, as seen in \citet{reynaud2010adaptive}, \citet{gaiffas2012high}, \citet*{hansen2015lasso}, \citet{bacry2020sparse}, and in our numerical experiments (see \cref{sec:Numerics}).

\subsubsection{Minimizing the LSE}\label{subsubsec:minimize_lse}
Directly minimizing the LSE, for example with exact first order methods, is particularly inefficient without further assumptions because of the auto-regressive structure of the conditional intensity. 
\paragraph{Evaluating the LSE is expensive}
Evaluating the  conditional intensity model $\lambda_i(t)$ based on \eqref{eq:linear_mhp} has a complexity in time which is roughly linear in the number of jumps of the MHP up to time $t$. Therefore, the time complexity to evaluate the second term
\begin{equation}\label{eq:term2_lse}
    \frac{2}{T}\sum_{k=1}^d\sum_{m=1}^{N^k_T}\lambda_k(t^k_m)
\end{equation}
in the LSE in \eqref{eq:lse_def_eq}, is roughly $\mathcal{O}(N_T^2)$ for a general kernel matrix $\Phi$. Now, consider the evaluation of the first term 
\begin{equation}\label{eq:term1_lse}
    \frac{1}{T}\sum_{k=1}^d \int_0^T \lambda_k(t)^2\mathrm{d}t
\end{equation}
in the LSE. If the integral in \eqref{eq:term1_lse} is approximated numerically using a quadrature rule, the cost of such approximation is at least linear since the conditional intensity has a discontinuity at each jump time. If we re-use the evaluations of the conditional intensity at jump times, which we computed for the evaluation of \eqref{eq:term2_lse}, we still face the problem that the conditional intensity varies between jump times. For general kernels, it is not clear how to interpolate the conditional intensity accurately between jump times, therefore, we might also have to evaluate $\boldsymbol{\lambda}$ between consecutive jump times. Another drawback of numerical integration is that the evaluation of the gradient would typically require a finite difference method or operator overloading, increasing the overall complexity.

In a nutshell, evaluating the LSE at a given $\theta$ has complexity $\mathcal{O}(N_T^2)$ in the general case, and the evaluation of the gradient in closed-form is roughly as expensive (if we ignore for now the curse of dimensionality coming from the number of parameters of the model). This makes exact first order methods impractical for general kernels, unless the kernels considered are sum-of-basis exponential functions (see \cref{subsubsec: sbf}).

\paragraph{Towards a SGD approach} A classic strategy to accelerate first order methods is the use of Stochastic Gradient Descent (SGD), which relies on an approximation of the gradient of the objective function; see \citet{robbins1951stochastic}. The application of SGD to the minimization of the LSE in \eqref{eq:lse_def_eq} faces some difficulties. 
For example, for $d=1$, write
\begin{equation}
    \mathcal{R}_T(\theta)=\frac{1}{T}(\mu^2-2\mu)+\frac{1}{T}\int_{t_{N_T}}^T \lambda(t)^2 \mathrm{d}t+\frac{1}{T}\sum_{m=1}^{N_T-1}\bigg( \int_{t_{m}}^{t_{m+1}}\lambda(t)^2 \mathrm{d}t -2\lambda(t_{m+1}) \bigg),
\end{equation}
and define
  \begin{equation}
    f_m(\theta) :=
    \begin{cases*}
      \int_{t_{m}}^{t_{m+1}}\lambda(t)^2 \mathrm{d}t -2\lambda(t_{m+1}) & if $m \in \llbracket 1,N_T-1 \rrbracket$, \\
      \int_{t_{N_T}}^T \lambda(t)^2 \mathrm{d}t        & if $m =N_T $,
    \end{cases*}
  \end{equation}
to yield the decomposition
\begin{equation}
  \mathcal{R}_T(\theta)= \frac{1}{T}(\mu^2-2\mu)+\frac{1}{T} \sum_{m=1}^{N_T} f_m (\theta).
\end{equation}
This decomposition preserves the chronological order of the data because, for all $m$, the function $f_m$ only depends on jump times up to $t_{m+1}$. However, computing $f_m (\theta)$ and its derivatives has complexity roughly of order $\mathcal{O}(m)$. Due to this linear cost, we propose a new additive decomposition of the LSE and build a fast, yet accurate, Monte Carlo estimator of large finite sums. 

\section{SGD for MHP estimation}\label{sec:The Method}
In this section, we state our decomposition of the LSE in \cref{theorem:lse}, which is at the heart of the method we propose. In \cref{subsec:grad_approx}, we build an adaptive stratified Monte Carlo sampling estimator of the gradient of the LSE based on this decomposition. Finally, \cref{subsec:sgd_procedure} presents our numerical scheme.

\subsection{Notation and definitions}
We start with some notation and definitions, noticing that in dimension $d>1$, the chronological ordering of jump times requires some care.  
\begin{definition}
For all $i \in [d]$ and for all $t> 0$, let
\begin{equation*}
\kappa(i,t):=N^i_{t^-},   
\end{equation*}
or equivalently, $\kappa(i,t):=\sup\{m:t^i_m<t \}$. We define, for all $n \in \mathbb{N}^*$, and $i,j \in [d]$,
\begin{equation*}
    \kappa(i,j,n):=\kappa(i,t^j_n);
\end{equation*}
that is, the index of the last jump of type $i$ before the $n$-th jump of type $j$.
\end{definition}
For $i=j$, it is clear that $\kappa(i,j,n)=n-1$. 

\begin{definition}
For all $h\in \mathbb{N}^*$ and $i,j \in [d]$, the upper-inverse of $\kappa(j,i,\cdot)$ is given by
\begin{equation*}
    \varpi(i,j,h):=\inf 	\{p \in \mathbb{N}^* :\kappa(j,i,p)\geq h  \};
\end{equation*}
that is, the index of the earliest jump of type $i$ preceded by at least $h$ jumps of type $j$. For simplicity, we write
\begin{equation*}
    \varpi(i,j):=\varpi(i,j,1).
\end{equation*}
\end{definition}
For $i=j$, it is clear that $\varpi(i,j,h)=h+1$. 

Next, we introduce some quantities of interest, which arise in our computations. For all $i\in [d]$, define the global event rate
\begin{equation}
    \eta^i_T:=\frac{N^i_T}{T}.
\end{equation}
This is both the maximum likelihood estimator and the least squares estimator of the intensity under a homogeneous Poisson model.
For all $i,j \in [d]$, and $t \geq 0$, define
\begin{equation}
    \psi_{ij}(t):=\int_0^t \phi_{ij}(u)\mathrm{d}u.
\end{equation}

\begin{rmk}
Since $\phi_{ij} \geq 0$, we have $\lim_{t\to +\infty}\psi_{ij}(t)=  \| \phi_{ij} \|_1$.
\end{rmk}
For all $i,j,k \in [d]$, and $t,s \geq 0$, define
\begin{equation}
    \Upsilon_{ijk}(t,s):=\int_{0}^{t}\phi_{ki}(u)\phi_{kj}(u+s)\mathrm{d}u.
\end{equation}
\begin{rmk}
Note that $\lim_{t\to +\infty} \Upsilon_{iik}(t,0)=  \| \phi_{ki} \|^2_2$. The function $\Upsilon_{ijk}$ quantifies the correlation (or auto-correlation) between $\phi_{ki}$ and $\phi_{kj}$. 
\end{rmk}

\subsection{Decomposition of the LSE}\label{subsec:lse_decomp}
Assume that the observed sample path of $\boldsymbol{N}$ is non-trivial, in particular, we observe at least one event of each type (i.e. for all $i\in [ d ]$, $N^i_T >1$) and, for each event type, the last observed event of that type is preceded by at least one event of every type (i.e. for all $i,j \in [ d ]$, $\varpi(i,j)<N^i_T$). We state our decomposition of the LSE as a sum involving the functions $(\phi_{ij})$, $(\psi_{ij})$, and $(\Upsilon_{ijk})$, as well as the background rates $(\mu_i)$. 

\begin{theoremEnd}[end, restate,text link={We give a proof of this expansion in \cref{app: proofs}.}]{theorem}[Least squares error]\label{theorem:lse}
The LSE $\mathcal{R}_T$ satisfies
\begin{equation}
\begin{split}
 \mathcal{R}_T(\theta) &= \frac{2}{T}\sum_{k=1}^d\sum_{i=1}^d\sum_{j=1}^d\sum_{m=\varpi(i,j)}^{N^i_T}\sum_{n=1}^{\kappa(j,i,m)}\Upsilon_{ijk}(T-t^i_m,t^i_m-t^j_n)   \\
&\quad -\frac{2}{T}\sum_{k=1}^d\sum_{j=1}^d\sum_{m=\varpi(k,j)}^{N^k_T}\sum_{n=1}^{\kappa(j,k,m)}\phi_{kj}(t^k_m-t^j_n) +\sum_{k=1}^d(\mu_k^2-2 \eta^k_T \mu_k)
\\
&\quad +\frac{2}{T} \sum_{k=1}^d \mu_k \sum_{i=1}^d \sum_{m=1}^{N^i_T} \psi_{ki}(T-t^i_m) +\frac{1}{T}\sum_{k=1}^d\sum_{i=1}^d \sum_{m=1}^{N^i_T}  \Upsilon_{iik}(T-t^i_m,0). 
\end{split}
\end{equation}
\end{theoremEnd}
\begin{proofEnd}
Fix $\theta \in \Theta$. By definition
\begin{equation*}
    \mathcal{R}_T(\theta):=\frac{1}{T}\sum_{k=1}^d \int_0^T \lambda_k(t)^2\mathrm{d}t-\frac{2}{T}\sum_{k=1}^d\sum_{m=1}^{N^k_T}\lambda_k(t^k_m).
\end{equation*}
It is clear that 
\begin{equation*}
    \sum_{k=1}^d\sum_{m=1}^{N^k_T}\lambda_k(t^k_m)= \sum_{k=1}^d N^k_T \mu_k+\sum_{k=1}^d\sum_{j=1}^d\sum_{m=1}^{N^k_T} \sum_{n=1}^{\kappa(j,k,m)}\phi_{kj}(t^k_m-t^j_n).
\end{equation*}
Let $j\in [d]$. By definition
\begin{equation*}
    \begin{split}
\int_0^T \lambda_j(t)^2\mathrm{d}t & = \int_0^T\Bigg(\mu_j^2+2\mu_j\sum_{l=1}^d \varphi_{jk}(t)+\bigg(\sum_{k=1}^d\varphi_{jk}(t)\bigg)^2 \Bigg) \mathrm{d}t , \\
 & = \mu_j^2T+2\mu_j \sum_{k=1}^d C_{jk}(T)+\int_{0}^T \bigg(\sum_{k=1}^d\varphi_{jk}(t)\bigg)^2\mathrm{d}t.
\end{split}
\end{equation*}
We note that 
\begin{equation*}
\int_{0}^T \bigg(\sum_{k=1}^d\varphi_{jk}(t)\bigg)^2\mathrm{d}t = \sum_{k=1}^d \sum_{m=1}^d \int_{0}^{T} \varphi_{jk}(t)\varphi_{jm}(t)\mathrm{d}t.
\end{equation*}
The result follows directly from the application of \cref{lemma:varphi and c} and \cref{lemma: varphi product}.
\end{proofEnd}
Note that the coupling between background rates $\boldsymbol{\mu}$ and kernels $\boldsymbol{\phi}$ is only through the term
\begin{equation*}
   \frac{2}{T}  \sum_{k=1}^d \mu_k \sum_{i=1}^d \sum_{m=1}^{N^i_T}   \psi_{ki}(T-t^i_m),
\end{equation*}
while the coupling between kernels $\boldsymbol{\phi}$ is only through the term
\begin{equation*}
\frac{2}{T}\sum_{k=1}^d\sum_{i=1}^d\sum_{j=1}^d\sum_{m=\varpi(i,j)}^{N^i_T}\sum_{n=1}^{\kappa(j,i,m)}\Upsilon_{ijk}(T-t^i_m,t^i_m-t^j_n)   . 
\end{equation*}
This decomposition guides our choice of parametric families for our algorithm, as we need the functions $(\psi_i)$ and $(\Upsilon_{ijk})$ and their derivatives to be computable efficiently. In \cref{appendix: kernel specific}, we give a selection of families where these functions are available in closed form.

\subsection{Model parameterization}\label{subsec: model param}
For all $i,j \in [d]$, denote the vector of parameters of the kernel $\phi_{ij}$ by $\vartheta_{ij}$, the dimension of $\vartheta_{ij}$ by $\rho_{ij}$, and the total number of parameters of the model by
\begin{equation*}
    n_{\textrm{param}}:=d+\sum_{i=1}^d\sum_{j=1}^d\rho_{ij}.
\end{equation*}
For all $k \in [d]$, concatenate these vectors as $\vartheta^\intercal_{k}=(\vartheta^\intercal_{k1}, \dots , \vartheta^\intercal_{kd})$, and for each $k\in [d]$, define $\theta^\intercal_k=(\mu_k,\vartheta_k^\intercal)$. Finally, define the vector of parameters $\theta$ of the $(\boldsymbol{\mu},\Phi)$-MHP by
\begin{equation*}
    \theta^\intercal=(\theta_1^\intercal, \dots, \theta_d^\intercal).
\end{equation*}
\subsubsection{Parallelization}
For all $k \in [d]$, the intensity $\lambda_k$ only depends on $\theta_k$ and the observed jumps. Define the partial LSE
\begin{equation}
    \mathcal{R}^{(k)}_T(\theta_k):=\frac{1}{T} \int_0^T \lambda_k(t)^2\mathrm{d}t-\frac{2}{T}\sum_{m=1}^{N^k_T}\lambda_k(t^k_m).
\end{equation}
It is clear from the definition of the LSE that $\mathcal{R}_T(\theta)=\sum_{k=1}^d\mathcal{R}^{(k)}_T(\theta_k)$.

Therefore, minimizing $\mathcal{R}_T$ is equivalent to $d$ independent minimization programs that can be solved in parallel. In the remainder of \cref{sec:The Method}, we fix $k\in [d]$ and focus on minimizing $\mathcal{R}^{(k)}_T$, which we re-write using \cref{theorem:lse}.
\begin{corollary}\label{corollary:LSE decomposition per dim}
For all $\theta_{k}$ we have
\begin{equation}\label{eq:lse_decomp_eq_partial}
\begin{split}
 \mathcal{R}^{(k)}_T(\theta_{k}) &= \sum_{i=1}^d\sum_{j=1}^d\sum_{m=\varpi(i,j)}^{N^i_T}\sum_{n=1}^{\kappa(j,i,m)}\frac{2}{T}\Upsilon_{ijk}(T-t^i_m,t^i_m-t^j_n)   \\
&\quad -\left(\frac{2}{T}\sum_{j=1}^d\sum_{m=\varpi(k,j)}^{N^k_T}\sum_{n=1}^{\kappa(j,k,m)}\phi_{kj}(t^k_m-t^j_n)\right) +\mu_k^2-2 \eta^k_T \mu_k
\\
&\quad +\frac{2\mu_k}{T}\sum_{i=1}^d \sum_{m=1}^{N^i_T}  \psi_{ki}(T-t^i_m) +\frac{1}{T}\sum_{i=1}^d \sum_{m=1}^{N^i_T} \Upsilon_{iik}(T-t^i_m,0). 
\end{split}
\end{equation}
\end{corollary}

\subsubsection{A quadratic program: SBF MHP}\label{subsubsec: sbf}
We highlight those MHP models where the kernels are a sum of basis functions (SBF); these are widely used in the literature. Our method applies to general linear MHP, and includes SBF MHP as a special case. In our numerical examples, we observe that SBF models often display better stability than more general parametric families, which may experience a form of `mode collapse' (see Appendix \ref{app:nonSBFmodecollapse}). 

Consider a $(\boldsymbol{\mu},\Phi)$-MHP and assume that for all $i,j \in [d]$
\begin{equation}
    \phi_{ij}=\sum_{l=1}^{r_{ij}}\omega_{ijl}\tilde{\phi}_{ijl},
\end{equation}
where $\tilde{\phi}_{ijl}$ are fixed (known) functions. The parameters of the model are the background intensities $\boldsymbol{\mu}=\{ \mu_i>0: \quad i \in [d] \}$, and the weights $\boldsymbol{\omega}=\{ \omega_{ijl}>0: \quad i,j \in [d] , l\in [r_{ij}]  \}$. 

We refer to this model as a $(\boldsymbol{\mu},\Phi)-$SBF. In this case, the minimization of $\mathcal{R}^{(k)}_T$ is a quadratic program (QP). One might be tempted to first compute the matrices in this QP formulation, then use standard solvers to estimate the model parameters (for example, dual-primal methods, see \citet{vandenberghe2010cvxopt}). The major difficulty is the pre-computation of the QP formulation, which in general, takes quadratic time. This complexity can be reduced by assuming $r_{ij}=r$, $\tilde{\phi}_{ijl}=\tilde{\phi}_{l}$,  for all $i,j,l$, and by choosing $\tilde{\phi}_{l}$ to be an exponential decay. In this case, the pre-computation of the QP formulation will still have linear complexity in time, which is slower than the method we develop in this paper.

\subsection{Gradient approximation}\label{subsec:grad_approx}
Next, for all $\theta_k$, we construct an unbiased estimator of the gradient $\nabla \mathcal{R}^{(k)}_T(\theta_{k})$ of the LSE \eqref{eq:lse_decomp_eq_partial}, which we use as an input to our optimization. To do this, we use variance reduction techniques to construct an efficient unbiased estimator using the additive structure of \eqref{eq:lse_decomp_eq_partial}.

\subsubsection{Monte Carlo approximation problem}
We first provide some notation to deal with the different types of sums involved in \cref{corollary:LSE decomposition per dim}. For the rest of this section, fix $i,j\in [d]$. Let 
\begin{align*} 
\mathcal{A}_T^i &:=  	\{  T-t^i_m : m \in  [ \, N^i_T	] \,	\} ,\\ 
\mathcal{B}_T^{ij} &:= \{  t^i_m-t^j_n : m \in  \llbracket \varpi(i,j) , N^i_T	\rrbracket, n \in [ \, \kappa(j,i,m)	] \,	\}  ,\\ 
\tilde{\mathcal{B}}_T^{ij} &:= \{(T-t^i_m,  t^i_m-t^j_n) : m \in  \llbracket \varpi(i,j) , N^i_T	\rrbracket, n \in [ \, \kappa(j,i,m)	] \,	\} .
\end{align*}
For some domain $\mathcal{E}$ consider a function $f_{\theta_k}:\mathcal{E}\to [0,+\infty)$ parameterized by $\theta_k$. Let $\mathcal{E}_S \subset \mathcal{E}$ be finite, with $N_S:=|\mathcal{E}_S |$, and consider the generic problem of estimating 
\begin{equation*}
    S_T(\theta_k):=\sum_{x\in \mathcal{E}_S} f_{\theta_k}(x).
\end{equation*}
In our problem, we need to address the following configurations:
\begin{enumerate}[i)]
    \item The single sums
\begin{itemize}
    \item $\sum_{m=1}^{N^i_T}  \psi_{ki}(T-t^i_m)$; corresponds to $f_{\theta_k}=\psi_{ki}$, $\mathcal{E}=\mathbb{R}$, $\mathcal{E}_S=\mathcal{A}_T^i $,
    \item $\sum_{m=1}^{N^i_T} \Upsilon_{iik}(T-t^i_m,0)$; corresponds to $f_{\theta_k}=\Upsilon_{iik}(\cdot,0)$, $\mathcal{E}=\mathbb{R}$, $\mathcal{E}_S=\mathcal{A}_T^i $.
\end{itemize}
    \item The double sums
\begin{itemize}
    \item $\sum_{m=\varpi(i,j)}^{N^i_T}\sum_{n=1}^{\kappa(j,i,m)}\phi_{ij}(t^i_m-t^j_n)$; corresponds to  $f_{\theta_k}=\phi_{ij}$, $\mathcal{E}=\mathbb{R}$, $\mathcal{E}_S=\mathcal{B}_T^{ij} $,
    \item $\sum_{m=\varpi(i,j)}^{N^i_T}\sum_{n=1}^{\kappa(j,i,m)}\Upsilon_{ijk}(T-t^i_m,t^i_m-t^j_n)$; corresponds to $f_{\theta_k}=\Upsilon_{ijk}$, $\mathcal{E}=\mathbb{R}^2$, $\mathcal{E}_S=\tilde{\mathcal{B}}_T^{ij} $.
\end{itemize}
\end{enumerate}
The vanilla Monte Carlo approach to this problem is to uniformly sample $N_{MC}<N_S$ elements of $\mathcal{E}_S$, denoted by $\mathcal{E}_{MC}$, and to consider the unbiased estimator
\begin{equation*}
    \hat{S}_T(\theta_k):=\frac{N_S}{N_{MC}}\sum_{x\in \mathcal{E}_{MC}} f_{\theta_k}(x).
\end{equation*}
In practice, only mild variations of this approach are needed to achieve satisfactory Monte Carlo estimation of the single sums, even in the case of a nearly unstable MHP. However, the estimation of the double sums is significantly more challenging and, in practice, vanilla Monte Carlo is too imprecise for our problem because it does not capture the variations of $f_{\theta_k}$ on the domain $\mathcal{E}_S$. For this reason, we develop a stratified sampling approach.
\subsubsection{Estimating the single sums}\label{subsubsec:estimating_the_single_sums}
Let $f_{\theta_k}$ denote $\psi_{ki}$ or $\Upsilon_{iik}(\cdot,0)$. In this case, we want to estimate 
\begin{equation*}
    S_T(\theta_k)=\sum_{m=1}^{N^i_T} f_{\theta_k}(T-t^i_m).
\end{equation*}
Fix $n_{\textrm{max}} \in \mathbb{N}$ and consider a fixed increasing sequence of integers 
\begin{equation*}
    b_{0} <b_{1} < \dots < b_{n_{\textrm{max}}},
\end{equation*}
with $b_{0}:=1$ and $b_{n_{\textrm{max}}}<N^i_T$. The integer intervals $\llbracket b_{p},b_{p+1}-1	\rrbracket$ are the strata of a stratified Monte Carlo estimator of 
\begin{equation*}
    \sum_{m=1}^{b_{n_{\textrm{max}}}-1} f_{\theta_k}(T-t^i_m).
\end{equation*}
Define an unbiased estimator of $S_T(\theta_k)$ by
\begin{equation}\label{eq:estimate_s_single_sum}
    \hat{S}_T(\theta_k):=\sum_{p=1}^{n_{\textrm{max}}-1}\frac{b_{p+1}-b_{p}}{q_p}Z_p+\sum_{m=b_{n_{\textrm{max}}}}^{N^i_T} f_{\theta_k}(T-t^i_m),
\end{equation}
where for each $p$, we sample uniformly $q_p$ integers $(m^{(p)}_1, \dots, m^{(p)}_{q_p})$ in the integer interval $\llbracket b_{p},b_{p+1}-1	\rrbracket$ without replacement and define
\begin{equation}\label{eq:z_p_singlesum}
    Z_p:=\sum_{l=1}^{q_p} f_{\theta_k}(T-t^i_{m^{(p)}_l}).
\end{equation}
We do not set $b_{n_{\textrm{max}}}=N^i_T$ because, even for a stable MHP, there are values of $m$ such that $t^i_m \sim	T$, and these values may contribute significantly to the sum. We summarize this procedure in \cref{algo:single_sum}.
\begin{algorithm}[h!]
\SetAlgoLined
\KwResult{Estimator $\hat{S}_T(\theta_k)$}
Initialize $\hat{S}_T(\theta_k)=\sum_{m=b_{n_{\textrm{max}}}}^{N^i_T} f_{\theta_k}(T-t^i_m)$\;
 \For{$p$ in $	[ \, n_{\textrm{max}-1} ] \,$ }{
  Sample $q_p$ integers $(m^{(p)}_1, \dots, m^{(p)}_{q_p})$\;
  Use \eqref{eq:z_p_singlesum} to compute $Z_p$ \;
  Increment $\hat{S}_T(\theta_k)$ by $\frac{b_{p+1}-b_{p}}{q_p}Z_p$\;
 }
 \caption{Estimation of a single sum}\label{algo:single_sum}
\end{algorithm}

\subsubsection{Estimating the double sums}\label{subsubsec:estimating_the_double_sums}
We now consider the estimation of the terms
\begin{equation*}
 \sum_{m=\varpi(i,j)}^{N^i_T}\sum_{n=1}^{\kappa(j,i,m)}\Upsilon_{ijk}(T-t^i_m,t^i_m-t^j_n),    
\end{equation*}
and
\begin{equation*}
\sum_{m=\varpi(k,j)}^{N^k_T}\sum_{n=1}^{\kappa(j,k,m)}\phi_{kj}(t^k_m-t^j_n),
\end{equation*}
which appear in \eqref{eq:lse_decomp_eq_partial}. We briefly outline our method, and give a detailed description in \cref{app_subsec:double_sums}. Approaches similar to those used to estimate the single sums fail because of the larger number of terms in these sums, many of which make very small contributions, leading to a high variance in a simple estimation. Ideally, one could use stratified sampling of the data by time differences $t^i_m-t^j_n$ to reduce the variance of the estimator $\hat{S}(\theta_k)$. However, constructing strata based directly on the time differences would require pre-computation of the density of time differences, which is computationally expensive, both in time and memory. Instead, we use stratified sampling by index differences as a proxy for time differences, which is significantly faster and memory efficient. Indeed, this approach does not store any additional data. 

For event indices $m,n$ such that $t^j_n<t^i_m$, the difference $\kappa(j,i,m)-n$ is the number of events of type $j$ in the interval $(t^j_n,t^i_m)$. We refer to the quantity $h=\kappa(j,i,m)-n+1$ as the lag between indices $m,n$.
For all $h \in 	[ \, \kappa(j,i,N^i_T) ] \,$, define the sets of event times with a given lag in their indices
\begin{align*} 
\mathcal{B}_T^{ij,h} &:= \{  t^i_m-t^j_n : m \in  \llbracket \varpi(i,j) , N^i_T	\rrbracket, n = \kappa(j,i,m)-h+1	\}  ,\\ 
\tilde{\mathcal{B}}_T^{ij,h} &:= \{(T-t^i_m,  t^i_m-t^j_n) : m \in  \llbracket \varpi(i,j) , N^i_T	\rrbracket, n = \kappa(j,i,m)-h+1	\} .
\end{align*}
For each element of the sets $\mathcal{B}_T^{ij,h}$ and $\tilde{\mathcal{B}}_T^{ij,h}$, the indices $(m,n)$ are such that $t^j_n$ is the $h-$th to last jump of type $j$ before $t^i_m$. Let $\mathcal{E}_T^{ij,h}$ denote $\mathcal{B}_T^{ij,h}$ or $\tilde{\mathcal{B}}_T^{ij,h}$ as appropriate for the sum under consideration. Define
\begin{equation*}
    S^h_T(\theta_k):=\sum_{x\in \mathcal{E}_T^{ij,h}} f_{\theta_k}(x), \quad \text{hence} \quad S_T(\theta_k):=\sum_{h=1}^{\kappa(j,i,N^i_T)} S^h_T(\theta_k).
\end{equation*}

\begin{figure}[h]
    \centering
\resizebox{0.55\textwidth}{!}{
\includegraphics{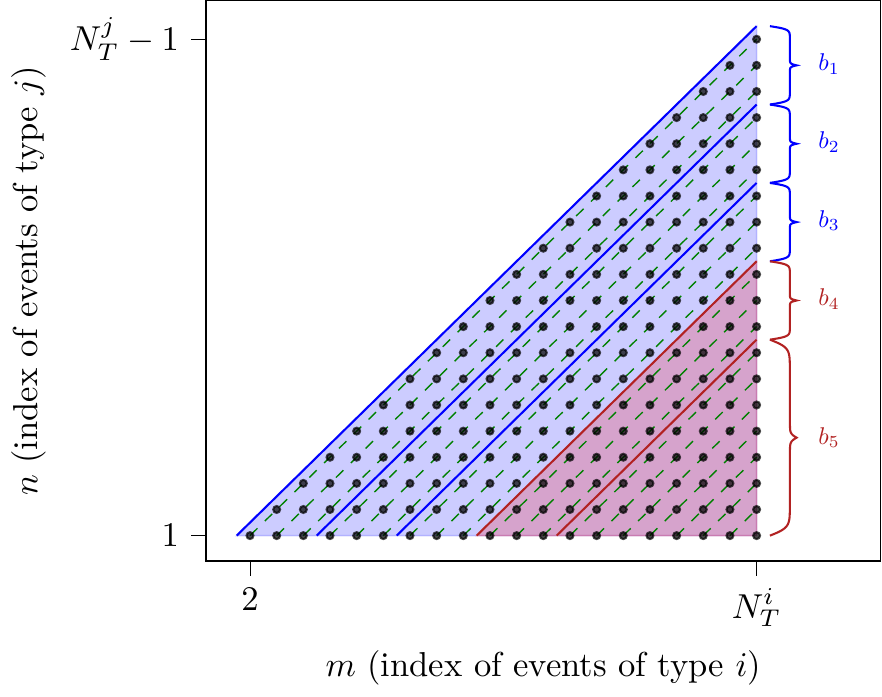}
}
    \caption{Illustration of our stratification.}
    
\medskip
\begin{minipage}{0.65\textwidth} 

{\footnotesize Here $i=j$ and $N^i_T=20$. Black points are the points $(m,n)$ such that $1\leq n < m \leq N^i_T$. Dashed green lines are the lines of equation $m-n=h$ for $h\in [N^i_T-1]$. With $h_{\mathrm{max}}=9$, the blue (respectively red) domain contains all the points $(m,n)$ such that $m-n \leq h_{\mathrm{max}}$ (respectively $m-n > h_{\mathrm{max}}$). In this example, we consider $3$ (respectively $2$) strata in the blue (respectively red) domain, $(b_1,b_2,b_3)$ (respectively $(b_4,b_5)$), delimited by solid blue (respectively red) lines. The Monte-Carlo estimator on the blue domain is constructed adaptively, whereas it is non adaptive in the red domain. \par}
\end{minipage}

    \label{fig:double_sum_stratification}
\end{figure}

 The sets $(\mathcal{E}_T^{ij,h})_h$ form a partition of $\mathcal{E}_T^{ij}$ of size $\kappa(j,i,N^i_T)$, which is typically still too large to use as a stratification.\footnote{For example, one can think of the case $j=i$, where $\kappa(j,i,N^i_T)=N^i_T-1$.} Our heuristic for this estimator is to first note that
 \begin{equation*}
 \lim_{t\to\infty} \phi_{ij}(t) = 0 \quad \text{and} \quad \lim_{s\to\infty} \Upsilon_{ijk}(t,s) = 0 .
 \end{equation*}
We expect the contribution of $S^h_T(\theta_k)$ to  $S_T(\theta_k)$ to decrease after a certain index $h_{\textrm{max}}$, as the sequence $(\min \mathcal{E}_T^{ij,h})_h$ is strictly decreasing. Hence, we focus on the estimation of 
\begin{equation*}
    S^{\mathrm{max}}_T(\theta_k):=\sum_{h=1}^{h_{\textrm{max}}} S^h_T(\theta_k)
\end{equation*}
separately from the estimation of the remainder
 \begin{equation*}
    S^{\mathrm{rest}}_T(\theta_k):=\sum_{h=h_{\textrm{max}}+1}^{\kappa(j,i,N^i_T)} S^h_T(\theta_k).
\end{equation*}
We group several index lag sets to reduce the number of strata. Formally, let $n_B \in \mathbb{N}^*$. Consider a partition $B=(\boldsymbol{b_1}, \dots , \boldsymbol{b_{n_B}} ) $ of $[ \, h_{\textrm{max}} 	] \,$. By abuse of notation, for all $\boldsymbol{b} \in B$ take the disjoint union and sum
\begin{equation*}
 \mathcal{E}_T^{ij,\boldsymbol{b}} := \bigcupdot_{h\in \boldsymbol{b}} \mathcal{E}_T^{ij,h} , \quad    S^{\boldsymbol{b}}_T(\theta_k) := \sum_{h\in \boldsymbol{b}} S^{h}_T(\theta_k).
\end{equation*}
\cref{fig:double_sum_stratification} illustrates the construction of this stratification in the case $i=j$.

For all $p\in [n_B]$, sample $q^p$ points $(x^{\boldsymbol{b_p}}_1, \dots,  x^{\boldsymbol{b_p}}_{q^p})$ uniformly and without replacement from $\mathcal{E}_T^{ij,\boldsymbol{b_p}}$ , and define an unbiased estimator of $S^{\boldsymbol{b_p}}_T(\theta_k)$ by
\begin{equation}
\hat{S}^{\boldsymbol{b_p},q_p}_T(\theta_k)=\frac{	|  \mathcal{E}_T^{ij,\boldsymbol{b}}	|}{q^p}\sum_{m=1}^{q^p} f_{\theta_k}(x^{\boldsymbol{b_p}}_m).    
\end{equation}
Fix in advance the total number $Q$ of points we want to sample
\begin{equation*}
    Q:=\sum_{p=1}^{n_B} q^p.
\end{equation*}
Let $\boldsymbol{q}:=(q^1, \dots, q^p)$ denote the absolute allocation. For all $p \in [ \, n_B ] \,$, define the relative allocation $\tilde{q}^p:=q^p/Q$, and write $\boldsymbol{\tilde{q}}=(\tilde{q}^1,\dots,\tilde{q}^{n_B})$. An unbiased estimator of $S^{\mathrm{max}}_T(\theta_k)$ is 
\begin{equation}
\hat{S}^{\boldsymbol{q}}_T(\theta_k)=\sum_{p=1}^{n_B} \hat{S}^{\boldsymbol{b_p},q_p}_T(\theta_k).    
\end{equation}
In practice, one could fix $\boldsymbol{q}$ a priori, and choose $h_{\mathrm{max}}$ of order $30$ to $50$, for example. We see in \cref{sec:Numerics} that this leads to satisfactory results in a variety of cases, particularly for monotonically decaying kernels. However, for general kernels, this approach is not sufficiently robust; we instead adaptively determine the allocation of sampled points per stratum in \cref{app_subsec:double_sums}. 
As in the single sum case, use a standard stratified Monte Carlo approach with a fixed allocation to estimate the remainder 
 \begin{equation*}
    S^{\mathrm{rest}}_T(\theta_k):=\sum_{h=h_{\textrm{max}}+1}^{\kappa(j,i,N^i_T)} S^h_T(\theta_k).
\end{equation*} 
We denote the estimator of the remainder by $\hat{S}^{\mathrm{rest}}_T(\theta_k)$, and the estimator of the sum $S_T(\theta_k)$ by
\begin{equation}\label{eq:double_sum_estimator}
\hat{S}_T(\theta_k)    =\hat{S}^{\boldsymbol{q}}_T(\theta_k)+\hat{S}^{\mathrm{rest}}_T(\theta_k).
\end{equation}
This procedure is summarized in \cref{algo:double_sum}.
\begin{algorithm}[h!]
\SetAlgoLined
\KwResult{Estimator $\hat{S}_T(\theta_k)$}
 Initialize $\boldsymbol{\tilde{q}^{(0)}_*}(\theta_k)$\;
 \For{$s$ in $	[ \, n_K ] \,	$}{

   \For{$p$ in $	[ \, n_B ] \,	$}{
   Set $\Delta q^{\boldsymbol{b_{p}}}_s =\tilde{q}^{\boldsymbol{b_{p}},(s-1)}_*(\theta_k)(n_B+\Delta Q_s)$\;
  Sample without replacement $\Delta q^{\boldsymbol{b_{p}}}_s$ points $(x^{\boldsymbol{b_p}}_m)_m$ in $\mathcal{E}_T^{ij,\boldsymbol{\boldsymbol{b_p}}}$\;
  Use \eqref{eq:delta_updates_double_sums} to compute $\hat{S}^{\boldsymbol{b_{p}},(\Delta s)}_T (\theta_k)$ and $\hat{\sigma}^{\boldsymbol{b_{p}},(\Delta s)}_T(\theta_k)^2$ \;
  Use \eqref{eq:updates_double_sums}  to compute $\hat{S}^{\boldsymbol{b_{p}},(s)}_T(\theta_k)$ and $ \hat{\sigma}^{\boldsymbol{b_{p}},(s)}_T(\theta_k)^2$ \;
 }
 }
 Use \eqref{eq:remainder_double_sums} to compute $\hat{S}^{\mathrm{rest}}_T(\theta_k)$  \;
 Use \eqref{eq:double_sum_estimator} to compute $\hat{S}_T(\theta_k)$ \;
 \caption{Estimation of a double sum}\label{algo:double_sum}
\end{algorithm}

\subsection{The procedure}\label{subsec:sgd_procedure}
\subsubsection{The gradient estimator} 
We use the sum estimators constructed above to define the gradient estimate $\boldsymbol{\mathcal{G}^{(k)}_T}(\theta_{k})$. For all $i \in [d]$, denote the sum estimates by
\begin{itemize}
    \item $\hat{S}_{\psi,ki,T}(\theta_{k})$ the estimator of $ \sum_{m=1}^{N^i_T}   \psi_{ki}(T-t^i_m)$,
    \item $\hat{S}_{\Upsilon,ik,T}(\theta_{k})$ the estimator of $ \sum_{m=1}^{N^i_T}   \Upsilon_{iik}(T-t^i_m)$,
    \item $\hat{S}_{\phi,kj,T}(\theta_{k})$ the estimator of $\sum_{m=\varpi(k,j)}^{N^k_T}\sum_{n=1}^{\kappa(j,k,m)}\phi_{kj}(t^k_m-t^j_n)$,
    \item $\hat{S}_{\Upsilon,ijk,T}(\theta_{k})$ the estimator of $\sum_{m=\varpi(i,j)}^{N^i_T}\sum_{n=1}^{\kappa(j,i,m)}\Upsilon_{ijk}(T-t^i_m,t^i_m-t^j_n)$.
\end{itemize}
We denote by $\boldsymbol{\mathcal{G}^{(k)}_T}|_{\mu_k}$ the $\mu_k$ component of $\boldsymbol{\mathcal{G}^{(k)}_T}$ (the estimator of the partial derivative of the partial LSE $\mathcal{R}^{(k)}_T$ with respect to $\mu_k$), defined by
\begin{equation}
\boldsymbol{\mathcal{G}^{(k)}_T}|_{\mu_k}(\theta_{k})=   2\bigg(\mu_k- \eta^k_T+\frac{1}{T}\sum_{i=1}^d \hat{S}_{\psi,ki,T}(\theta_{k}) \bigg).
\end{equation}
For $p \in [d]$ and $l\in [ \,  \rho_{ij} ] \,$, let $\vartheta_{kpl}$ be the $l$-th parameter of $\phi_{kp}$. Denote by $\boldsymbol{\mathcal{G}^{(k)}_T}|_{\vartheta_{kpl}}$ the $\vartheta_{kpl}$ component of $\boldsymbol{\mathcal{G}^{(k)}_T}$ (the estimator of the partial derivative of the partial LSE $\mathcal{R}^{(k)}_T$ with respect to $\vartheta_{kpl}$) defined by
\begin{equation}
\begin{split}
\boldsymbol{\mathcal{G}^{(k)}_T}|_{\vartheta_{kpl}}(\theta_{k}) = \frac{2}{T}\sum_{i=1,i\neq p}^d & \frac{\partial \hat{S}_{\Upsilon,ipk,T}}{\partial \vartheta_{kpl}}+\frac{\partial \hat{S}_{\Upsilon,pik,T}}{\partial \vartheta_{kpl}}  +\frac{2}{T}\frac{\partial \hat{S}_{\Upsilon,ppk,T}}{\partial \vartheta_{kpl}}-\frac{2}{T}\frac{\partial \hat{S}_{\phi,kp,T}}{\partial \vartheta_{kpl} }\\
&+ \frac{2\mu_k}{T}   \frac{\partial \hat{S}_{\psi,kp,T}}{\partial \vartheta_{kpl}} +\frac{1}{T}\frac{\partial \hat{S}_{\Upsilon,pk,T}}{\partial \vartheta_{kpl}}.       
\end{split}
\end{equation}

\begin{rmk}
To give a concise expression for the complexity in time of the computation of the estimate $\boldsymbol{\mathcal{G}^{(k)}_T}(\theta_{k})$, suppose that
\begin{itemize}
    \item the number of parameters per kernel is constant; i.e.  there exists $\rho\in \mathbb{N}$ such that for all $i,j \in [ d ]$, we have $\rho_{ij}=\rho$,
    \item the total sample size for the estimation of each single sum is constant, denoted $Q^{(1)}$,
    \item the total sample size for the estimation of each double sum is constant, denoted $Q^{(2)}$.
\end{itemize}
Under these assumptions, the complexity in time of the computation of a gradient estimate is roughly of order $\mathcal{O}(\rho d^2 Q^{(2)}+\rho dQ^{(1)})$.
\end{rmk}
\subsubsection{Numerical scheme}
Let $n_{\textrm{iter}} \in \mathbb{N}^*$ denote the number of iterations in the optimization procedure. We initialize the parameters at a value $\theta_k^{(0)}$, which can be chosen in a deterministic way or randomly sampled. 

We use classic stochastic gradient methods to build a sequence $\big(\theta_k^{(t)}\big)_{t \in [n_{\textrm{iter}}] }$. Consider methods of the form 
\begin{equation*}
   \theta^{(t+1)}_k=\mathrm{proj}_{\Theta_k}\left(  \theta^{(t)}_k+\Delta \theta^{(t+1)}_k \right),
\end{equation*}
where $\Delta \theta^{(0)}_k=0$ and $\big(\Delta \theta_k^{(t)}\big)_{t \in [n_{\textrm{iter}}] }$ is a sequence that depends on the gradient estimate $\boldsymbol{\mathcal{G}^{(k)}_T}\big(\theta^{(t)}_k\big)$. In the SGD literature, there exists a variety of constructions of the sequence $\big(\Delta \theta_k^{(t)}\big)_{t \in [n_{\textrm{iter}}]}$, leading to numerical schemes with very different properties. In this work, we consider the ADAM algorithm, proposed by \citet{kingma2015adam}, which is an adaptive learning rate method known to outperform most other state of the art algorithms on several machine learning tasks; see \cref{app:insights_on_grad_estimator} for more details.

\subsection{Complexity comparisons with other methods}
\cref{table:recap_methods} summarizes some features of our algorithm in comparison with existing techniques. The \textbf{Algorithm} column contains the reference of the algorithm, and its name if it has been named in the publication. \textbf{Parametric} specifies whether this is a parametric or non-parametric method. \textbf{Complexity} is the time complexity of a method, using the notation of this paper. Some algorithms consider a discretization of the kernels, we denote by $n_{\mathrm{res}}$ the resolution of this discretization. In case the algorithms contain an inner loop, we denote its number of iterations by $n^\prime_{\textrm{iter}}$. We denote by $n_{\textrm{samples}}$ the number of observed sample paths of the MHP in case the method considers several sample paths. \textbf{Assumptions} refers to the additional assumptions made on the MHP (\texttt{SBF exp.} for an SBF MHP with exponential kernels, \texttt{SBF uni.} for an SBF MHP with $r=1$ and a unique basis function $\tilde{\phi}_{ij}=\tilde{\phi}$, \texttt{exp.} for an MHP with exponential kernels). \textbf{Type} is either the type of objective function used (\emph{LSE} for Least Squares Error, \emph{LL} for log-likelihood, \emph{LL-EM} for marginal likelihood in an EM framework) or \emph{MM} in the case of the method of moments. \textbf{Regularization} refers to the type of penalty used in the algorithm, if any. In case no penalty is used in the paper, but could be incorporated to the method with mild modifications, we write an asterisk. 

\begin{table}[p]
\centering
\renewcommand{\arraystretch}{1.25}
\begin{adjustbox}{width=0.4\textwidth,center=\textwidth}
\begin{sideways}
\begin{tabular}{||c c c c c c||} 
 \hline
\textbf{Algorithm} & \textbf{Parametric} & \textbf{Complexity} & \textbf{Assumptions} & \textbf{Type} & \textbf{Regularization} \\ [0.5ex] 
 \hline\hline
ASLSD (This paper) & param. & $\mathcal{O}(n_{\textrm{iter}}\rho d^2 Q^{(2)}+n_{\textrm{iter}}\rho dQ^{(1)})$ & - & LSE & * \\ 
   \hline
 \cite{bacry2020sparse} & param. & $\mathcal{O}(N_T \rho^2 d+n_{\textrm{iter}} \rho^2 d^3)$ & SBF exp. & LSE & Sparsity, Low Rank\\
  \hline
MF \cite{bacry2016mean}& param. & $\mathcal{O}(d^{2} r \Lambda T \times \max (\Lambda T, d \rho)+d^4 \rho^3) $ & SBF, Stable, & LL & * \\
  & &  & Mean field & & \\
  \hline
 (SumExp) \cite{bompaire2018dual} & param. & $\mathcal{O}(\rho N_T^2)$ & SBF exp. & LL &  *\\
  \hline

 \cite{veen2008estimation} & param. & $\mathcal{O}(n_{\textrm{iter}}N_T^2)$ & exp. & LL-EM & - \\
  \hline
 MPLE \cite{lewis2011nonparametric} & non-param. & $\mathcal{O}(n_{\textrm{iter}}N_T^2)$ & -  & LL-EM & Good's penalty \\
  \hline
 ADM4 \cite{zhou2013adm4} & param. & $\mathcal{O}(d^3n_{\textrm{iter}}+d^2N_Tn_{\textrm{samples}}n_{\textrm{iter}}n^\prime_{\textrm{iter}}+n_{\textrm{samples}}N_T^2)$ & SBF uni. & LL-EM & Sparsity, Low rank\\ 
  \hline
 MMEL \cite{zhou2013learning} & non-param. & $\mathcal{O}(\rho N_T^3 d^2 n_{\textrm{iter}} +\rho n_{\mathrm{res}}n_{\textrm{iter}} (d N_T + N_T^2) )$ & - & LL-EM & Kernel smoothing\\
  \hline
 MLE-SGLP \cite{xu2016learning} & param. & $\mathcal{O}(\rho N_T^3 d^2 n_{\textrm{iter}})$ & SBF & LL-EM & Sparse, Lasso, \\
 & & & & & Pairwise similarity \\
  \hline
(WH) \cite{bacry2016first} & non-param. & $\mathcal{O}(N_T d^{2} n_{\mathrm{res}}+d^{4}n_{\mathrm{res}}^{3})$ & Stable &  Autocovariance & - \\

  \hline

  NPHC \cite{achab2017uncovering} & param. & $\mathcal{O}(N_T d^2+n_{\textrm{iter}} d^3)$ & Stable & Integrated & - \\ 
  &  & &  & cumulants &  \\  [1.ex] 

 \hline
\end{tabular}
\end{sideways}
\end{adjustbox}

\caption{Comparison of the computational complexity of our algorithm \texttt{ASLSD} with state of the art estimation of MHP. Our two baseline cases are denoted by \texttt{SumExp} and \texttt{WH}, both here and in subsequent sections.}
    \medskip
\begin{minipage}{0.65\textwidth} 
{\footnotesize  \par}
\end{minipage}

\label{table:recap_methods}
\end{table}

\section{Numerical experiments }\label{sec:Numerics}
In this section, we evaluate our estimation procedure on data simulated with an exact cluster based algorithm. To reproduce the results in this section, see the code in the \emph{Experiments} folder of \url{https://github.com/saadlabyad/aslsd}.
\paragraph{Evaluation metrics}
Consider a $(\boldsymbol{\mu^{\diamond}},\Phi^{\diamond})-$MHP observed on a window $[ \, 0,T ] \,	$, and a model $(\boldsymbol{\mu},\Phi)$. We define metrics to evaluate the performance of our algorithms:
\begin{itemize}
    \item \texttt{L2RelErr} : We define this metric by
\begin{equation}
    \texttt{L2RelErr}:=\frac{\norm{ \boldsymbol{\mu^{\diamond}}-\boldsymbol{\mu} }^2_2}{\norm{ \boldsymbol{\mu^{\diamond}} }^2_2}+\frac{\norm{ \boldsymbol{\Phi^{\diamond}}-\Phi }^2_2}{\norm{ \boldsymbol{\Phi^{\diamond}} }^2_2}, \quad \textrm{where} \quad  \norm{ \boldsymbol{\Phi} }^2_2:=\sum_{i,j}\int_0^{+\infty}\phi_{ij}^2(t)\mathrm{d} t.
\end{equation}
    \item \texttt{WassErr}: The (first) Wasserstein distance between probability measures $f$ and $g$ is given by
    \begin{equation*}
        \mathbb{W}_1(f,g):=\inf _{\pi \in \Gamma(f, g)} \int_{[0,+\infty) \times [0,+\infty)}|x-y| \mathrm{d} \pi(x, y),
    \end{equation*}
where $\Gamma(f, g)$ is the space of measures on $[0,+\infty)\times [0,+\infty)$ with marginals $f,g$. Define 
\begin{equation} 
 \texttt{WassErr} :=  \sum_{i=1}^d 	| \mu_i^\diamond-\mu_i 	| + \sum_{i=1}^d \sum_{j=1}^d  \mathbb{W}_1 \left(\frac{\phi^{\diamond}_{ij}}{||\phi^{\diamond}_{ij} ||_1},\frac{\phi_{ij}}{||\phi_{ij} ||_1} \right)  +\sum_{i=1}^d \sum_{j=1}^d  \left| ||\phi^{\diamond}_{ij} ||_1-||\phi_{ij} ||_1 \right| .
\end{equation}
\end{itemize}

\paragraph{Goodness-of-fit} 
Residual analysis is the state of the art goodness-of-fit test for MHP. This analysis relies on the time transformation property given by \citet[Section 3.3]{ogata1988statistical}. For all $m\in \mathbb{N}^*, k\in [ d ] $, let
\begin{equation*}
    s^k_{m}=\Lambda_k(t^k_m).
\end{equation*}
For each $k\in [ d ] $, define the point process $\mathcal{S}^k:=\{s^k_m: m\in \mathbb{N}^*\}$; then $(\mathcal{S}^k)_{k\in [ d ] }$ are independent standard Poisson processes. The inter-arrival times of $\mathcal{S}^k$ (`residuals'), for a model that fits the data well must therefore be close to a standard exponential distribution. To assess if fitted residuals satisfy this property, we display the Q-Q plots of the residuals against a standard exponential distribution. As visual comparison of the fit of different models can be difficult, we also use the probability plot of residuals, which are defined by
\begin{equation*}
    z^k_{m}=1-\exp{(-s^k_{m})}, \quad \textrm{and} \quad  \mathcal{Z}^k:=\{z^k_m: m\in \mathbb{N}^*\}.
\end{equation*}
If the residuals $\mathcal{S}^k$ are exponentially distributed, then $(\mathcal{Z}^k)_{k\in [ d ] }$ are independent random variables, uniformly distributed on $[0,1]$. For improved clarity, we subtract the $y=x$ line from the probability plots and rescale $z^k_{m}$ with a multiplicative factor $\sqrt{N^k_T-1}$; for large $N^k_T$, Donsker's theorem indicates that this results in a process which is approximately a Brownian bridge. In each probability plot, we plot dashed lines corresponding to the $99\%$ critical value for the Kolmogorov--Smirnov test.

\paragraph{Benchmarks} We compare the performance of our algorithm to the following state of the art methods:
\begin{itemize}
    \item \texttt{SumExp}: an SBF exponential MHP model, fitted using the algorithm in \citet{bompaire2018dual}. This is an interesting benchmark to evaluate the quality of the fit for the decay parameter $\beta$ in a non SBF exponential using our method.
    \item \texttt{WH}: the algorithm proposed in \citet{bacry2016first}, a non-parametric estimation method which solves a  Wiener--Hopf system derived from the autocovariance of the MHP. This method applies to any stable MHP.
\end{itemize}
These two algorithms are implemented in the python \texttt{tick} package \cite{bacry2017tick}, which we use for comparison.
In \cref{subsec:data}, we specify the parameter values for the various MHP that we consider, and we plot paths of the solver to illustrate the performance of our estimation algorithm. In \cref{subsec:results}, we compare the results of our algorithm with those of our benchmarks for each evaluation metric. 

\subsection{Data}\label{subsec:data}
For each kernel type, we simulate one path of each MHP and fit the corresponding models, to illustrate the trajectory of our solver.
\subsubsection{Exponential kernels}
\begin{figure}[htp]
    \centering
    \includegraphics[width=0.6\textwidth]{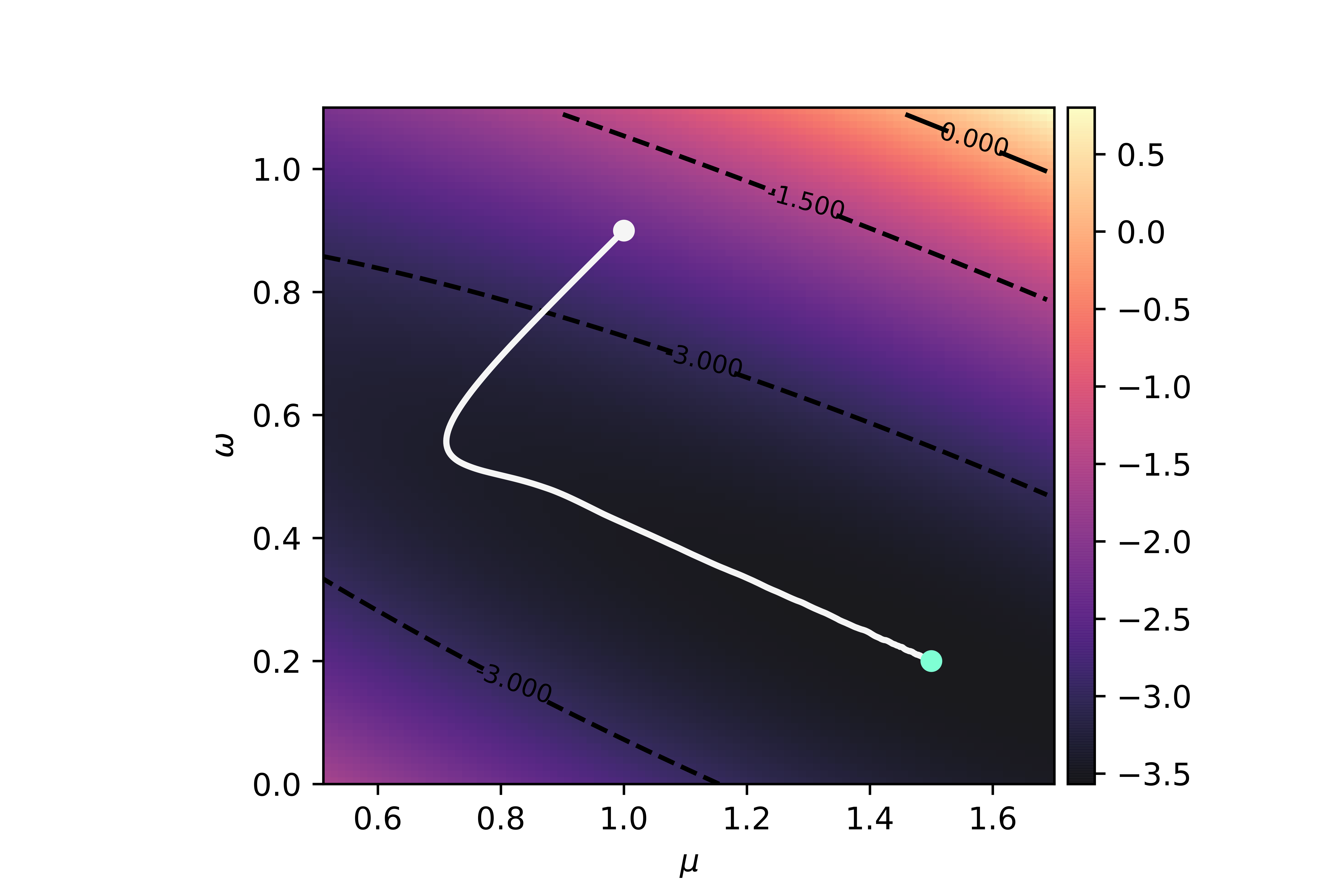}
    \caption{Contour plot of the LSE and SGD updates in the \texttt{SbfExp1D} model.}
\begin{minipage}{0.65\textwidth} 
{\footnotesize White circle is the randomly chosen initial point. Green circle is the true value of the parameters used for simulation. White line is the trajectory of parameter estimates using our algorithm. \par}
\end{minipage}

    \label{fig:SbExp1D_contour_plot}
\end{figure}

\begin{figure}[h]
    \centering
\resizebox{1.\textwidth}{!}{
\includegraphics{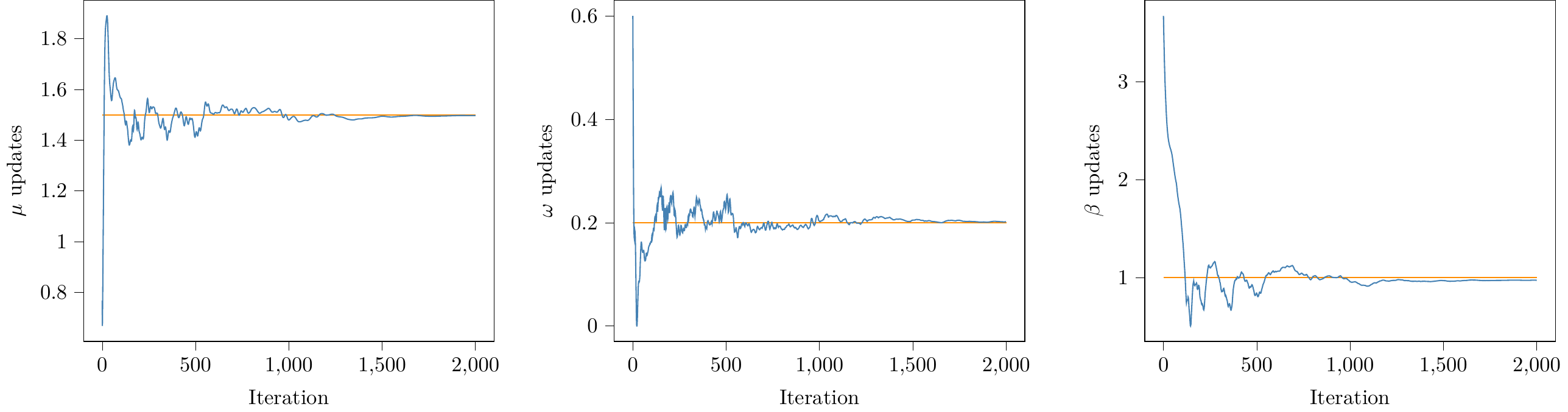}
}
    \caption{SGD updates in the \texttt{Exp1D} model.}
\begin{minipage}{0.65\textwidth} 
{\footnotesize Orange lines correspond to the true parameter values in the parameter updates column (left), and to zero in the gradient updates column (right). \par}
\end{minipage}
    \label{fig:summary_stats_1d_exp}
\end{figure}

\paragraph{Univariate case}
Consider a univariate MHP with exponential kernel, with true parameters
\begin{equation*}
    \mu^{\diamond}=1.5, \quad \omega^{\diamond}=0.2, \quad \beta^{\diamond}=1.
\end{equation*}
For this ground truth MHP, we consider two estimation models:
\begin{itemize}
    \item \texttt{SbfExp1D}: SBF exponential model as in \cref{subsubsec: sbf}, with only one basis function ($r=1$), and fix $\beta=\beta^{\diamond}$. We only estimate $\mu^{\diamond},\omega^{\diamond}$.  
    \item \texttt{Exp1D}: exponential model as in \cref{subsubsec:Some kernels}, and estimate $\mu^{\diamond},\omega^{\diamond},\beta^{\diamond}$.
\end{itemize}
We simulate one path of this process up to $T=10^7$, which results in 18,754,765 jumps. We fit the \texttt{SbfExp1D} model and plot the path of our solver in \cref{fig:SbExp1D_contour_plot}.
We fit the model \texttt{Exp1D} and plot the updates of the parameters and the estimates of the partial derivatives of the LSE in \cref{fig:summary_stats_1d_exp}.

\begin{figure}[htp]
    \centering
\resizebox{0.8\textwidth}{!}{
\includegraphics{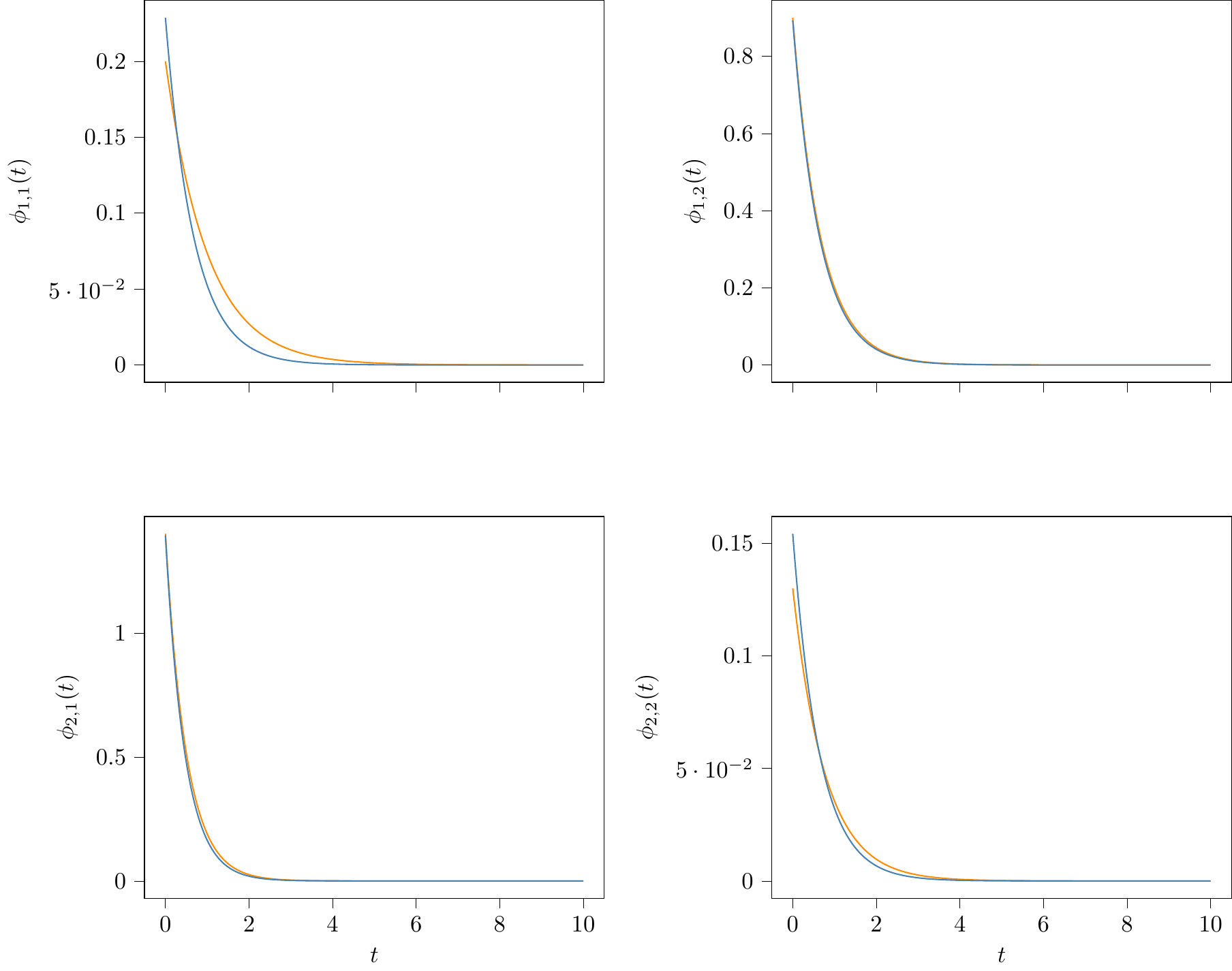}
}
    \caption{Fitted kernels for the \texttt{Exp2D} model.}

    \label{fig:Exp2D_phis}
\end{figure}

\paragraph{Multivariate case}
Consider a bivariate MHP with exponential kernels, and true parameters
\begin{equation*}
 \boldsymbol{\mu^{\diamond}}=\begin{pmatrix}
1.5  \\
1 
\end{pmatrix}, \quad
\boldsymbol{\omega^\diamond} =   \begin{pmatrix}
0.2 & 0.6 \\
0.7 & 0.1 
\end{pmatrix}, \quad \boldsymbol{\beta^\diamond} =   \begin{pmatrix}
1 & 1.5 \\
2 & 1.3 
\end{pmatrix}.
\end{equation*}
For this ground truth MHP, we consider an estimation model \texttt{Exp2D} which is an exponential model as in \cref{subsubsec:Some kernels}. We simulate one path of this process up to $T=10^5$, which results in $1,098,456$ jumps. We fit the \texttt{Exp2D} model; \cref{fig:Exp2D_phis} plots the estimated kernels.

\subsubsection{Gaussian kernels}\label{subsubsec:numexp_gaussker}
\begin{figure}[htp]
    \centering
\resizebox{0.8\textwidth}{!}{
\includegraphics{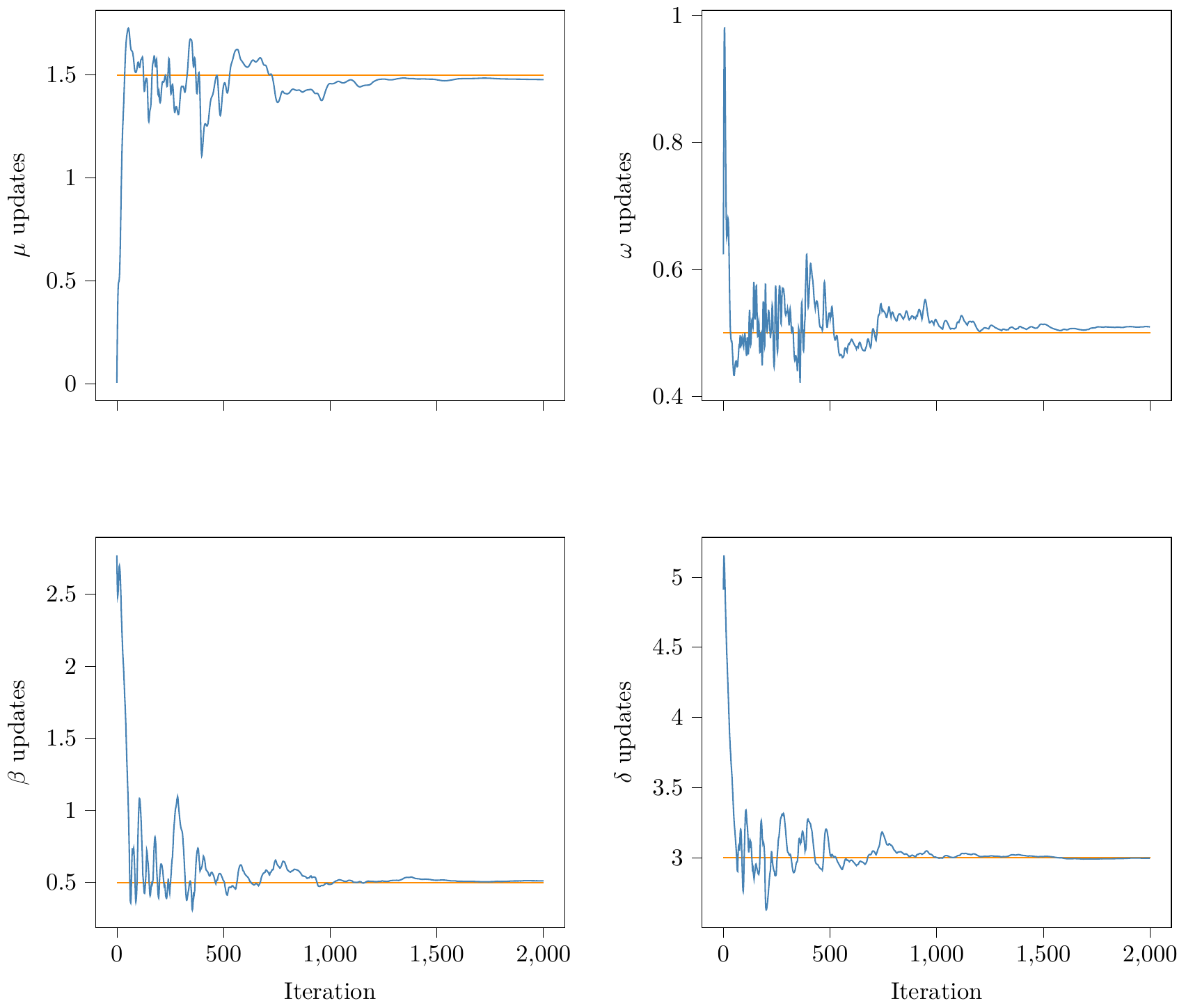}
}
    \caption{SGD updates in the \texttt{Gauss1D} model.}
\begin{minipage}{0.65\textwidth} 
{\footnotesize Orange lines correspond to the true parameter values. \par}
\end{minipage}
    \label{fig:param_updates_1d_gauss}
\end{figure}

\paragraph{Univariate unimodal case} Consider a univariate MHP with Gaussian kernel, and with true parameters
\begin{equation*}
    \mu^{\diamond}=1.5, \quad \omega^{\diamond}=0.5, \quad \beta^{\diamond}=0.5, \quad \delta^{\diamond}=3.
\end{equation*}
For this ground truth MHP, we consider the \texttt{Gauss1D} estimation model, which is a Gaussian model as in \cref{subsubsec:Some kernels}. We simulate one path of this process up to $T=10^6$, which results in $3,005,742$ jumps, and use our procedure to estimate $\mu^{\diamond},\omega^{\diamond},\beta^{\diamond},\delta^{\diamond}$.  We fit the \texttt{Gauss1D} model and provide more insight on the estimation of each parameter in \cref{fig:param_updates_1d_gauss}.

\begin{figure}%
    \centering
\resizebox{1.\textwidth}{!}{
\includegraphics{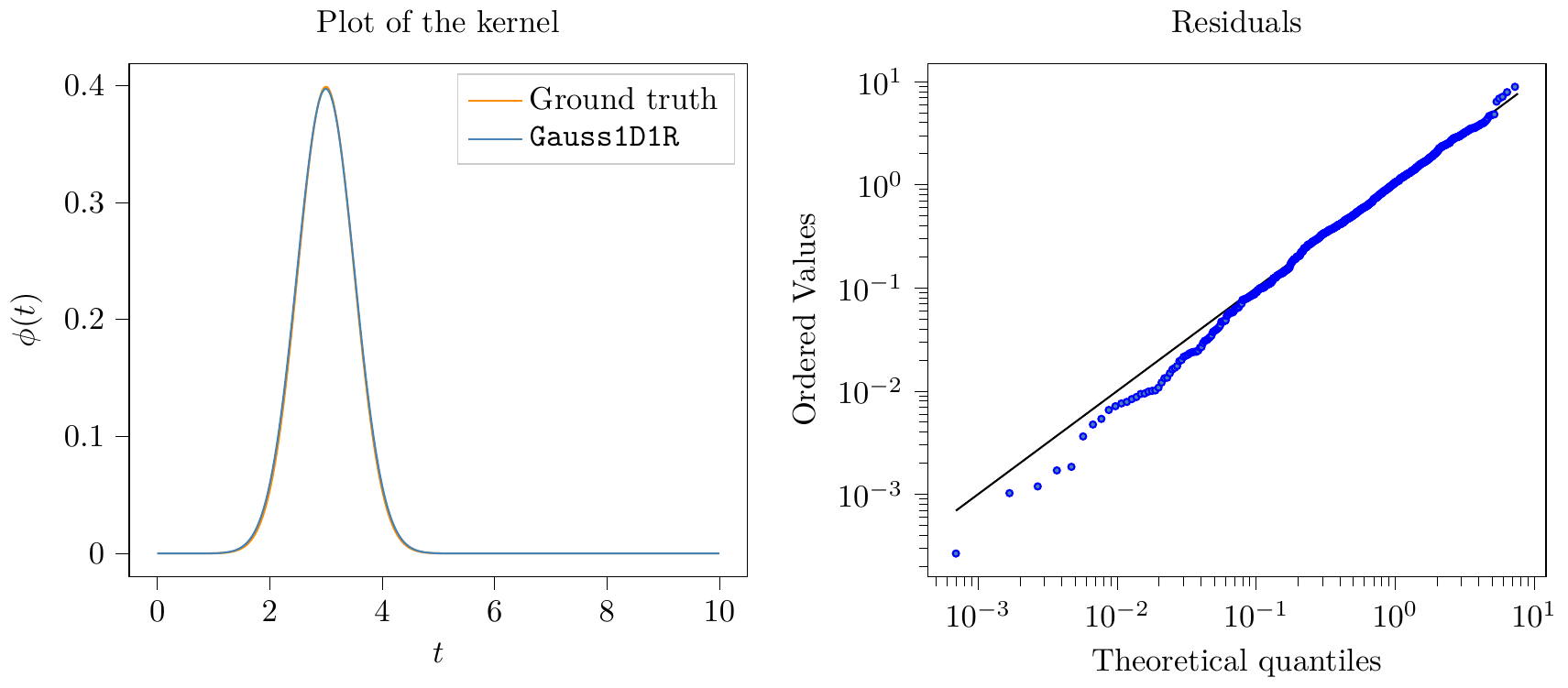}
}
    \caption{Fit results for the \texttt{Gauss1D} model.}%

    \medskip
\begin{minipage}{0.65\textwidth} 
{\footnotesize   \par}
\end{minipage}

    \label{fig:phi_and_residuals_gauss1d}%
\end{figure}

\paragraph{Univariate multimodal case}
Consider a univariate MHP with a basis of three Gaussian kernels, with true background rate $\mu^{\diamond}=0.01$ and with true kernel parameters
\begin{align*}
\omega^{\diamond}_{1}&=0.2 & \beta^{\diamond}_{1}&=0.4  &  \delta^{\diamond}_{1}&=1\\
\omega^{\diamond}_{2}&=0.3 & \beta^{\diamond}_{2}&=0.6  &  \delta^{\diamond}_{2}&=3\\
\omega^{\diamond}_{3}&=0.4 & \beta^{\diamond}_{3}&=0.8  &  \delta^{\diamond}_{3}&=8.
\end{align*}

For this ground truth MHP, we consider the \texttt{SbfGauss1D10R} estimation model, which is a SBF Gaussian model as in \cref{subsubsec: sbf}, with ten basis function ($r=10$), where for all $l \in [10]$ set $\beta_{l} =0.5$ and $\delta_{l}=l-1$. To reflect typical usage, the model is somewhat misspecified, as the true kernel mixture cannot be expressed in terms of the SBF family.

\begin{figure}%
    \centering
\resizebox{1.\textwidth}{!}{
\includegraphics{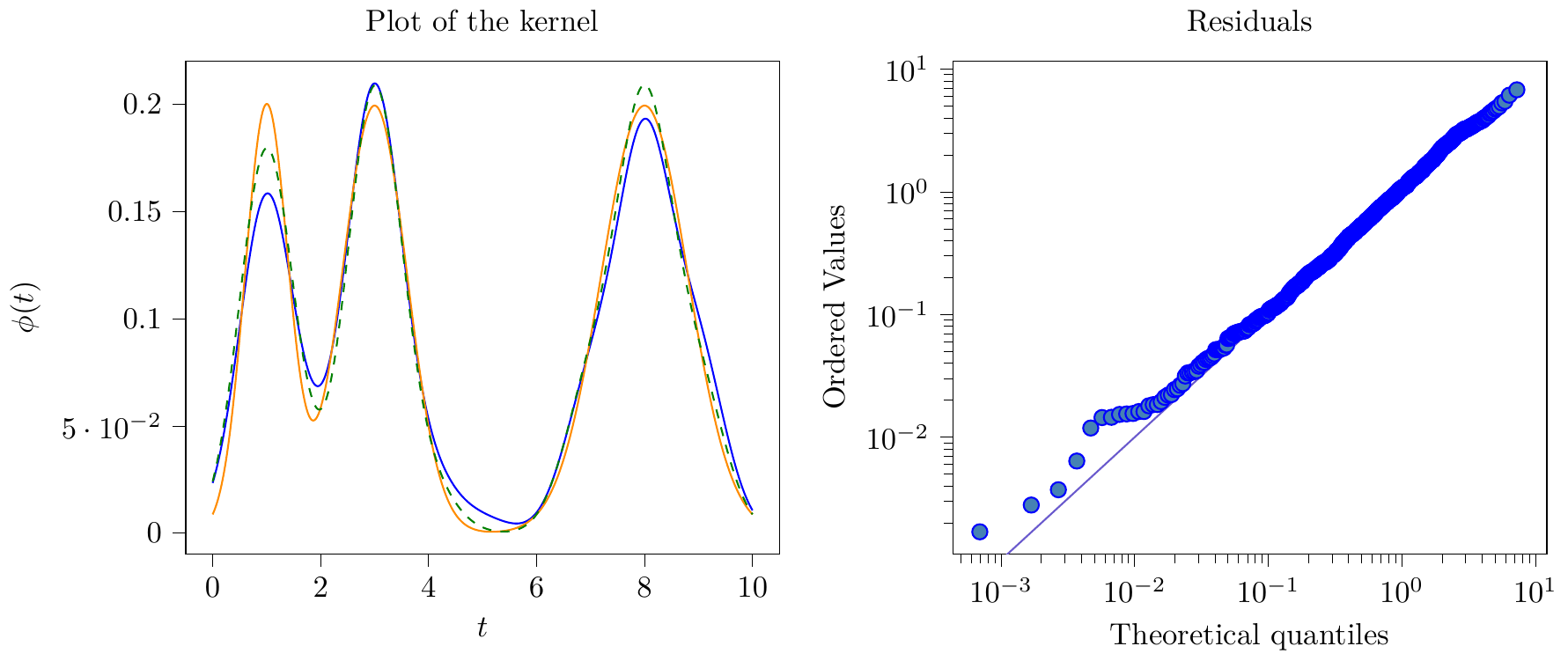}
}
    \caption{Fit results for the  \texttt{SbfGauss1D10R} model.}%

    \medskip
\begin{minipage}{0.65\textwidth} 
{\footnotesize Left: blue line corresponds to the fitted \texttt{SbfGauss1D10R} model using \texttt{ASLSD}, orange lines to the ground truth, and green lines to the $L^2$ projection of the ground truth MHP on the parametric space of kernels of \texttt{SbfGauss1D10R}. Right: Q-Q plot of fitted residuals. \par}
\end{minipage}

    \label{fig:fit_gauss1d2r}%
\end{figure}

We simulate one path of this process up to $T=10^7$, which results in $986,996$ jumps, and use our procedure to estimate the background rate $\mu^{\diamond}$ and the kernel parameters $\left(\omega^{\diamond}_l\right)_{l\in [3]}$.  We fit the \texttt{SbfGauss1D10R} model and provide more insight on the fitted kernel and residuals in \cref{fig:fit_gauss1d2r}.

We have also considered fitting this model using a non-SBF approach. This exhibited a persistent problem of mode collapse -- after sufficient training, all the components of the mixture coalesce, leading to a poorly fitting model. We discuss this in more detail in \cref{app:nonSBFmodecollapse}.

\subsection{Performance for various sample sizes}\label{subsec:results}
\begin{figure}[htp]
    \centering
\resizebox{1.\textwidth}{!}{
\includegraphics{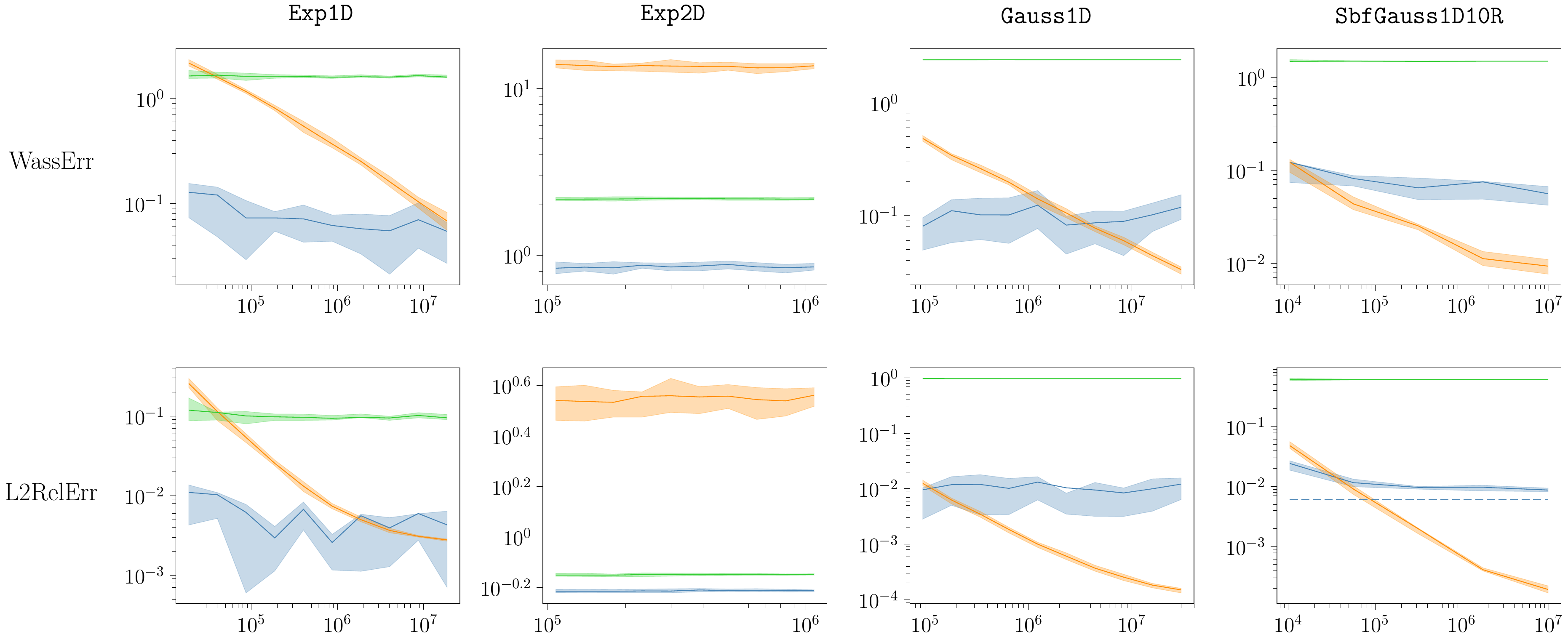}
}

    \caption{Performance of \texttt{ASLSD} and benchmarks.}
\begin{minipage}{0.65\textwidth} 
{\footnotesize Error plots for the various ground truth MHP. Blue lines correspond to our algorithm, orange lines to the \texttt{WH} benchmark, and green lines to the \texttt{SumExp} benchmark. In the lower-rightmost plot, dashed blue line is the lower bound of the \texttt{L2RelErr} between the ground truth MHP and a \texttt{SbfGauss1D10R} MHP. This lower bound corresponds to the \texttt{L2RelErr} between the ground truth MHP and the $L^2$ projection of the ground truth MHP on the parametric space of kernels of \texttt{SbfGauss1D10R}. \par}
\end{minipage}
    \label{fig:error_plots_final}
\end{figure}

For each ground truth MHP $(\boldsymbol{\mu^{\diamond}},\Phi^\diamond)$ specified earlier, we simulate $n_{\textrm{paths}}$ paths $(\mathcal{T}^{(p)})_{p\in [ n_{\textrm{paths}} 	] }$ of the process up to a terminal horizon $T$. We consider $(T_q)_{q\in [ n_{\textrm{times}} 	] }$, an increasing sequence of times. For each $(\boldsymbol{\mu^{\diamond}},\Phi^\diamond)$, for each simulated path $\mathcal{T}^{(p)}$, and for each integer $q\in [ n_{\textrm{times}} 	] $, we define
\begin{equation*}
    \mathcal{T}^{(p,q)}:=\mathcal{T}^{(p)}\cap 	[ \,0, T_q	] \,^d ;
\end{equation*}
i.e., the path of the process truncated at $T_q$, containing $N^{(p)}_{T_q}$ jumps. Finally, for each ground truth MHP, for each evaluation metric, and for each of the three algorithms considered (our method and the two benchmarks), we fit a given MHP model to the observations $\mathcal{T}^{(p,q)}$ and compute the error $\epsilon^{(p,q)}$. For each time discretization step $q$, we compute the empirical mean (resp. $25$th percentile and $75$th percentile) of $(\epsilon^{(p,q)})_{p\in [ n_{\textrm{paths}} 	] }$ and denote it by $M^{(q)}$ (resp. $Q_{0.25}^{(q)}$ and $Q_{0.75}^{(q)}$). Define the mean number of jumps per path 
\begin{equation*}
 \tilde{N}_q:=\frac{1}{n_{\textrm{paths}}}\sum_{p=1}^{n_{\textrm{paths}}}N^{(p)}_{T_q}.  
\end{equation*}
\cref{fig:error_plots_final} plots the mean errors $(M^{(q)})_{q\in [n_{\textrm{times}}] }$ against the mean number of jumps $(\tilde{N}_q)_{q\in [n_{\textrm{times}}] }$, and the shaded area between the lower quartiles $(Q_{0.25}^{(q)})_{q\in [n_{\textrm{times}} ] }$ and the upper quartiles $(Q_{0.75}^{(q)})_{q\in [n_{\textrm{times}}] }$. We see that our algorithm outperforms the \texttt{WH} benchmark with respect to each evaluation metric in the two exponential cases, $\texttt{Exp1D}$ and $\texttt{Exp2D}$. In the unimodal Gaussian case $\texttt{Gauss1D}$, \texttt{ASLSD} outperforms \texttt{WH} for smaller datasets (under $10^6$ jumps) in the Wasserstein metric \texttt{WassErr}. For the multimodal Gaussian example \texttt{SbGauss1D10R}, \texttt{WH} typically outperforms \texttt{ASLSD}, however this is due to the (deliberate) misspecification of the SBF Gaussian model which is seen to approach its lower $L^2$ error bound. Our procedure consistently outperforms the \texttt{SumExp} benchmark for all ground truths.

\section{Applications}\label{sec:applications}
In this section, we apply our estimation procedure to real world data. To reproduce the results in this section, see code in the \emph{Applications} folder of \url{https://github.com/saadlabyad/aslsd}.

\subsection{News propagation}
\paragraph{Data}
In this application, we are interested in the diffusion of information across different media platforms. \citet*{gomez2013structure} compiled news articles from several websites that mention a selection of keywords into the MemeTracker dataset. There exist different versions of the MemeTracker dataset, with different data and different structures; we are interested in the one proposed by \citet{gomez2012memetrackerurl}. Each file lists the times of occurrences of a given keyword in a variety of media  outlets. The primary focus of the MemeTracker dataset is the analysis of event cascades: the occurrences of keywords happen through generic sentences shared in websites (referred to as memes). In this application, we do not take an event cascade viewpoint on the data in MemeTracker, but rather model the excitation in the occurrences of a given keyword irrespective of the meme to which it belongs to.

We model mentions of the keyword related to the British Royal family, a few months after the wedding of Prince William and Catherine Middleton on $29$ April $2011$.\footnote{The exact keyword is \emph{prince william-william-kate middleton-kate-middleton-westminster-watch-marriage-queen-king-elizabeth-charles}.} We limit ourselves to data between $1$ November $2011$ at midnight UTC and $1$ March $2012$ at midnight UTC. The data is timestamped in Unix time in hours with a second resolution. When two or more posts have the same timestamp, we only keep the event that appears first in the dataset.

\begin{figure}[htp]
    \centering
\resizebox{0.8\textwidth}{!}{
\includegraphics{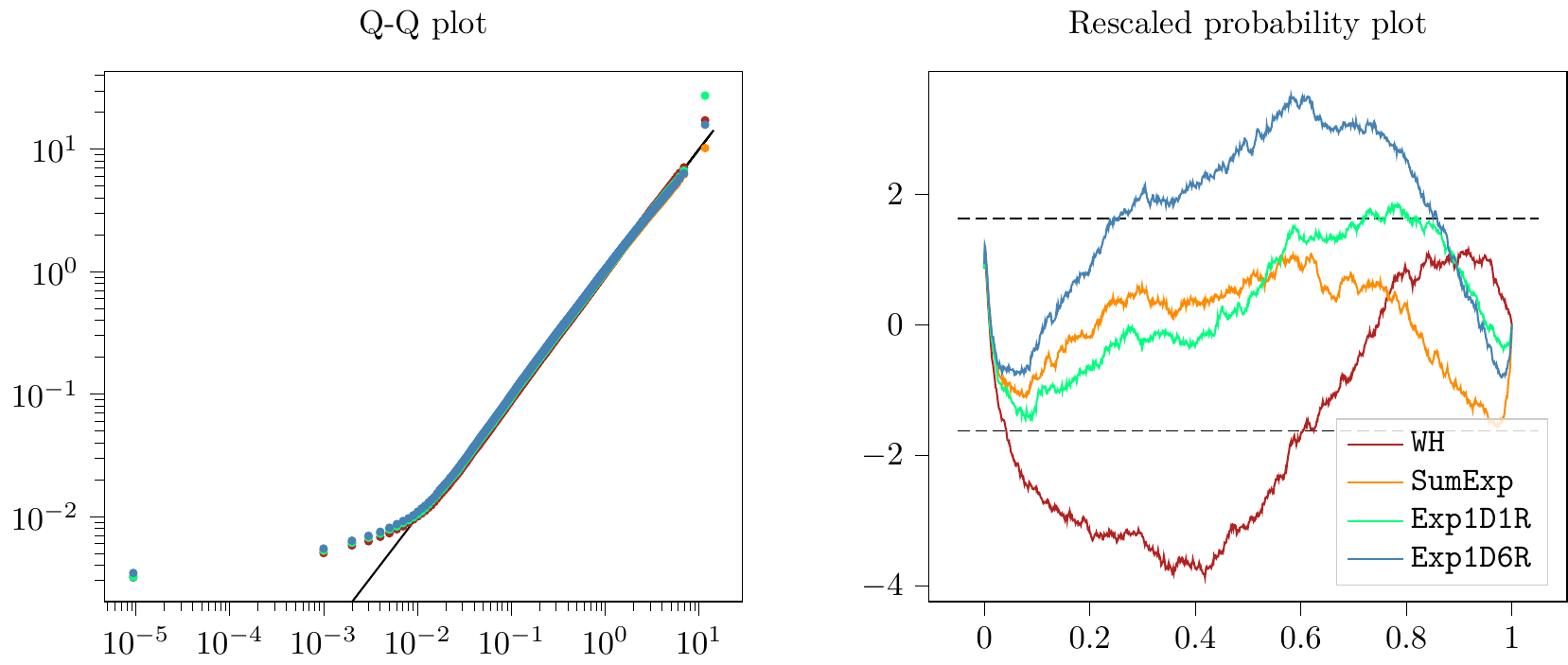}
}
    \caption{Residual plots for the fitted models.}%
    \label{fig:memetracker1d_residual_plots}%
\end{figure}

\begin{figure}[htp]
    \centering
\resizebox{0.45\textwidth}{!}{
\includegraphics{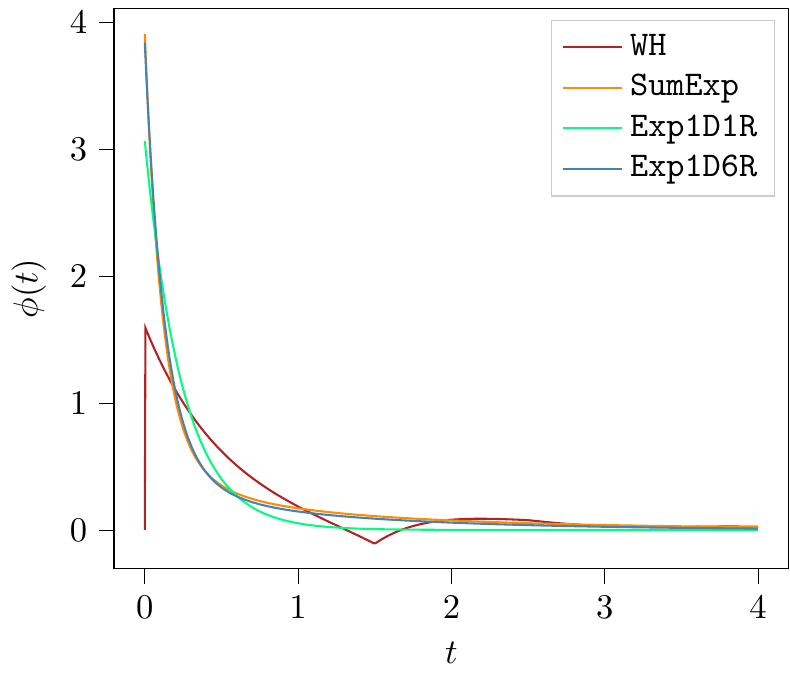}
}
    \caption{Fitted kernels.}
    \label{fig:memetracker1d_fitted_kernels}
\end{figure}

\paragraph{Univariate} First, we aggregate all publication timestamps into a path of a one-dimensional point process. We use our method to fit two exponential models: with one exponential (\texttt{Exp1D1R}) and with six exponentials (\texttt{Exp1D6R}). For benchmark purposes, we fit \texttt{SumExp} (with 6 fixed exponential decays) and \texttt{WH}. \cref{fig:memetracker1d_residual_plots} displays the probability plots of the residuals for each model, showing that our method is competitive with state of the art algorithms. \cref{fig:memetracker1d_fitted_kernels} plots the fitted kernels.

\begin{figure}[htp]
    \centering
\resizebox{0.8\textwidth}{!}{
\includegraphics{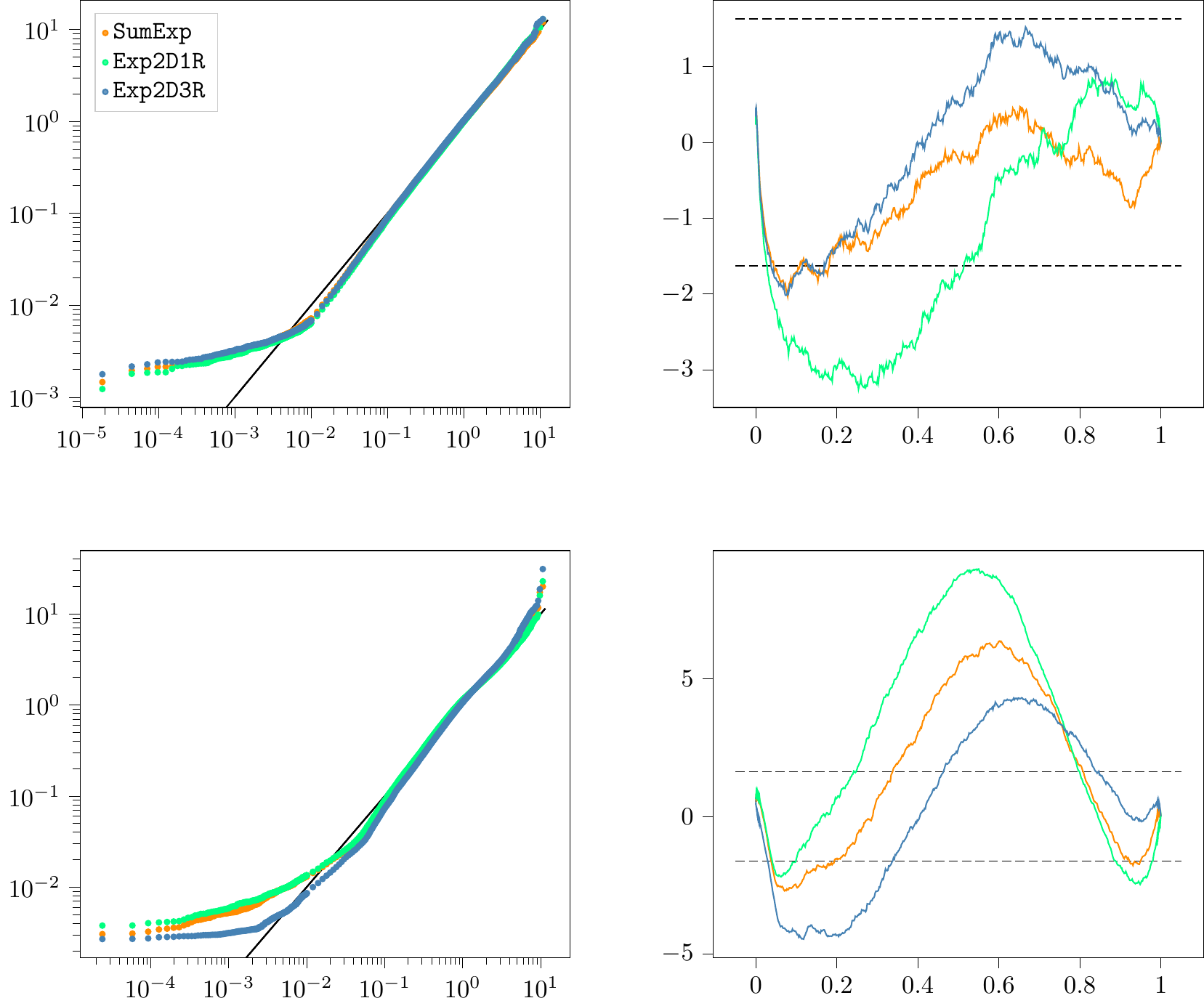}
}
    \caption{Residual plots.}
\begin{minipage}{0.65\textwidth} 
{\footnotesize Q-Q plots (left column) and rescaled probability plots (right column) for the US (upper row) and UK publications (lower row). \texttt{WH} displays very poor goodness of fit and is omitted for clarity.\par}
\end{minipage}
    \label{fig:app_rw2d_residuals}
\end{figure}

\begin{figure}[ht]
    \centering
\resizebox{0.8\textwidth}{!}{
\includegraphics{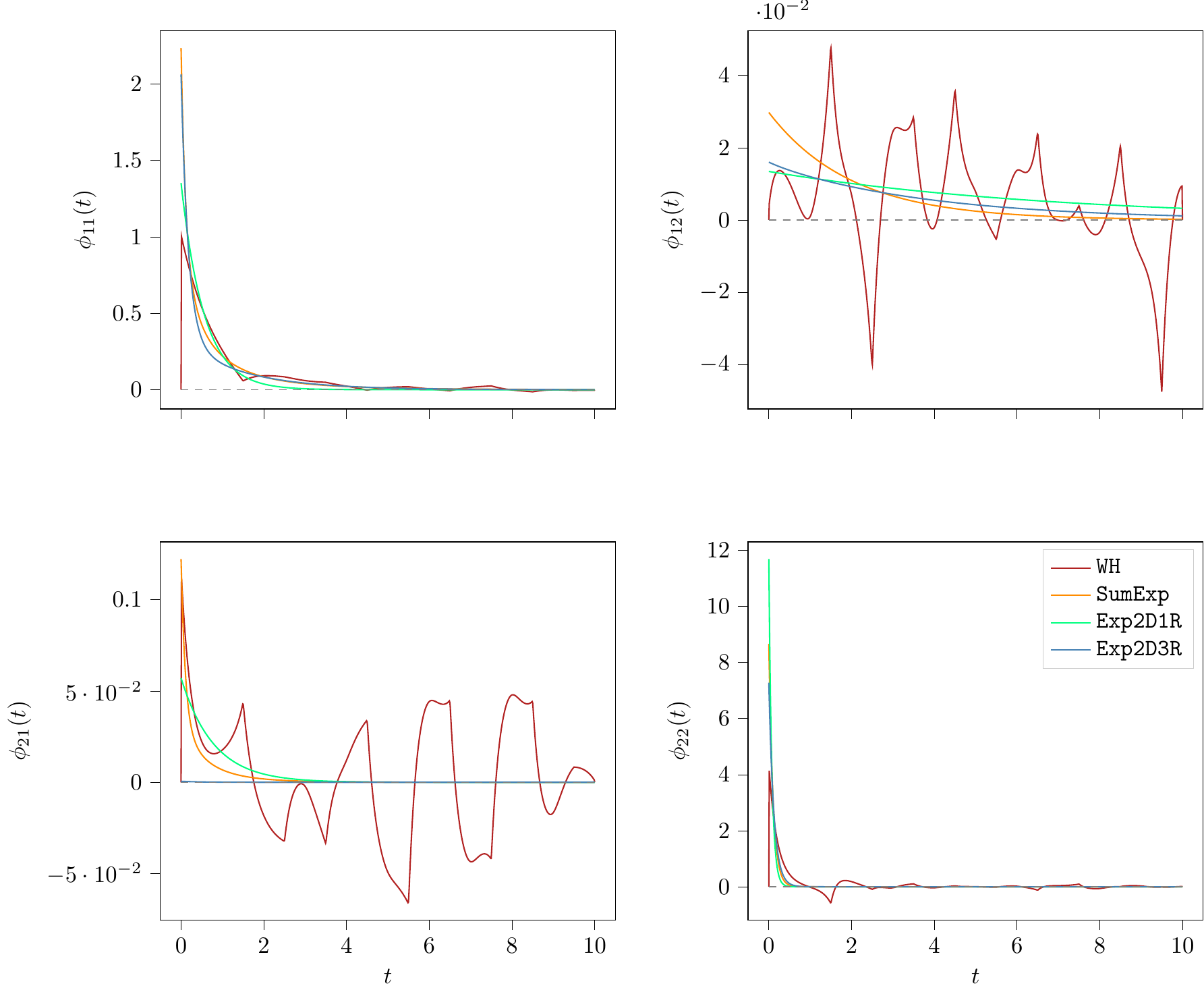}
}
    \caption{Fitted kernels.}
\begin{minipage}{0.65\textwidth} 
{\footnotesize The line $y=0$ is in dashed grey. \par}
\end{minipage}
    \label{fig:app_rw2d_phis}
\end{figure}

\paragraph{Distinction by media nationality} 
Diffusion dynamics of news related to the British Royal family might significantly differ between British news outlets, North American and Australian media, and those from other nationalities. In the $5,000$ news websites that appear in the MemeTracker dataset, it is not sufficient to use the top-level domain of the website to deduce its country.\footnote{For example, several major news websites had a .com top-level domain at the time of data collection of MemeTracker, such as \emph{The Economist} (British), \emph{El País} (Spanish) or \emph{Globo} (Brazilian).}  We manually verify the nationality of the media sources; the list of media nationalities is available as a \emph{csv} file in the \emph{Applications} folder of our repository. We model publication times of US and UK articles related to this keyword as a bi-dimensional MHP (dimension $i=1$ corresponds to US articles and $i=2$ to UK articles). We use our method to fit two exponential models: with one exponential (\texttt{Exp2D1R}) and three exponentials (\texttt{Exp2D3R}); as well as one Gaussian model (\texttt{Gauss2D1R}). For benchmark purposes, we fit \texttt{SumExp} (with 6 fixed exponential decays) and \texttt{WH}. \cref{fig:app_rw2d_residuals} displays the Q-Q plots and probability plots for these models, showing that our algorithms compete with those of the benchmarks.  \cref{fig:app_rw2d_phis} plots the fitted kernels.

In this case, the data seems well modeled by an Exponential MHP. The hyperparameters for the \texttt{WH} algorithm were selected by hand to give reasonable performance, however further tuning may be possible, and may reduce the oscillatory behaviour we observe, particularly in the cross-excitation kernels $\phi_{12}$ and $\phi_{21}$.
\subsection{Epidemic propagation}
\paragraph{Data}
In this second application, we model the propagation of Malaria in China. \citet*{unwin2021using} use a slightly modified univariate linear Hawkes model with a delayed Rayleigh kernel to study the transmission in this context. To fit their model, given the typically small number of observations in the applications they consider, \citet{unwin2021using} compute exactly the log-likelihood of their observations and input it to a standard optimization solver. We use their data for the propagation of malaria in the Yunan province between $1$ January $2011$ and $24$ September $2013$, which is available in \citet{unwin2021replication}.

\paragraph{Model} We fit two models using \texttt{ASLSD}: \texttt{SbfGauss1D10R} is a SBF Gaussian model with ten Gaussians (with uniformly spaced means in $[0,20]$ and standard deviations equal to $1.9$), and \texttt{Gauss1D1R} is a non-SBF Gaussian model. In addition to this, we consider two benchmarks: \texttt{SumExp}, and \texttt{Poisson}, which is a naive homogeneous Poisson model. (Given the small number of observations, the standard implementation of the \texttt{WH} method would not run on this dataset.)
\begin{figure}[ht]
    \centering
\resizebox{0.45\textwidth}{!}{
\includegraphics{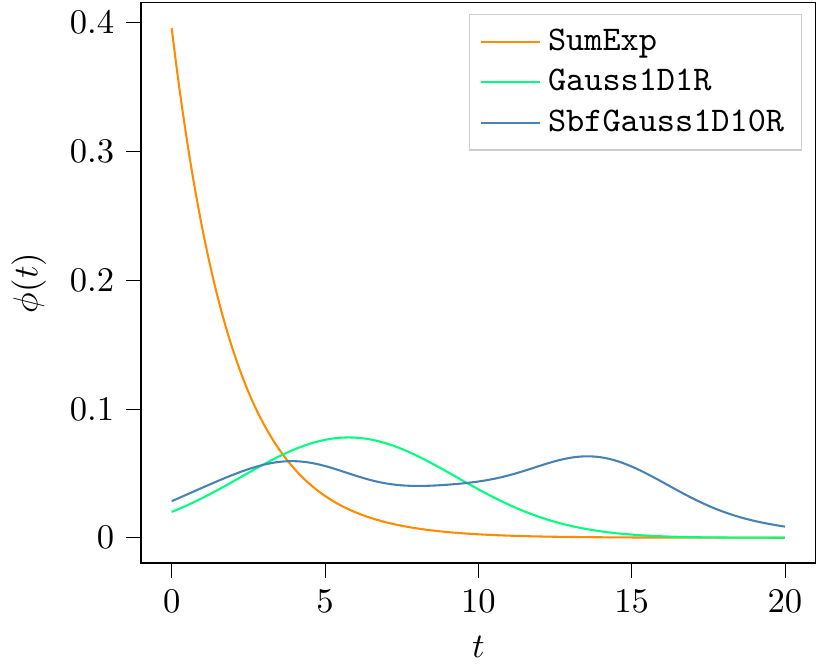}
}
    \caption{Fitted kernels.}
\begin{minipage}{0.65\textwidth} 
{\footnotesize . \par}
\end{minipage}
    \label{fig:app_malaria_1d_phi}
\end{figure}

\begin{figure}[ht]
    \centering
\resizebox{0.8\textwidth}{!}{
\includegraphics{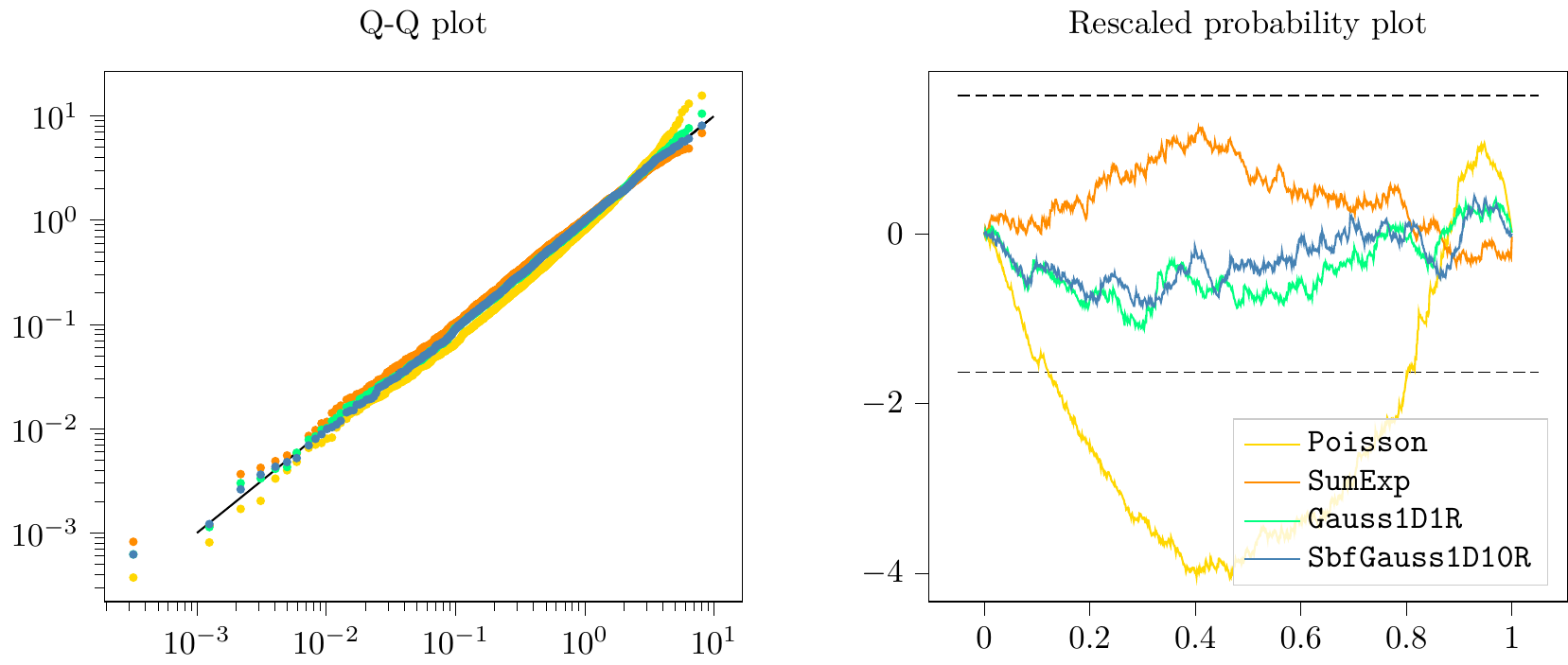}
}
    \caption{Residual plots for the fitted models.}%
    \label{fig:results_malaria}%
\end{figure}

For this data, the \texttt{SumExp} model gives residuals which pass a KS-test, with a p-value of $7.8\%$. The \texttt{Gauss1D1R} gives a better fit, passing the KS test with a p-value of $14.6\%$. \citet{unwin2021using} propose that the self-excitation of malaria infection is better modelled by a non-monotonic kernel, because of the delay in contagions due to the mosquito phase of the disease. They force this by using a delayed Rayleigh kernel, with a delay of $15$ days. It is therefore not surprising that \texttt{Gauss1D1R} outperforms \texttt{SumExp}. However, the fitted mean of the Gaussian kernel in \texttt{Gauss1D1R} is around $5$ days, well below the $15$ days delay in \citet{unwin2021using}. The \texttt{SbfGauss1D10R} model gives a significantly better p-value, at $40.5\%$. In addition to the mode at $5$, we see a second mode appearing slightly below $15$. 

Given the delays in contagion caused by the malaria life cycle (in particular during the mosquito phase,  as discussed, for example, in \citet*{stopard2021} or \citet{unwin2021using}), we might expect the true kernel to have negligible mass below $10$ days. However, we speculate that this dataset may not be based on  full observation of malaria cases (for example, due to some cases remaining unreported). This can introduce significant biases in the estimated self-excitation structure, which may explain the observed primary mode with a delay of $5$ days. 

\section{Conclusion}
In this work, we proposed a new estimation method for linear multivariate Hawkes processes, which applies to large datasets and general kernels. Our method is a fast stochastic method that can be used with rich parametric classes of kernels, such as Gaussian, Rayleigh, Exponential,  and triangular mixtures, in order to represent larger families of kernels. To the best of our knowledge, the algorithm we propose in this paper is the first stochastic optimization algorithm for the estimation of MHP that applies to general kernels. The numerical experiments we conduct in \cref{sec:Numerics} show that the precision of this algorithm competes with state of the art methods with significant computational advantages. The major difficulty in the implementation of our method is to compute the correlation functions $\left(\Upsilon_{ijk}\right)$ and their partial derivatives in closed-form; we do this for the cases mentioned in \cref{appendix: kernel specific}. Our algorithm does not formally require the MHP to be stable (but the efficiency of our sampling scheme suffers if this is not the case). Below, we briefly discuss orientations for our future work.

\paragraph{Model selection} In the procedure we proposed, we did not consider robustness to model error. Therefore, a model selection procedure appears crucial when applying this algorithm to real world data. Unfortunately, model selection is still an unexplored area in the literature of MHP. Cross validation methods are one of the major model selection frameworks, but it is not clear how to apply them to MHP due to the auto-regressive form of the conditional intensity. Our methods can easily be extended to include regularization terms, for example to encourage sparsity, opening the way for a wide variety of modelling approaches to be considered.

\paragraph{Non-linear Hawkes} The conditional intensity of a point process has to remain positive, and the fact that the kernels of a linear MHP are positive is sufficient to satisfy this condition. However, because their kernels are positive, linear MHP are unable to model inhibition between event types. In high-frequency limit order book data for example, inhibitory behaviours are well known: for example, \citet*{lu2018high} observe empirically that order cancellations on one side that change the spread may inhibit submission of contraside market orders. Non-linear Hawkes processes (\citet{bremaud1996stability}) can model inhibitory interactions, but the estimation of non-linear Hawkes processes is a hard problem and the literature is scarce, with the notable exceptions of the work of \citet*{wang2016isotonic} and \citet*{menon2018proper}. To the best of our knowledge, there exists no fast method to calibrate non-linear Hawkes processes with general kernels. The additive decomposition we propose in \cref{theorem:lse} cannot be directly transposed to the LSE of a non-linear Hawkes process; in fact it seems improbable to obtain any useful additive decomposition of the LSE in this case. Following the approach of \citet{menon2018proper}, it would be interesting to investigate the use of other contrast functions than the LSE for the estimation of non-linear Hawkes processes.
\newpage 

\appendix
\section{Proof of Theorem 1}\label{app: proofs}
In this appendix, we prove \cref{theorem:lse}. 

\begin{lemma}\label{lemma:varphi and c}
For all $i,j\in [d]$ and all $t\geq 0$, define
\begin{equation*}
    \varphi_{ij}(t):=\sum_{n=1}^{\kappa(j,t)}\phi_{ij}(t-t^j_n).
\end{equation*}
Note that for all $t \leq t^j_1$, we have $\varphi_{ij}(t)=0$. For all $i,j\in [d]$ and all $t\geq 0$, define 
\begin{equation*}
    C_{ij}(t):=\int_{0}^t \varphi_{ij}(u)\mathrm{d}u.
\end{equation*}
Then, 
\begin{equation}\label{eq:c_jk primitive}
      C_{ij}(T)=\sum_{n=1}^{N^j_T} \psi_{ij}(T-t^j_n) .    
\end{equation}
\end{lemma}
\begin{proof}
Fix $i,j\in [d]$. By definition 
\begin{equation*} 
C_{ij}(T) = \int_{0}^T \sum_{n=1}^{\kappa(j,t)}\phi_{ij}(t-t^j_n)\mathrm{d}t  = \int_{t^j_1}^T \sum_{n=1}^{\kappa(j,t)}\phi_{ij}(t-t^j_n)\mathrm{d}t.
\end{equation*}
Split the integral
\begin{equation*} 
C_{ij}(T) =\int_{t^j_{N^j_T}}^T \sum_{n=1}^{\kappa(j,t)}\phi_{ij}(t-t^j_n)\mathrm{d}t+\sum_{m=1}^{N^j_T-1} \int_{t^j_m}^{t^j_{m+1}} \sum_{n=1}^{\kappa(j,t)}\phi_{ij}(t-t^j_n)\mathrm{d}t.
\end{equation*}
For all $t\in [ \, t^j_{N^j_T},T ] \,$ we have $\kappa(j,t)=N^j_T$. For all $m\in [ \,N^j_T-1  ] \,$, for all $t\in [ \, t^j_{m},t^j_{m+1} ] \,$ we have $\kappa(j,t)=m$. Hence,
\begin{equation*} 
C_{ij}(T) =\int_{t^j_{N^j_T}}^T \sum_{n=1}^{N^j_T}\phi_{ij}(t-t^j_n)\mathrm{d}t+\sum_{m=1}^{N^j_T-1} \int_{t^j_m}^{t^j_{m+1}} \sum_{n=1}^{m}\phi_{ij}(t-t^j_n)\mathrm{d}t.
\end{equation*}
Use Fubini's Theorem to write
\begin{equation*} 
C_{ij}(T) =\sum_{n=1}^{N^j_T}\int_{t^j_{N^j_T}}^T \phi_{ij}(t-t^j_n)\mathrm{d}t+\sum_{m=1}^{N^j_T-1}\sum_{n=1}^{m} \int_{t^j_m}^{t^j_{m+1}} \phi_{ij}(t-t^j_n)\mathrm{d}t.
\end{equation*}
Re-index the sums of the second term and use a change of variable to conclude
\begin{equation*} 
\begin{split}
C_{ij}(T) & = \sum_{n=1}^{N^j_T}\int_{t^j_{N^j_T}-t^j_n}^{T-t^j_n} \phi_{ij}(u)\mathrm{d}u+\sum_{n=1}^{N^j_T-1}\sum_{m=n}^{N^j_T-1} \int_{t^j_m-t^j_n}^{t^j_{m+1}-t^j_n} \phi_{ij}(u)\mathrm{d}u, \\
 & = \sum_{n=1}^{N^j_T}\int_{t^j_{N^j_T}-t^j_n}^{T-t^j_n} \phi_{ij}(u)\mathrm{d}u+\sum_{n=1}^{N^j_T-1} \int_{0}^{t^j_{N^j_T}-t^j_n} \phi_{ij}(u)\mathrm{d}u, \\
 & = \sum_{n=1}^{N^j_T} \int_{0}^{T-t^j_n} \phi_{ij}(u)\mathrm{d}u.
\end{split}
\end{equation*}
\end{proof}

\begin{lemma}\label{lemma: varphi product}
For all $i,j,k \in [d]$ define
\begin{equation*}
    I_{ijk}(T):=\int_0^T \varphi_{ki}(t) \varphi_{kj}(t) \mathrm{d}t.
\end{equation*}
Then,
\begin{equation}
    I_{iik}(T)=\sum_{m=1}^{N^i_T}\Upsilon_{iik}(T-t^i_m,0)+2\sum_{m=1}^{N^i_T}\sum_{n=1}^{\kappa(i,i,n)}\Upsilon_{iik}(T-t^i_m,t^i_m-t^i_n),
\end{equation}
and if $i\neq j$
\begin{equation}
    I_{ijk}(T)=\sum_{m=1}^{N^i_T}\sum_{n=1}^{\kappa(j,i,m)}\Upsilon_{ijk}(T-t^i_m,t^i_m-t^j_n)+\sum_{n=1}^{N^j_T}\sum_{m=1}^{\kappa(i,j,n)}\Upsilon_{jik}(T-t^j_n,t^j_n-t^i_m).
\end{equation}
\end{lemma}
\begin{proof}
Fix $j,k,m \in [d]$. Let $N=N^k_T+N^m_T$. Define 
\begin{equation*}
    (t_q)_{q\in \llbracket 1,N \rrbracket }=\{ t^p_i, p \in \{k,m\}, i \in \llbracket 1, N^p_T \rrbracket   	\},
\end{equation*}
such that for all $q \in \llbracket 1, N-1 \rrbracket$, $t_q<t_{q+1}$. For all $q\in [N]$, define $\varsigma(q)\in \{k,m\}$ and $\iota(q) \in [ \, N^{\varsigma(q)}_T ] \, $  such that $t_q=t^{\varsigma(q)}_{\iota(q)}$. Define for all $s \geq 0$
\begin{equation*}
    I_{jkm}(s)=\int_0^s \varphi_{jk}(t) \varphi_{jm}(t) \mathrm{d}t.
\end{equation*}
We prove by induction that for all $q \in [ N ]$, 
\begin{align*}
I_{jkm}(t_q)&=\sum_{i=1}^{\kappa(k,\varsigma(q),\iota(q))}\sum_{n=1}^{\kappa(m,k,i)}\int_{t^k_i}^{t_q}\phi_{jk}(t-t^k_i)\phi_{jm}(t-t^m_n)\mathrm{d}t \\&+ \sum_{n=1}^{\kappa(n,\varsigma(q),\iota(q))}\sum_{i=1}^{\kappa(k,m,n)}\int_{t^m_n}^{t_q}\phi_{jk}(t-t^k_i)\phi_{jm}(t-t^m_n)\mathrm{d}t.    
\end{align*}
For $q=1$, the sum is empty so the result is true. Let $q \in [ N-1 ]$, assume the property is true for $q$. Define 
\begin{equation*}
    \kappa_k:= \kappa(k,\varsigma(q+1),\iota(q+1)), \quad \kappa_m:=  \kappa(m,\varsigma(q+1),\iota(q+1)).
\end{equation*}
The proof that the property is true for $q+1$ results directly from the decomposition
\begin{equation*}
\begin{split}
I_{jkm}(t_{q+1})-I_{jkm}(t_q) &= \int_{t_q}^{t_{q+1}} \varphi_{jk}(t)\varphi_{jm}(t)\mathrm{d}t  \\
  & = \sum_{i=1}^{\kappa_k}\sum_{n=1}^{\kappa_m} \int_{t_q}^{t_{q+1}} \phi_{jk}(t-t^k_i)\phi_{jm}(t-t^m_n)  .
\end{split}
\end{equation*}
Therefore, we have
\begin{equation*}
I_{jkm}(T)=\sum_{i=1}^{N^k_T}\sum_{n=1}^{\kappa(m,k,i)}\int_{t^k_i}^{T}\phi_{jk}(t-t^k_i)\phi_{jm}(t-t^m_n)\mathrm{d}t + \sum_{n=1}^{N^m_T}\sum_{i=1}^{\kappa(k,m,n)}\int_{t^m_n}^{T}\phi_{jk}(t-t^k_i)\phi_{jm}(t-t^m_n)\mathrm{d}t.  
\end{equation*}
Use the change of variables $u:=t-t^k_i$ and $u:=t-t^m_n$, to write
\begin{align*}
I_{jkm}(T)&=\sum_{i=1}^{N^k_T}\sum_{n=1}^{\kappa(m,k,i)}\int_{0}^{T-t^k_i}\phi_{jk}(u)\phi_{jm}(u+t^k_i-t^m_n)\mathrm{d}u \\&\quad + \sum_{n=1}^{N^m_T}\sum_{i=1}^{\kappa(k,m,n)}\int_{0}^{T-t^m_n}\phi_{jm}(u)\phi_{jk}(u+t^m_n-t^k_i)\mathrm{d}u.
\end{align*}
The result then follows from the definitions of $\Upsilon_{jkm}$ and $\Upsilon_{jmk}$.
\end{proof}

\printProofs

\section{Insights on the gradient estimator}\label{app:insights_on_grad_estimator}
In this appendix, we bring additional insight to the construction of the estimators of single sums from \cref{subsubsec:estimating_the_single_sums}, and the estimators of double sums from \cref{subsubsec:estimating_the_double_sums}.
\subsection{Estimating the single sums}\label{app_subsec:single_sums}
Our heuristic for the construction of this estimator is first to note that the sequence $m \mapsto f_{\theta_k}(T-t^i_m)$ is decreasing, because the functions $\psi_{ki}(\cdot)$ and $\Upsilon_{iik}(\cdot,0)$ are increasing. Second, qualitatively, we expect $t^i_m \ll T$ for a stationary process, except for the largest values of index $m$. Finally, we expect the variance of $f_{\theta_k}(T-t^i_m)$ to increase with $m$. The hyper-parameters of this estimator are the bounds of the strata, $b_p$, and the number of points sampled in the strata, $q_p$. In our numerical experiments, we chose the index $ b_{n_{\textrm{max}}}$ such that $N^i_T- b_{n_{\textrm{max}}}\sim 10^3$. One can imagine schemes where $ b_{n_{\textrm{max}}}$ is chosen adaptively, but our experiments suggested this was not useful in practice. Given $ b_{n_{\textrm{max}}}$, we choose $ b_{n_{\textrm{max}}-1}=b_{n_{\textrm{max}}}-\delta$ where $\delta\in \mathbb{N}$ is a hyper-parameter, and chose the other bounds $(b_p)$, such that
\begin{equation*}
    b_{p+1}-b_{p} \sim (b_{p+2}-b_{p+1})^2.
\end{equation*}
\subsection{Estimating the double sums}\label{app_subsec:double_sums}
\paragraph{Constructing a stratification}
For a given number of sample points $Q$, we build an adaptive strategy to allocate our points in $\boldsymbol{q}$. A naive criterion is to minimize the total variance of the estimator. Define the total standard deviation in bucket $\boldsymbol{b}$ by
\begin{equation*}
\sigma^{\boldsymbol{b}}_T(\theta_k):=  \sqrt{\frac{1}{|  \mathcal{E}_T^{ij,\boldsymbol{b}}	|}\sum_{x \in \mathcal{E}_T^{ij,\boldsymbol{b}}}f_{\theta_k}^2(x)-\frac{1}{|  \mathcal{E}_T^{ij,\boldsymbol{b}}	|^2}\bigg( \sum_{x \in \mathcal{E}_T^{ij,\boldsymbol{b}}} f_{\theta_k}(x) \bigg)^2 }.  
\end{equation*}
For all $p\in [n_B]$, define 
\begin{equation}\label{eq:optimal allocation}
    \tilde{q}^p_*(\theta_k):=\frac{\sqrt{\frac{|  \mathcal{E}_T^{ij,\boldsymbol{\boldsymbol{b_p}}}	|}{|  \mathcal{E}_T^{ij,\boldsymbol{\boldsymbol{b_p}}}	|-1}} \sigma^{\boldsymbol{b_p}}_T(\theta_k)}{\sum_{p^\prime=1}^{n_B}\sqrt{\frac{|  \mathcal{E}_T^{ij,\boldsymbol{\boldsymbol{b_{p^\prime}}}}	|}{|  \mathcal{E}_T^{ij,\boldsymbol{\boldsymbol{b_{p^\prime}}}}	|-1}} \sigma^{\boldsymbol{b_{p^\prime}}}_T(\theta_k)},
\end{equation}
and let 
\begin{equation*}
 \boldsymbol{\tilde{q}_*}(\theta_k)=\bigg(\tilde{q}^1_*(\theta_k),\dots,\tilde{q}^{n_B}_*(\theta_k) \bigg)    .
\end{equation*}
The variances of the estimators, accounting for sampling without replacement, are
\begin{align*} 
\Var{ \bigg[ \, \hat{S}^{\boldsymbol{b_p},q^p}_T(\theta_k) 	\bigg] \,} &= \frac{|  \mathcal{E}_T^{ij,\boldsymbol{b_p}}	|^2}{q^p} \bigg(1-\frac{q^p-1}{|  \mathcal{E}_T^{ij,\boldsymbol{\boldsymbol{b_p}}}	|-1}\bigg) \sigma^h_T(\theta_k)^2 ,\\ 
 \Var{ \bigg[ \, \hat{S}^{\boldsymbol{q}}_T(\theta_k) 	\bigg] \,}&=  \sum_{p=1}^{n_B} \frac{|  \mathcal{E}_T^{ij,\boldsymbol{b_p}}	|^2}{q^p}\bigg(1-\frac{q^p-1}{|  \mathcal{E}_T^{ij,\boldsymbol{\boldsymbol{b_p}}}	|-1}\bigg) \sigma^h_T(\theta_k)^2.
\end{align*}
The variance of the estimator, as a function of $\boldsymbol{\tilde{q}}$, is
\begin{equation*}
    \Var{ \bigg[ \, \hat{S}^{\boldsymbol{q}}_T(\theta_k) 	\bigg] \,}=  \frac{1}{Q}\sum_{p=1}^{n_B} \frac{|  \mathcal{E}_T^{ij,\boldsymbol{\boldsymbol{b_p}}}	|^3}{(|  \mathcal{E}_T^{ij,\boldsymbol{\boldsymbol{b_p}}}	|-1)\tilde{q}^p}\sigma^{\boldsymbol{b_p}}_T(\theta_k)^2 -\sum_{p=1}^{n_B}\frac{|  \mathcal{E}_T^{ij,\boldsymbol{b_p}}	|^2}{|  \mathcal{E}_T^{ij,\boldsymbol{b_p}}	|-1}\sigma^{\boldsymbol{b}_p}_T(\theta_k)^2.
\end{equation*}
Use Jensen's inequality to write
\begin{equation}\label{eq:jensen optimal allocation}
\Var{ \bigg[ \, \hat{S}^{\boldsymbol{q}}_T(\theta_k) 	\bigg] \,} \geq  \frac{1}{Q}\Bigg(\sum_{p=1}^{n_B} |  \mathcal{E}_T^{ij,\boldsymbol{\boldsymbol{b_p}}}	|\sqrt{\frac{|  \mathcal{E}_T^{ij,\boldsymbol{\boldsymbol{b_p}}}	|}{|  \mathcal{E}_T^{ij,\boldsymbol{\boldsymbol{b_p}}}	|-1}} \sigma^{\boldsymbol{b_p}}_T(\theta_k)\Bigg)^2 -\sum_{p=1}^{n_B}\frac{|  \mathcal{E}_T^{ij,\boldsymbol{b_p}}	|^2}{|  \mathcal{E}_T^{ij,\boldsymbol{b_p}}	|-1}\sigma^{\boldsymbol{b}_p}_T(\theta_k)^2.
\end{equation}
The inequality in \eqref{eq:jensen optimal allocation} is tight, and this lower bound is attained for $\boldsymbol{\tilde{q}}=\boldsymbol{\tilde{q}_*(\theta_k)}$.
For a given number of sample points $Q$, the allocation $\boldsymbol{q}=\boldsymbol{\tilde{q}_*}(\theta_k)$ is the allocation that minimizes the variance of the estimator $\hat{S}^{\boldsymbol{q}}_T(\theta_k)$. However, the computation of the optimal allocation $\boldsymbol{\tilde{q}_*}(\theta_k)$ is expensive, because the computation of the vector of variances $ \big(\sigma^{\boldsymbol{b_p}}_T(\theta_k)\big)_p$ has quadratic worst-case complexity. 

In \cref{sec:Numerics}, we discuss cases where it is not necessary to chose  $\boldsymbol{\tilde{q}}=\boldsymbol{\tilde{q}_*(\theta_k)}$ for the estimation procedure to converge, notably for some decreasing kernels. But for general kernels, setting an arbitrary allocation $\boldsymbol{\tilde{q}}$ does not lead to satisfactory estimates in practice, this is why we propose an adaptive estimator of the optimal allocation $\boldsymbol{\tilde{q}_*(\theta_k)}$ below.

\paragraph{Efficient adaptive estimation of $\boldsymbol{\tilde{q}_*}(\theta_k)$}
We slightly modify the work of \citet{etore2010adaptive} on adaptive stratified Monte Carlo sampling to the case of simple random sampling without replacement. Fix $\theta_k$ and an initial allocation guess $\boldsymbol{\tilde{q}^{(0)}_*}(\theta_k)$. Let $n_K$ denote the number of iterations used to estimate $\boldsymbol{\tilde{q}_*(\theta_k)}$, and $Q$ the total number of points that we sample in this procedure. Fix $(\Delta Q_s)_{s\in [ \, n_K 	] \,}$ such that at each step $s \in [ \, n_K ] \,$ we sample $Q_s:=n_B+\Delta Q_s$ points. Denote the points sampled in stratum $\boldsymbol{b}$ at step $s$ by
\begin{equation*}
    \Delta q^{\boldsymbol{b}}_s:=1+\delta q^{\boldsymbol{b}}_s,
\end{equation*}
so that we sample at least one point in each stratum at each step $k$. Denote by $q^{\boldsymbol{b}}_s:=\sum_{s^\prime=1}^s \Delta q^{\boldsymbol{b}}_{s^\prime}$ the total number of points sampled in stratum $\mathcal{E}_T^{ij,\boldsymbol{\boldsymbol{b}}}$ up to and including step $s$. Begin the procedure with an initial guess of the optimal allocation $\boldsymbol{\tilde{q}^{(0)}_*}(\theta_k)$. For all $p\in [ \, n_B	] \,$, denote by $\hat{S}^{\boldsymbol{b_{p}},(s)}_T(\theta_k)$, $\hat{\sigma}^{\boldsymbol{b_{p}},(s)}_T(\theta_k)$, and $\boldsymbol{\tilde{q}^{(s)}_*}(\theta_k)$, our estimates constructed inductively, based on all the samples up to step $s$ inclusive. At step $s$, for all $p$, compute
\begin{equation*}
\Delta q^{\boldsymbol{b_{p}}}_s =\tilde{q}^{\boldsymbol{b_{p}},(s-1)}_*(\theta_k)(n_B+\Delta Q_s).
\end{equation*}
Next, for all $p$, sample without replacement $\Delta q^{\boldsymbol{b_{p}}}_s$ points $(x^{\boldsymbol{b_p}}_m)_m$ in $\mathcal{E}_T^{ij,\boldsymbol{\boldsymbol{b_p}}}$, and obtain
\begin{align*} 
\hat{S}^{\boldsymbol{b_{p}}}_T (\theta_k)&=\hat{S}^{\boldsymbol{b_{p}},(n_B)}_T (\theta_k)  ,\\ 
\hat{q}^{\boldsymbol{b_{p}}}_*(\theta_k)&= \hat{q}^{\boldsymbol{b_{p}},(n_B)}_*(\theta_k) ,\\ 
\hat{\sigma}^{\boldsymbol{b_{p}}}_T(\theta_k)^2 &=  \hat{\sigma}^{\boldsymbol{b_{p}},(\Delta s)}_T(\theta_k)^2.
\end{align*}

In some cases, using a fixed, preset allocation $\boldsymbol{\tilde{q}}$ is sufficient to get satisfactory numerical results. In particular, for decaying kernels, we note that the sequence $\left(S^h_T(\theta_k)\right)_h$ is decreasing with the lag $h$. For a sufficiently fine stratification,  a monotonically decaying allocation leads to good results. This is not the case for more general kernels, hence the importance of constructing an adaptive estimator $\boldsymbol{\hat{q}_*}(\theta_k)$ of the optimal allocation $\boldsymbol{\tilde{q}_*}(\theta_k)$ without increasing significantly the complexity of the procedure. 

At each step $s$, compute the sample mean and sample variance of the corresponding batches
\begin{align}\label{eq:delta_updates_double_sums}
\hat{S}^{\boldsymbol{b_{p}},(\Delta s)}_T (\theta_k)&= \frac{	|  \mathcal{E}_T^{ij,\boldsymbol{b_p}}	|}{\Delta q^{\boldsymbol{b_{p}}}_s}\sum_{m=1}^{\Delta q^{\boldsymbol{b_{p}}}_s} f_{\theta_k}(x^{\boldsymbol{b_p}}_m) ,\\ 
\hat{\sigma}^{\boldsymbol{b_{p}},(\Delta s)}_T(\theta_k)^2 &= \frac{|  \mathcal{E}_T^{ij,\boldsymbol{b_p}}	|-1}{|  \mathcal{E}_T^{ij,\boldsymbol{b_p}}	|}\frac{1}{\Delta q^{\boldsymbol{b_{p}}}_s-1}\sum_{m=1}^{\Delta q^{\boldsymbol{b_{p}}}_s} \bigg( f_{\theta_k}(x^{\boldsymbol{b_p}}_m)- \hat{S}^{\boldsymbol{b_{p}},(\Delta s)}_T (\theta_k)\bigg)^2 .
\end{align}
To avoid unnecessary computations, update the running estimates in batches with
\begin{align}\label{eq:updates_double_sums}
\hat{S}^{\boldsymbol{b_{p}},(s)}_T(\theta_k) ={}& \frac{ q^{\boldsymbol{b_{p}}}_{s-1} }{ q^{\boldsymbol{b_{p}}}_s} \hat{S}^{\boldsymbol{b_{p}},(s-1)}_T(\theta_k)+\frac{\Delta q^{\boldsymbol{b_{p}}}_s}{q^{\boldsymbol{b_{p}}}_s}\hat{S}^{\boldsymbol{b_{p}},(\Delta s)}_T (\theta_k),\\ 
\begin{split}
 \hat{\sigma}^{\boldsymbol{b_{p}},(s)}_T(\theta_k)^2 ={}&  \frac{ q^{\boldsymbol{b_{p}}}_{s-1}-1 }{ q^{\boldsymbol{b_{p}}}_s-1} \hat{\sigma}^{\boldsymbol{b_{p}},(s-1)}_T(\theta_k)^2 +\frac{\Delta q^{\boldsymbol{b_{p}}}_s}{q^{\boldsymbol{b_{p}}}_s-1} \hat{\sigma}^{\boldsymbol{b_{p}},(\Delta s)}_T(\theta_k)^2
 \\ & +\frac{	|  \mathcal{E}_T^{ij,\boldsymbol{b_p}}	|-1}{	|  \mathcal{E}_T^{ij,\boldsymbol{b_p}}	|} \frac{q^{\boldsymbol{b_{p}}}_{s-1}\Delta q^{\boldsymbol{b_{p}}}_{s}}{(q^{\boldsymbol{b_{p}}}_{s}-1)q^{\boldsymbol{b_{p}}}_{s}}\bigg( \hat{S}^{\boldsymbol{b_{p}},(s-1)}_T(\theta_k)-\hat{S}^{\boldsymbol{b_{p}},(\Delta s)}_T (\theta_k) \bigg)^2.    
\end{split}
\end{align}

\paragraph{Sensitivity of $\boldsymbol{\tilde{q}_*}(\theta_k)$ to $\theta_k$}
Note that for a SBF MHP such that each kernel $\phi_{ij}$ is represented by only one basis function; i.e. only one corresponding parameter $\omega_{ij}>0$, we have $\nabla \boldsymbol{\tilde{q}_*}(\theta_k)=0$. In the general case though, the optimal allocation $\boldsymbol{\tilde{q}_*}(\theta_k)$ depends on $\theta_k$. In the context of a gradient-based optimization procedure, consider a sequence of parameters 
    \begin{equation*}
       \bigg(\theta_k^{(1)},\dots,\theta_k^{(t)} \bigg). 
    \end{equation*}
Use an exponential moving average (EMA) of the estimates 
\begin{equation*}
    \boldsymbol{\tilde{q}_*}\bigg(\theta^{(1)}_{k}\bigg),\dots,\boldsymbol{\tilde{q}_*}\bigg(\theta^{(t)}_{k}\bigg)
\end{equation*}
as an initial guess for this procedure in step $t+1$. Denote by $w$ the EMA weight, fixed at the beginning of the procedure, and write
\begin{equation*}
    \boldsymbol{\tilde{q}^{(0)}_*}\bigg(\theta^{(t+1)}_{k}\bigg)=w \cdot \boldsymbol{\tilde{q}_*}\bigg(\theta^{(t)}_{k}\bigg) +(1-w)\cdot\boldsymbol{\tilde{q}_*}\bigg(\theta^{(t-1)}_{k}\bigg) .
\end{equation*}
The role of the EMA step in our procedure is to try to use the corresponding sequence of optimal share estimates
\begin{equation*}
\bigg( \boldsymbol{\hat{q}_*}\bigg(\theta_k^{(1)}\bigg), \dots, \boldsymbol{\hat{q}_*}\bigg(\theta_k^{(t)}\bigg)\bigg), 
\end{equation*}    
to get a heuristic for $\boldsymbol{\hat{q}_*}\bigg(\theta_k^{(t+1)}\bigg)$.

\paragraph{Estimation of the remainder $S^{\mathrm{rest}}_T(\theta_k)$} As in the single sum case, use a standard stratified Monte Carlo approach with a fixed allocation to estimate the remainder 
 \begin{equation*}
    S^{\mathrm{rest}}_T(\theta_k):=\sum_{h=h_{\textrm{max}}+1}^{\kappa(j,i,N^i_T)} S^h_T(\theta_k).
\end{equation*} 
Let $n_R \in \mathbb{N}^*$ denote the number of strata we use for this estimation, and consider a partition $B^\prime=(\boldsymbol{b^\prime_{1}}, \dots , \boldsymbol{b^\prime_{n_R}} ) $ of $\llbracket h_{\textrm{max}}+1, \kappa(j,i,N^i_T)	\rrbracket$. Sample $q_p$ points $(x^{\boldsymbol{b^\prime_p}}_1, \dots,  x^{\boldsymbol{b^\prime_p}}_{q_p})$ uniformly and without replacement from $\mathcal{E}_T^{ij,\boldsymbol{b^\prime_p}}$ for all $p\in 	[ \, n_R 	] \,$. Fix in advance the number of sample points per stratum, which we denote by
\begin{equation*}
    \boldsymbol{q}=(q_1, \dots, q_p).
\end{equation*}
Next, use the unbiased estimator of $S^{\mathrm{rest}}_T(\theta_k)$ defined by 
\begin{equation}\label{eq:remainder_double_sums}
\hat{S}^{rest}_T(\theta_k)=\sum_{p=1}^{n_R} \frac{|\boldsymbol{b^\prime_{p}} |}{q_p} \sum_{n=1}^{q_p} f_{\theta_k}(x^{\boldsymbol{b^\prime_{p}}}_{q_p} ).
\end{equation}
\paragraph{Adaptive learning rate}
Denote the element-wise product between vectors or matrices by $\odot$; the learning rate hyper-parameter by $t \mapsto a_{\textrm{rate}}(t)>0$; and two moment hyper-parameters by $ a_{\textrm{M}1}>0$ and $ a_{\textrm{M}2}>0$. Denote a last hyper-parameter $a_{\textrm{E}}>0$, which is used to avoid division by zero. The scheme we consider is defined by
\begin{align} 
g^{(t)}_1  &= \frac{ a_{\textrm{M}1} \cdot g^{(t-1)}_1+\left(1-a_{\textrm{M}1}\right) \cdot \boldsymbol{\mathcal{G}^{(k)}_T}\bigg(\theta_{k}^{(t)}\bigg)}{1-a_{\textrm{M}1}^t}  ,\\ 
g^{(t)}_2 &=   \frac{ a_{\textrm{M}2} \cdot g^{(t-1)}_2+\left(1-a_{\textrm{M}2}\right) \cdot \boldsymbol{\mathcal{G}^{(k)}_T}\odot\boldsymbol{\mathcal{G}^{(k)}_T}\bigg(\theta_{k}^{(t)}\bigg)}{1-a_{\textrm{M}2}^t} ,\\
\Delta \theta^{(t)}_k &=  -a_{\textrm{rate}}(t) \cdot \frac{g^{(t)}_1}{\sqrt{g^{(t)}_2 }+a_{\textrm{E}}} .
\end{align}
This scheme differs from the standard ADAM scheme by the non-constant learning rate $a_{\textrm{rate}}(t)$. We start this rate at a fixed value $a_{\textrm{rate}}(0) > 0$ and then use
  \begin{equation}
    a_{\textrm{rate}}(t) := \frac{a_{\textrm{rate}}(0) }{2^{\lfloor t/200 \rfloor}}.
  \end{equation}
\section{Mode collapse}\label{app:nonSBFmodecollapse}
In our numerical experiments, we observe that when a given kernel is modelled as a multimodal mixture,  and the calibration fits all the parameters of the mixture model (rather than being a sum-of-basis-functions model), the estimates are prone to numerical instability.

\begin{figure}[htp]
    \centering
\resizebox{1.\textwidth}{!}{
\includegraphics{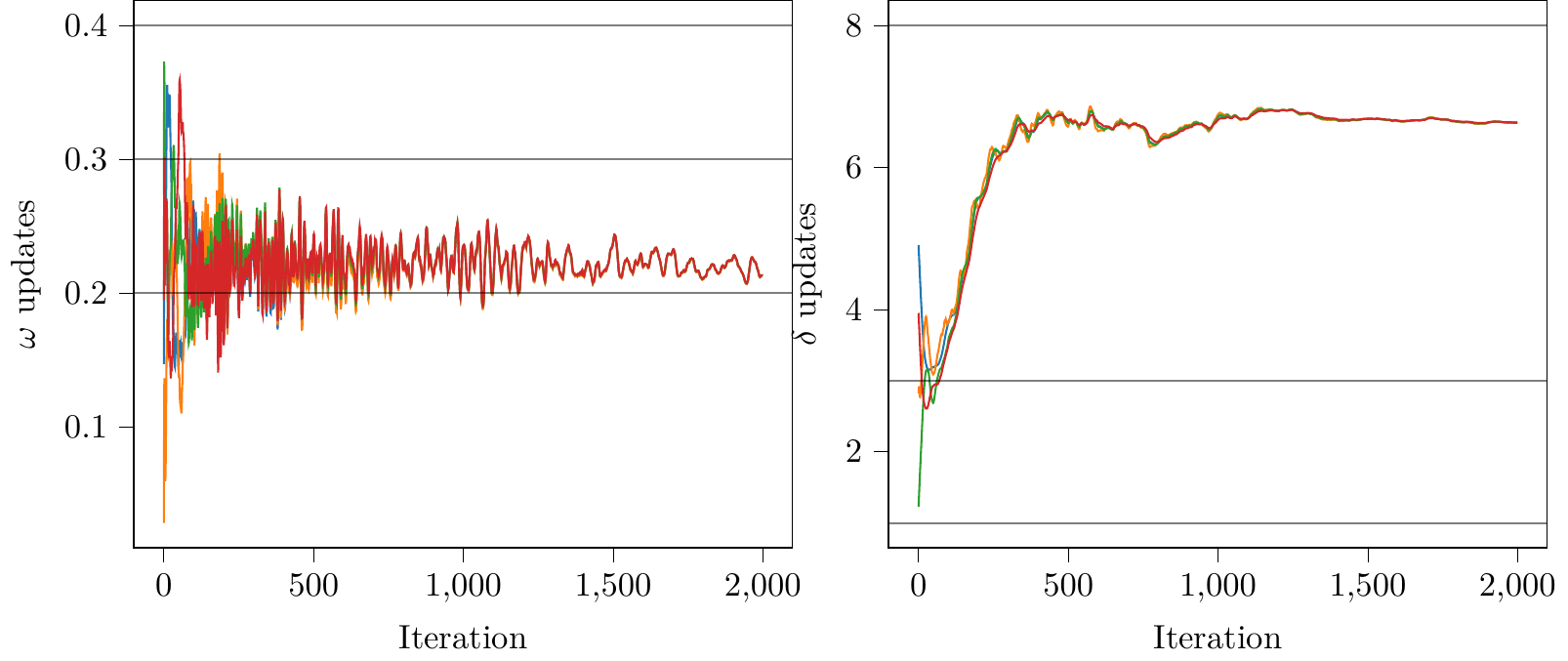}
}
 \caption{Updates of the model parameters per gradient iteration.}%

    \label{fig:param_updates_fixed_beta_modecollapse}%
\end{figure}

For example, consider the Gaussian MHP in the univariate multimodal case from \cref{subsubsec:numexp_gaussker}. Consider fitting a \texttt{Gauss4} be a univariate Gaussian model as in \cref{subsubsec:Some kernels}, with four Gaussian functions ($r=4$), where for all $l \in [10]$ set $\beta_{l}=0.5$. \cref{fig:param_updates_fixed_beta_modecollapse} plots the updates of model parameters per gradient iterations, we see a clear mode collapse towards the lowest of the true means. 
The SBF Gaussian model helps circumvent this issue, which may be explained through the fact the LSE is a quadratic function of the weights $\omega$. We leave the detailed study of this problem for future work.

\section{Kernel-specific results}\label{appendix: kernel specific}
{\scriptsize
In this appendix, we derive the formulas of the associated $\psi$ and $\Upsilon$ functions and their partial derivatives for some parametric classes of kernels; as we need them in closed-form in our algorithm.

Consider a $(\boldsymbol{\mu},\Phi)-$linear MHP. Fix $i,j,k \in [d]$, and denote by $\boldsymbol{\theta_{ki}},\boldsymbol{\theta_{kj}}$ the vectors of parameters of the kernels $\phi_{ki}$ and $\phi_{kj}$. 

\subsection{Monotonic kernels}
\subsubsection{Exponential}
We assume that the kernel
\begin{equation*}
    \phi_{ki}:=\phi^{\mathcal{E}}_{(\boldsymbol{\omega_{ki}},\boldsymbol{\beta_{ki}})}
\end{equation*}
is an exponential kernel with parameters $\boldsymbol{\theta_{ki}}=\big( \boldsymbol{\omega_{ki}},\boldsymbol{\beta_{ki}} \big)  $, but do not make any assumptions on $\phi_{k^\prime,i^\prime}$ for $(k^\prime,i^\prime)\neq (k,i)$. 
\begin{prop}
For all $x\geq 0$, we have
\begin{equation}\label{eq:psi_jk delayed exp}
    \psi_{ki}(x)=\sum_{l=1}^{r_{ki}} \omega_{kil} \big( 1-e^{-\beta_{kil} x} \big).
\end{equation}
Fix $p \in [ \, r_{ki}  ]$. The partial derivatives of $\phi_{ki}$ and $\psi_{ki}$ with respect to model parameters are 
\begin{align} 
\frac{\partial \phi_{ki} }{\partial \omega_{kip}} (x) ={}& \beta_{kip} e^{-\beta_{kip} x}   ,\\ 
 \frac{\partial \phi_{ki} }{\partial \beta_{kip}} (x)={}& \omega_{kip}(1-\beta_{kip} x) e^{-\beta_{kip} x},\\ 
\begin{split}
\frac{\partial \psi_{ki} }{\partial \omega_{kip}} (x) ={}& 1-e^{-\beta_{kip} x} ,   
\end{split}\\ 
\begin{split}
\frac{\partial \psi_{ki} }{\partial \beta_{kip}} (x) ={}& \omega_{kip} x e^{-\beta_{kip} x}.    
\end{split}
\end{align}
\end{prop}
We give closed-form formulas for $\Upsilon_{ijk}$ and its partial derivatives given various parametric classes for kernel $\phi_{kj}$.

\begin{prop}[Exponential with exponential]
We assume that the kernel
\begin{equation*}
\phi_{kj}:=\phi^{\mathcal{E}}_{(\boldsymbol{\omega_{kj}},\boldsymbol{\beta_{kj}})}
\end{equation*}
is an exponential kernel with parameters $\boldsymbol{\theta_{kj}}=\big( \boldsymbol{\omega_{kj}},\boldsymbol{\beta_{kj}} \big)  $. Then for all $t,s \geq 0$
\begin{equation}
    \Upsilon_{ijk}(t,s)=\sum_{l=1}^{r_{ki}}\sum_{l^\prime=1}^{r_{kj}}\omega_{kil}\omega_{kjl^\prime}\frac{\beta_{kil}\beta_{kjl^\prime}}{\beta_{kil}+\beta_{kjl^\prime}}e^{-\beta_{kjl^\prime}s}(1-e^{-(\beta_{kil}+\beta_{kjl^\prime})t}).
\end{equation}
Let $p \in [ \, r_{ki}  ]$ and $q \in [ \, r_{kj}  ]$. If $i \neq j$, the partial derivatives of $\Upsilon_{ijk}$ with respect to the parameters are 
\begin{align} 
\frac{\partial \Upsilon_{ijk} }{\partial \omega_{kip}} (t,s) ={}& \sum_{l^\prime=1}^{r_{kj}}\omega_{kjl^\prime}\frac{\beta_{kip}\beta_{kjl^\prime}}{\beta_{kip}+\beta_{kjl^\prime}}e^{-\beta_{kjl^\prime}s}(1-e^{-(\beta_{kip}+\beta_{kjl^\prime})t})  ,\\ 
\frac{\partial \Upsilon_{ijk} }{\partial \omega_{kjq}} (t,s) ={}& \sum_{l=1}^{r_{ki}}  \omega_{kil}\frac{\beta_{kil}\beta_{kjq}}{\beta_{kil}+\beta_{kjq}}e^{-\beta_{kjq}s}(1-e^{-(\beta_{kil}+\beta_{kjq})t}),\\ 
\begin{split}
\frac{\partial \Upsilon_{ijk} }{\partial \beta_{kip}} (t,s) ={}& \omega_{kip} \sum_{l^\prime=1}^{r_{kj}}\omega_{kjl^\prime}e^{-\beta_{kjl^\prime}s} \bigg( \frac{\beta_{kjl^\prime}^2}{(\beta_{kip}+\beta_{kjl^\prime})^2}(1-e^{-(\beta_{kip}+\beta_{kjl^\prime})t})\\ &+ \frac{\beta_{kip}\beta_{kjl^\prime}}{\beta_{kip}+\beta_{kjl^\prime}}te^{-(\beta_{kip}+\beta_{kjl^\prime})t} \bigg),   
\end{split}\\ 
\begin{split}
\frac{\partial \Upsilon_{ijk} }{\partial \beta_{kjq}} (t,s) ={}& \omega_{kjq}e^{-\beta_{kjq}s}  \sum_{l=1}^{r_{ki}}\omega_{kil}\bigg( \frac{\beta_{kil}^2}{(\beta_{kil}+\beta_{kjq})^2}(1-e^{-(\beta_{kil}+\beta_{kjq})t})\\ &+ \frac{\beta_{kil}\beta_{kjq}}{\beta_{kil}+\beta_{kjq}}\big( (t+s)e^{-(\beta_{kil}+\beta_{kjq})t} -s\big) \bigg) .    
\end{split}
\end{align}
If $i=j$, the partial derivatives of $\Upsilon_{ijk}$ with respect to the parameters are given by
\begin{align} 
\begin{split}
\frac{\partial \Upsilon_{iik} }{\partial \omega_{kip}} (t,s) ={}& \omega_{kip}\beta_{kip}e^{-\beta_{kip}s}(1-e^{-2\beta_{kip}t})\\ &+\sum_{l^\prime \in [ \, r_{ki}  ], l^\prime \neq p}\omega_{kil^\prime}\frac{\beta_{kip}\beta_{kil^\prime}}{\beta_{kip}+\beta_{kil^\prime}}e^{-\beta_{kil^\prime}s}(1-e^{-(\beta_{kip}+\beta_{kil^\prime})t})\\ &+\sum_{l \in [ \, r_{ki}  ], l \neq p}\omega_{kil}\frac{\beta_{kip}\beta_{kil}}{\beta_{kip}+\beta_{kil}}e^{-\beta_{kip}s}(1-e^{-(\beta_{kip}+\beta_{kil})t}),   
\end{split}\\
\begin{split}
\frac{\partial \Upsilon_{iik} }{\partial \beta_{kip}} (t,s) ={}& \frac{1}{2}\omega_{kip}^2e^{-\beta_{kip}s}\bigg( 1- \beta_{kip}s+e^{-2\beta_{kip}t}(\beta_{kip}(2t+s)-1) \bigg)
\\ &+\omega_{kip} \sum_{l^\prime \in [ \, r_{ki}  ], l^\prime \neq p}\omega_{kil^\prime} e^{-\beta_{kil^\prime}s}\bigg( (\frac{\beta_{kil^\prime}}{\beta_{kil}+\beta_{kil^\prime}})^2\\ &+e^{-(\beta_{kip}+\beta_{kil^\prime})t}\frac{\beta_{kil^\prime}}{\beta_{kil}+\beta_{kil^\prime}}(\beta_{kip}t-\frac{\beta_{kil^\prime}}{\beta_{kil}+\beta_{kil^\prime}}) \bigg)\\ &+ \omega_{kip} \sum_{l \in [ \, r_{ki}  ], l \neq p}\omega_{kil} e^{-\beta_{kip}s}\bigg( (\frac{\beta_{kil}}{\beta_{kil}+\beta_{kip}})^2-s\beta_{kip}\frac{\beta_{kil}}{\beta_{kil}+\beta_{kip}}\\ &+e^{-(\beta_{kip}+\beta_{kil})t}\frac{\beta_{kil}}{\beta_{kil}+\beta_{kip}}(\beta_{kip}(t+s)-\frac{\beta_{kil}}{\beta_{kil}+\beta_{kip}}) \bigg).  
\end{split}
\end{align}
\end{prop}

\subsection{Non monotonic kernels}
\subsubsection{Delayed exponential}
Given a monotonically decaying function $\phi$, a naive way to model a delayed response is to consider kernels of the form
\begin{equation*}
   x \mapsto \mathbbm{1}_{ 	\{ x>\delta \} }\phi\bigg( x-\delta  \bigg),
\end{equation*}
where $\delta>0$ controls the delay. We define the delayed exponential kernel using this basic approach.
\begin{definition}[Delayed exponential kernel]
Let $r\in \mathbb{N}^*$. For $x\geq 0$, the delayed exponential kernel $\phi^{\mathcal{DE}}_{(\boldsymbol{\omega},\boldsymbol{\beta},\boldsymbol{\delta})}$ is
\begin{equation*}
    \phi^{\mathcal{DE}}_{(\boldsymbol{\omega},\boldsymbol{\beta},\boldsymbol{\delta})}(x):=\sum_{l=1}^r \omega_l \beta_l e^{-\beta_l (x-\delta_l)}\mathbbm{1}_{ 	\{ x>\delta_l \} },
\end{equation*}
where the parameters are the vector of weights $\boldsymbol{\omega}:=(\omega_l)_{l\in \llbracket 1, r \rrbracket}\in [0,+\infty)^r$, the vector of decays $\boldsymbol{\beta}:=(\beta_l)_{l\in \llbracket 1, r \rrbracket}\in (0,+\infty)^r$, and the vector of delays $\boldsymbol{\delta}:=(\delta_l)_{l\in \llbracket 1, r \rrbracket}\in (0,+\infty)^r$.We have
\begin{equation*}
    \| \phi^{\mathcal{DE}}_{(\boldsymbol{\omega},\boldsymbol{\beta},\boldsymbol{\delta})} \|_1=\sum_{l=1}^r\omega_l.
\end{equation*}
\end{definition}

\begin{rmk}
Given the discontinuity of the kernel when the delay $\boldsymbol{\delta}$ is varied, gradient methods do not seem appropriate for calibrating $\boldsymbol{\delta}$. If $\boldsymbol{\delta}$ is unknown, an alternative model family (eg. a mixture of Rayleigh or Gaussian kernels) may be more appropriate.
\end{rmk}
\begin{prop}
For all $x\geq 0$, we have
\begin{equation}\label{eq:psi_jk delayed exp}
    \psi_{ki}(x)=\sum_{l=1}^{r_{ki}} \omega_{kil} \big( 1-\exp(-\beta_{kil}(x-\delta_{kil})_+) \big).
\end{equation}
Fix $p \in [ \, r_{ki}  ]$. The partial derivatives of $\phi_{ki}$ and $\psi_{ki}$ with respect to model parameters are given by 
\begin{align} 
\frac{\partial \phi_{ki} }{\partial \omega_{kip}} (x) ={}& \beta_{kip} e^{-\beta_{kip} (x-\delta_{kip})}\mathbbm{1}_{ 	\{ x>\delta_{kip} \}}   ,\\ 
 \frac{\partial \phi_{ki} }{\partial \beta_{kip}} (x)={}& \omega_{kip}(1-\beta_{kip} (x-\delta_{kip})) e^{-\beta_{kip} (x-\delta_{kip})} \mathbbm{1}_{ 	\{ x>\delta_{kip} \}},\\ 
\begin{split}
\frac{\partial \psi_{ki} }{\partial \omega_{kip}} (x) ={}& 1-e^{-\beta_{kip} (x-\delta_{kip})_+} ,   
\end{split}\\ 
\begin{split}
\frac{\partial \psi_{ki} }{\partial \beta_{kip}} (x) ={}& \omega_{kip} (x-\delta_{kip})_+ e^{-\beta_{kip} (x-\delta_{kip})_+}.    
\end{split}
\end{align}
\end{prop}

\begin{rmk}
In terms of the intensity $\lambda_k$ and the compensator $\Lambda_k$, the fact that $\phi_{ki}$ is a delayed exponential kernel is similar to having $r_{ki}$ additional event types with jump times $(t_i^{(l)})$ and standard exponential decay kernels. Nonetheless, since these jump times $(t_i^{(l)})$ are obtained by translation of the jump times $(t^k_i)$, the delayed exponential kernel cannot be replicated by any linear MHP with exponential kernels only  because the $r_{ki}$ fictional jump types do not self or cross excite with the other jump types. 
\end{rmk}
We give closed-form formulas for $\Upsilon_{ijk}$ and its partial derivatives given various parametric classes  for kernel $\phi_{kj}$.

\begin{prop}[Delayed exponential with delayed exponential]
We assume that the kernel
\begin{equation*}
    \phi_{kj}:=\phi^{\mathcal{DE}}_{(\boldsymbol{\omega_{kj}},\boldsymbol{\beta_{kj}},\boldsymbol{\delta_{kj}})}
\end{equation*}
is a delayed exponential kernel with parameters $\boldsymbol{\theta_{kj}}=(\boldsymbol{\omega_{kj}},\boldsymbol{\beta_{kj}},\boldsymbol{\delta_{kj}})$. Define 
  \begin{equation}
    b^s_{ijk,ll^{\prime}} =
    \begin{cases*}
      \beta_{kjl^{\prime}} & if $\delta_{kjl^{\prime}}-\delta_{kil} < s$, \\
      \beta_{kil} & if $\delta_{kjl^{\prime}}-\delta_{kil} > s$, \\
      0        & if $\delta_{kjl^{\prime}}-\delta_{kil} = s$.
    \end{cases*}
  \end{equation}
 Then for all $t,s \geq 0$
\begin{equation} 
    \Upsilon_{ijk}(t,s) = \sum_{l=1}^{r_{ki}}\sum_{l^\prime=1}^{r_{kj}}\omega_{kil}\omega_{kjl^\prime}\frac{\beta_{kil}\beta_{kjl^\prime}}{\beta_{kil}+\beta_{kjl^\prime}}(e^{- b^s_{ijk,ll^{\prime}} |s-(\delta_{kjl^{\prime}}-\delta_{kil})|} -e^{-\beta_{kil}(t-\delta_{kil})-\beta_{kjl^\prime}(t+s-\delta_{kjl^{\prime}})}) .
\end{equation}
Let $p \in [ \, r_{ki}  ]$ and $q \in [ \, r_{kj}  ]$. If $i \neq j$, the partial derivatives of $\Upsilon_{ijk}$ with respect to the parameters are given by
\begin{align} 
\begin{split}
\frac{\partial \Upsilon_{ijk} }{\partial \omega_{kip}} (t,s) ={}& \sum_{l^\prime=1}^{r_{kj}}\omega_{kjl^\prime}\frac{\beta_{kip}\beta_{kjl^\prime}}{\beta_{kip}+\beta_{kjl^\prime}}(e^{- b^s_{ijk,pl^{\prime}} |s-(\delta_{kjl^{\prime}}-\delta_{kip})|} -e^{-\beta_{kip}(t-\delta_{kip})-\beta_{kjl^\prime}(t+s-\delta_{kjl^{\prime}})}),    
\end{split}\\ 
\begin{split}
 \frac{\partial \Upsilon_{ijk} }{\partial \omega_{kjq}} (t,s) ={}& \sum_{l=1}^{r_{ki}}\omega_{kil}\frac{\beta_{kil}\beta_{kjq}}{\beta_{kil}+\beta_{kjq}}(e^{- b^s_{ijk,lq} |s-(\delta_{kjq}-\delta_{kil})|} -e^{-\beta_{kil}(t-\delta_{kil})-\beta_{kjq}(t+s-\delta_{kjq})}),    
\end{split}\\ 
\begin{split}
\frac{\partial \Upsilon_{ijk} }{\partial \beta_{kip}} (t,s) ={}& \omega_{kip}\sum_{l^\prime=1}^{r_{kj}} \omega_{kjl^\prime}\frac{\beta_{kjl^\prime}}{\beta_{kip}+\beta_{kjl^\prime}}\bigg(  \Big(\frac{\beta_{kjl^\prime}}{\beta_{kip}+\beta_{kjl^\prime}}-\beta_{kip}((\delta_{kjl^\prime}-\delta_{kip})-s)_+ \Big)e^{- b^s_{ijk,pl^\prime} |s-(\delta_{kjq}-\delta_{kil})|}\\ &\qquad+\Big(\frac{\beta_{kjl^\prime}}{\beta_{kip}+\beta_{kjl^\prime}}+\beta_{kip}(t-\delta_{kip}) \Big)e^{-\beta_{kip}(t-\delta_{kip})-\beta_{kjl^\prime}(t+s-\delta_{kjl^\prime})} \bigg),   
\end{split}\\ 
\begin{split}
\frac{\partial \Upsilon_{ijk} }{\partial \beta_{kjq}} (t,s) ={}& \omega_{kjq}\sum_{l^=1}^{r_{ki}} \omega_{kil}\frac{\beta_{kil}}{\beta_{kil}+\beta_{kjq}}\bigg(  \Big(\frac{\beta_{kil}}{\beta_{kip}+\beta_{kjl^\prime}}-\beta_{kjq}(s-(\delta_{kjq}-\delta_{kil}))_+ \Big)e^{- b^s_{ijk,lq} |s-(\delta_{kjq}-\delta_{kil})|}\\ &\qquad+\Big(\frac{\beta_{kil}}{\beta_{kil}+\beta_{kjq}}+\beta_{kjq}(t+s-\delta_{kjq}) \Big)e^{-\beta_{kil}(t-\delta_{kil})-\beta_{kjq}(t+s-\delta_{kjq})} \bigg) .    
\end{split}
\end{align}
If $i=j$, the partial derivatives of $\Upsilon_{ijk}$ with respect to the parameters are given by
\begin{align} 
\begin{split}
\frac{\partial \Upsilon_{iik} }{\partial \omega_{kip}} (t,s) ={}& \omega_{kip}\beta_{kip}e^{-\beta_{kip}s}(1-e^{-2\beta_{kip}t})\\ &+\sum_{l^\prime \in [ \, r_{ki}  ], l^\prime \neq p}\omega_{kil^\prime}\frac{\beta_{kip}\beta_{kil^\prime}}{\beta_{kip}+\beta_{kil^\prime}}\left(e^{- b^s_{iik,pl^{\prime}} |s-(\delta_{kil^{\prime}}-\delta_{kip})|}-e^{-\beta_{kip}(t-\delta_{kip})-\beta_{kil^\prime}(t+s-\delta_{kil^{\prime}})}\right)\\ &+\sum_{l \in [ \, r_{ki}  ], l \neq p}\omega_{kil}\frac{\beta_{kip}\beta_{kil}}{\beta_{kip}+\beta_{kil}}\left(e^{- b^s_{iik,lp} |s-(\delta_{kip}-\delta_{kil})|}-e^{-\beta_{kil}(t-\delta_{kil})-\beta_{kip}(t+s-\delta_{kip})}\right),   
\end{split}\\
\begin{split}
\frac{\partial \Upsilon_{iik} }{\partial \beta_{kip}} (t,s) ={}& \frac{\omega_{kip}^2}{2}\bigg( 1+ \big( \beta_{kip}(2t+s-2\delta_{kip})-1 \big)e^{-\beta_{kip}(2t+s-2\delta_{kip}) } \bigg)
\\ &+\omega_{kip}\sum_{l^\prime \in [ \, r_{ki}  ], l^\prime \neq p} \omega_{kil^\prime}\frac{\beta_{kil^\prime}}{\beta_{kip}+\beta_{kil^\prime}} \bigg[  \bigg(\frac{\beta_{kil^\prime}}{\beta_{kip}+\beta_{kil^\prime}}
\\ &\qquad \qquad \qquad \qquad -\beta_{kip}\Big((\delta_{kil^\prime}-\delta_{kip})-s\Big)_+ \bigg)e^{- b^s_{iik,pl^\prime} |s-(\delta_{kiq}-\delta_{kil})|}\\ &\qquad \qquad \qquad \qquad+\bigg(\frac{\beta_{kil^\prime}}{\beta_{kip}+\beta_{kil^\prime}}+\beta_{kip}(t-\delta_{kip}) \bigg)e^{-\beta_{kip}(t-\delta_{kip})-\beta_{kil^\prime}(t+s-\delta_{kil^\prime})} \bigg]\\ &+ \omega_{kip}\sum_{l \in [ \, r_{ki}  ], l \neq p} \omega_{kil}\frac{\beta_{kil}}{\beta_{kil}+\beta_{kip}}\bigg[  \bigg(\frac{\beta_{kil}}{\beta_{kip}+\beta_{kjl^\prime}}\\ &\qquad \qquad \qquad \qquad-\beta_{kip}\Big(s-(\delta_{kip}-\delta_{kil})\Big)_+ \bigg)e^{- b^s_{iik,lp} |s-(\delta_{kip}-\delta_{kil})|}\\ &\qquad \qquad \qquad \qquad+\bigg(\frac{\beta_{kil}}{\beta_{kil}+\beta_{kip}}+\beta_{kip}(t+s-\delta_{kip}) \bigg)e^{-\beta_{kil}(t-\delta_{kil})-\beta_{kip}(t+s-\delta_{kip})} \bigg] .  
\end{split}
\end{align}
\end{prop}

\subsubsection{Gaussian}
We assume that the kernel
\begin{equation*}
    \phi_{ki}:=\phi^{\mathcal{N}}_{(\boldsymbol{\omega_{ki}},\boldsymbol{\beta_{ki}},\boldsymbol{\delta_{ki}})}
\end{equation*}
is a Gaussian kernel with parameters $\boldsymbol{\theta_{ki}}=(\boldsymbol{\omega_{ki}},\boldsymbol{\beta_{ki}},\boldsymbol{\delta_{ki}})$, but do not make any assumptions on $\phi_{k^\prime,i^\prime}$ for $(k^\prime,i^\prime)\neq (k,i)$. We write $f_{\mathcal{N}}$ and $F_{\mathcal{N}}$ for the standard Gaussian density and distribution function respectively.

\begin{prop}
For all $x\geq 0$, we have
\begin{equation}\label{eq:psi_jk delayed exp}
    \psi_{ki}(x)=\sum_{l=1}^{r_{ki}} \omega_{kil} \left( F_{\mathcal{N}}\left(\frac{x-\delta_{kil}}{\beta_{kil}}\right)-F_{\mathcal{N}}\left(-\frac{\delta_{kil}}{\beta_{kil}}\right) \right).
\end{equation}
Fix $p \in [ \, r_{ki}  ]$. The partial derivatives of $\phi_{ki}$ and $\psi_{ki}$ with respect to model parameters are given by 
\begin{align} 
\frac{\partial \phi_{ki} }{\partial \omega_{kip}} (x) ={}& \frac{1}{\beta_{kip} \sqrt{2 \pi}} \exp \left(-\frac{(x-\delta_{kip})^{2}}{2 \beta_{kip}^{2}}\right)   ,\\ 
\frac{\partial \phi_{ki} }{\partial \beta_{kip}} (x) ={}& \frac{\omega_{kip}}{\beta_{kip}^2 \sqrt{2 \pi}} \exp \left(-\frac{(x-\delta_{kip})^{2}}{2 \beta_{kip}^{2}}\right)\left(\left(\frac{x-\delta_{kip}}{\beta_{kip}}\right)^2-1\right)   ,\\ 
 \frac{\partial \phi_{ki} }{\partial \delta_{kip}} (x)={}& \frac{\omega_{kip}}{\beta_{kip}^3 \sqrt{2\pi}}(x-\delta_{kip})\exp\left(-\frac{(x-\delta_{kip})^2}{2\beta_{kip}^2}\right),\\ 
\begin{split}
\frac{\partial \psi_{ki} }{\partial \omega_{kip}} (x) ={}& F_{\mathcal{N}}\bigg(\frac{x-\delta_{kip}}{\beta_{kip}}\bigg)-F_{\mathcal{N}}\bigg(-\frac{\delta_{kip}}{\beta_{kip}}\bigg) ,   
\end{split}\\ 
\begin{split}
\frac{\partial \psi_{ki} }{\partial \beta_{kip}} (x) ={}& \frac{\omega_{kip}}{\beta_{kip}^2} \bigg( (\delta_{kip}-x) f_{\mathcal{N}}\bigg(\frac{x-\delta_{kip}}{\beta_{kip}}\bigg)-\delta_{kip}f_{\mathcal{N}}\bigg(-\frac{\delta_{kip}}{\beta_{kip}}\bigg) \bigg) ,   
\end{split}\\ 
\begin{split}
\frac{\partial \psi_{ki} }{\partial \delta_{kip}} (x) ={}& -\frac{\omega_{kip}}{\beta_{kip}}\bigg( f_{\mathcal{N}}\bigg(\frac{x-\delta_{kip}}{\beta_{kip}}\bigg)-f_{\mathcal{N}}\bigg(-\frac{\delta_{kip}}{\beta_{kip}}\bigg) \bigg).    
\end{split}
\end{align}
\end{prop}
We give closed-form formulas for $\Upsilon_{ijk}$ and its partial derivatives given various parametric classes  for kernel $\phi_{kj}$.

\begin{prop}[Gaussian with Gaussian]
We assume that the kernel
\begin{equation*}
\phi_{kj}:=\phi^{\mathcal{N}}_{(\boldsymbol{\omega_{kj}},\boldsymbol{\beta_{kj}},\boldsymbol{\delta_{kj}})}
\end{equation*}
is a Gaussian kernel with parameters $\boldsymbol{\theta_{kj}}=(\boldsymbol{\omega_{kj}},\boldsymbol{\beta_{kj}},\boldsymbol{\delta_{kj}})$. Define 
\begin{align} 
b_{ijk,ll^\prime} &=  \frac{\beta_{kil}\beta_{kjl^\prime}}{\sqrt{\beta_{kil}^2+\beta_{kjl^\prime}^2}}, \\ 
d^s_{ijk,ll^\prime} &=  \frac{\beta_{kjl^\prime}^2}{\beta_{kil}^2+\beta_{kjl^\prime}^2}\delta_{kil}+\frac{\beta_{kil}^2}{\beta_{kil}^2+\beta_{kjl^\prime}^2}(\delta_{kjl^\prime}-s).
\end{align}
Then for all $t,s \geq 0$
\begin{equation}
    \Upsilon_{ijk}(t,s)=\sum_{l=1}^{r_{ki}}\sum_{l^\prime=1}^{r_{kj}}\omega_{kil}\omega_{kjl^\prime}\frac{\exp\left(-\frac{1}{2}\frac{(s-(\delta_{kjl^\prime}-\delta_{kil}))^2}{\beta_{kil}^2+\beta_{kjl^\prime}^2}\right)}{\sqrt{2\pi(\beta_{kil}^2+\beta_{kjl^\prime}^2)}}\left[F_{\mathcal{N}}\Big(\frac{t-d^s_{ijk,ll^\prime}}{b_{ijk,ll^\prime}}\Big)-F_{\mathcal{N}}\Big(-\frac{d^s_{ijk,ll^\prime}}{b_{ijk,ll^\prime}}\Big)\right].
\end{equation}
Let $p \in [ \, r_{ki}  ]$ and $q \in [ \, r_{kj}  ]$. If $i \neq j$, the partial derivatives of $\Upsilon_{ijk}$ with respect to the parameters are given by
\begin{align} 
\frac{\partial \Upsilon_{ijk} }{\partial \omega_{kip}} (t,s) ={}& \sum_{l^\prime=1}^{r_{kj}}\omega_{kjl^\prime} \frac{\exp(-\frac{1}{2}\frac{(s-(\delta_{kjl^\prime}-\delta_{kip}))^2}{\beta_{kip}^2+\beta_{kjl^\prime}^2})}{\sqrt{2\pi(\beta_{kip}^2+\beta_{kjl^\prime}^2)}}\left[F_{\mathcal{N}}\left(\frac{t-d^s_{ijk,pl^\prime}}{b_{ijk,pl^\prime}}\right)-F_{\mathcal{N}}\left(-\frac{d^s_{ijk,pl^\prime}}{b_{ijk,pl^\prime}}\right)\right] ,\\ 
\frac{\partial \Upsilon_{ijk} }{\partial \omega_{kjq}} (t,s) ={}& \sum_{l=1}^{r_{ki}}  \omega_{kil}\frac{\exp(-\frac{1}{2}\frac{(s-(\delta_{kjq}-\delta_{kil}))^2}{\beta_{kil}^2+\beta_{kjq}^2})}{\sqrt{2\pi(\beta_{kil}^2+\beta_{kjq}^2)}}\left[F_{\mathcal{N}}\left(\frac{t-d^s_{ijk,lq}}{b_{ijk,lq}}\right)-F_{\mathcal{N}}\left(-\frac{d^s_{ijk,lq}}{b_{ijk,lq}}\right)\right],\\ 
\begin{split}
\frac{\partial \Upsilon_{ijk} }{\partial \beta_{kip}} (t,s) ={}& \omega_{kip}\sum_{l^\prime=1}^{r_{kj}}\omega_{kjl^\prime}\frac{\exp(-\frac{1}{2}\frac{(s-(\delta_{kjl^\prime}-\delta_{kip}))^2}{\beta_{kip}^2+\beta_{kjl^\prime}^2})}{\sqrt{2\pi(\beta_{kip}^2+\beta_{kjl^\prime}^2)}} \\ &\qquad \times \Bigg[ \frac{\beta_{kip}}{\beta_{kip}^2+\beta_{kjl^\prime}^2}\left(\frac{(s-\delta_{kjl^\prime}+\delta_{kip})^2}{\beta_{kip}^2+\beta_{kjl^\prime}^2}-1\right)\bigg(F_{\mathcal{N}}\Big(\frac{t-d^s_{ijk,pl^\prime}}{b_{ijk,pl^\prime}}\Big)-F_{\mathcal{N}}\Big(-\frac{d^s_{ijk,pl^\prime}}{b_{ijk,pl^\prime}}\Big)\bigg) \\
&\qquad-\frac{\beta_{kjl^\prime}}{\beta_{kip} \sqrt{\beta_{kip}^2+\beta_{kjl^\prime}^2}}\bigg( \Big( \frac{2\beta_{kip}(\delta_{kjl^\prime}-\delta_{kip}-s) }{\beta_{kip}^2+\beta_{kjl^\prime}^2}+\frac{t-d^s_{ijk,pl^\prime}}{\beta_{kip}}\Big) f_{\mathcal{N}}\Big(\frac{t-d^s_{ijk,pl^\prime}}{b_{ijk,pl^\prime}}\Big)\\
&\qquad-\Big( \frac{2\beta_{kip}(\delta_{kjl^\prime}-\delta_{kip}-s) }{\beta_{kip}^2+\beta_{kjl^\prime}^2}-\frac{d^s_{ijk,pl^\prime}}{\beta_{kip}}\Big)f_{\mathcal{N}}\Big(-\frac{d^s_{ijk,pl^\prime}}{b_{ijk,pl^\prime}}\Big)\bigg) \Bigg],   
\end{split}\\ 
\begin{split}
\frac{\partial \Upsilon_{ijk} }{\partial \beta_{kjq}} (t,s) ={}& \omega_{kjq}\sum_{l=1}^{r_{ki}}\omega_{kil}\frac{\exp(-\frac{1}{2}\frac{(s-(\delta_{kjq}-\delta_{kil}))^2}{\beta_{kil}^2+\beta_{kjq}^2})}{\sqrt{2\pi(\beta_{kil}^2+\beta_{kjq}^2)}} \\ &\qquad \times \Bigg[ \frac{\beta_{kjq}}{\beta_{kil}^2+\beta_{kjq}^2}\bigg(\frac{(s-\delta_{kjq}+\delta_{kil})^2}{\beta_{kil}^2+\beta_{kjq}^2}-1\bigg)\bigg(F_{\mathcal{N}}\Big(\frac{t-d^s_{ijk,lq}}{b_{ijk,lq}}\Big)-F_{\mathcal{N}}\Big(-\frac{d^s_{ijk,lq}}{b_{ijk,lq}}\Big)\bigg) \\
&\qquad-\frac{\beta_{kil}}{\beta_{kjq} \sqrt{\beta_{kil}^2+\beta_{kjq}^2}}\bigg( \Big( \frac{2\beta_{kjq}(\delta_{kil}-\delta_{kjq}+s) }{\beta_{kil}^2+\beta_{kjq}^2}-\frac{t-d^s_{ijk,lq}}{\beta_{kjq}}\Big) f_{\mathcal{N}}\Big(\frac{t-d^s_{ijk,lq}}{b_{ijk,lq}}\Big)\\
&\qquad-\Big( \frac{2\beta_{kjq}(\delta_{kil}-\delta_{kjq}+s) }{\beta_{kil}^2+\beta_{kjq}^2}+\frac{d^s_{ijk,lq}}{\beta_{kjq}}\Big)f_{\mathcal{N}}\Big(-\frac{d^s_{ijk,lq}}{b_{ijk,lq}}\Big)\bigg) \Bigg],   
\end{split}\\ 
\begin{split}
\frac{\partial \Upsilon_{ijk} }{\partial \delta_{kip}} (t,s) ={}& \omega_{kip}\sum_{l^\prime=1}^{r_{kj}}\omega_{kjl^\prime}\frac{\exp(-\frac{1}{2}\frac{(s-(\delta_{kjl^\prime}-\delta_{kip}))^2}{\beta_{kip}^2+\beta_{kjl^\prime}^2})}{\sqrt{2\pi(\beta_{kip}^2+\beta_{kjl^\prime}^2)}} \\ &\qquad\times\Bigg[\frac{\delta_{kjl^\prime}-s-\delta_{kip}}{\beta_{kip}^2+\beta_{kjl^\prime}^2}\bigg(F_{\mathcal{N}}\Big(\frac{t-d^s_{ijk,pl^\prime}}{b_{ijk,pl^\prime}}\Big)-F_{\mathcal{N}}\Big(-\frac{d^s_{ijk,pl^\prime}}{b_{ijk,pl^\prime}}\Big)\bigg)\\ &\qquad\qquad-\frac{\beta_{kjl^\prime}}{\beta_{kip}\sqrt{\beta_{kip}^2+\beta_{kjl^\prime}^2}}\left(f_{\mathcal{N}}\left(\frac{t-d^s_{ijk,pl^\prime}}{b_{ijk,pl^\prime}}\right)-f_{\mathcal{N}}\left(-\frac{d^s_{ijk,pl^\prime}}{b_{ijk,pl^\prime}}\right)\right) \Bigg],   
\end{split}\\ 
\begin{split}
\frac{\partial \Upsilon_{ijk} }{\partial \delta_{kjq}} (t,s) ={}& \sum_{l=1}^{r_{ki}}\omega_{kil}\omega_{kjq}\frac{\exp(-\frac{1}{2}\frac{(s-(\delta_{kjq}-\delta_{kil}))^2}{\beta_{kil}^2+\beta_{kjq}^2})}{\sqrt{2\pi(\beta_{kil}^2+\beta_{kjq}^2)}} \\ &\qquad\times\Bigg[\frac{s+\delta_{kil}-\delta_{kjq}}{\beta_{kil}^2+\beta_{kjq}^2}\left(F_{\mathcal{N}}\bigg(\frac{t-d^s_{ijk,lq}}{b_{ijk,lq}}\bigg)-F_{\mathcal{N}}\bigg(-\frac{d^s_{ijk,lq}}{b_{ijk,lq}}\bigg)\right)\\ &\qquad\qquad-\frac{\beta_{kil}}{\beta_{kjq}\sqrt{\beta_{kil}^2+\beta_{kjq}^2}}\left(f_{\mathcal{N}}\bigg(\frac{t-d^s_{ijk,lq}}{b_{ijk,lq}}\bigg)-f_{\mathcal{N}}\bigg(-\frac{d^s_{ijk,lq}}{b_{ijk,lq}}\bigg)\right) \Bigg] .    
\end{split}
\end{align}
If $i=j$, the partial derivatives of $\Upsilon_{ijk}$ with respect to the parameters are given by
\begin{align} 
\begin{split}
\frac{\partial \Upsilon_{iik} }{\partial \omega_{kip}} (t,s) ={}& 2\omega_{kip}\frac{\exp(-\frac{s^2}{4\beta_{kip}^2})}{\sqrt{4\beta_{kip}^2\pi}}\Big(F_{\mathcal{N}}\Big(\frac{\sqrt{2}(t-\delta_{kip})}{\beta_{kip}}\Big)-F_{\mathcal{N}}\Big(-\frac{\sqrt{2}\delta_{kip}}{\beta_{kip}}\Big)\Big)
\\ &+\sum_{l^\prime \in [ \, r_{ki}  ], l^\prime \neq p}\omega_{kil^\prime}\frac{\exp(-\frac{1}{2}\frac{(s-(\delta_{kil^\prime}-\delta_{kip}))^2}{\beta_{kip}^2+\beta_{kil^\prime}^2})}{\sqrt{2\pi(\beta_{kip}^2+\beta_{kil^\prime}^2)}}\Big(F_{\mathcal{N}}\Big(\frac{t-d^s_{iik,pl^\prime}}{b_{iik,pl^\prime}}\Big)-F_{\mathcal{N}}\Big(-\frac{d^s_{iik,pl^\prime}}{b_{iik,pl^\prime}}\Big)\Big)
\\ &+\sum_{l \in [ \, r_{ki}  ], l \neq p}\omega_{kil}\frac{\exp(-\frac{1}{2}\frac{(s-(\delta_{kip}-\delta_{kil}))^2}{\beta_{kil}^2+\beta_{kip}^2})}{\sqrt{2\pi(\beta_{kil}^2+\beta_{kip}^2)}}\Big(F_{\mathcal{N}}\Big(\frac{t-d^s_{iik,lp}}{b_{iik,lp}}\Big)-F_{\mathcal{N}}\Big(-\frac{d^s_{iik,lp}}{b_{iik,lp}}\Big)\Big)  ,   
\end{split}\\
\begin{split}
\frac{\partial \Upsilon_{iik} }{\partial \beta_{kip}} (t,s) ={}& \omega_{kip}^2\frac{\exp(-\frac{s^2}{4\beta_{kip}^2})}{\sqrt{4\pi\beta_{kip}^2}}\Bigg[ \frac{s^2-2\beta_{kil}^2}{2\beta_{kil}^3}  \Bigg(F_{\mathcal{N}}\left(\frac{\sqrt{2}(t-\delta_{kip}+s/2)}{\beta_{kip}}\right)-F_{\mathcal{N}}\left(-\frac{\sqrt{2}(\delta_{kip}-s/2)}{\beta_{kip}}\right) \Bigg)
\\ &-\Bigg(\frac{\sqrt{2}(t-\delta_{kip}+s/2)}{\beta_{kip}^2} f_{\mathcal{N}}\left(\frac{\sqrt{2}(t-\delta_{kip}+s/2)}{\beta_{kip}}\right)-\frac{\sqrt{2}(-\delta_{kip}+s/2)}{\beta_{kip}^2}f_{\mathcal{N}}\left(-\frac{\sqrt{2}(\delta_{kip}-s/2)}{\beta_{kip}}\right) \Bigg) \Bigg]
\\ &+\omega_{kip}\sum_{l^\prime \in [ \, r_{ki}  ], l^\prime \neq p}\omega_{kil^\prime}\frac{\exp(-\frac{1}{2}\frac{(s-(\delta_{kil^\prime}-\delta_{kip}))^2}{\beta_{kip}^2+\beta_{kil^\prime}^2})}{\sqrt{2\pi(\beta_{kip}^2+\beta_{kil^\prime}^2)}} \\ &\qquad \qquad \qquad \times \Bigg[ \frac{\beta_{kip}}{\beta_{kip}^2+\beta_{kil^\prime}^2}\left(\frac{(s-\delta_{kjl^\prime}+\delta_{kip})^2}{\beta_{kip}^2+\beta_{kil^\prime}^2}-1\right)\bigg(F_{\mathcal{N}}\Big(\frac{t-d^s_{iik,pl^\prime}}{b_{iik,pl^\prime}}\Big)-F_{\mathcal{N}}\Big(-\frac{d^s_{iik,pl^\prime}}{b_{iik,pl^\prime}}\Big)\bigg) \\
&\qquad \qquad \qquad-\frac{\beta_{kjl^\prime}}{\beta_{kip} \sqrt{\beta_{kip}^2+\beta_{kil^\prime}^2}}\bigg( \Big( \frac{2\beta_{kip}(\delta_{kil^\prime}-\delta_{kip}-s) }{\beta_{kip}^2+\beta_{kil^\prime}^2}+\frac{t-d^s_{iik,pl^\prime}}{\beta_{kip}}\Big) f_{\mathcal{N}}\Big(\frac{t-d^s_{iik,pl^\prime}}{b_{iik,pl^\prime}}\Big)\\
&\qquad \qquad \qquad-\Big( \frac{2\beta_{kip}(\delta_{kil^\prime}-\delta_{kip}-s) }{\beta_{kip}^2+\beta_{kil^\prime}^2}-\frac{d^s_{iik,pl^\prime}}{\beta_{kip}}\Big)f_{\mathcal{N}}\Big(-\frac{d^s_{iik,pl^\prime}}{b_{iik,pl^\prime}}\Big)\bigg) \Bigg] \\ &+ \omega_{kip}\sum_{l \in [ \, r_{ki}  ], l \neq p}\omega_{kil}\frac{\exp(-\frac{1}{2}\frac{(s-(\delta_{kip}-\delta_{kil}))^2}{\beta_{kil}^2+\beta_{kip}^2})}{\sqrt{2\pi(\beta_{kil}^2+\beta_{kip}^2)}} \\ &\qquad \qquad \qquad \times \Bigg[ \frac{\beta_{kip}}{\beta_{kil}^2+\beta_{kip}^2}\left(\frac{(s-\delta_{kip}+\delta_{kil})^2}{\beta_{kil}^2+\beta_{kip}^2}-1 \right)\bigg(F_{\mathcal{N}}\Big(\frac{t-d^s_{iik,lp}}{b_{iik,lp}}\Big)-F_{\mathcal{N}}\Big(-\frac{d^s_{iik,lp}}{b_{iik,lp}}\Big)\bigg) \\
&\qquad \qquad \qquad-\frac{\beta_{kil}}{\beta_{kip} \sqrt{\beta_{kil}^2+\beta_{kip}^2}}\bigg( \Big( \frac{2\beta_{kip}(\delta_{kil}-\delta_{kip}+s) }{\beta_{kil}^2+\beta_{kip}^2}-\frac{t-d^s_{iik,lp}}{\beta_{kip}}\Big) f_{\mathcal{N}}\Big(\frac{t-d^s_{iik,lp}}{b_{iik,lp}}\Big)\\
&\qquad \qquad \qquad-\Big( \frac{2\beta_{kip}(\delta_{kil}-\delta_{kip}+s) }{\beta_{kil}^2+\beta_{kip}^2}+\frac{d^s_{iik,lp}}{\beta_{kip}}\Big)f_{\mathcal{N}}\Big(-\frac{d^s_{iik,lp}}{b_{iik,lp}}\Big)\bigg) \Bigg],  
\end{split}\\
\begin{split}
\frac{\partial \Upsilon_{iik} }{\partial \delta_{kip}} (t,s) ={}& -\omega_{kip}^2\frac{\exp(-\frac{s^2}{4\beta_{kip}^2})}{\sqrt{2\pi}\beta_{kip}^2}\bigg(f_{\mathcal{N}}(\frac{\sqrt{2}(t-\delta_{kip}+s/2)}{\beta_{kip}})-f_{\mathcal{N}}(-\sqrt{2}\frac{\delta_{kip}-s/2}{\beta_{kip}}) \bigg)
\\ &+\omega_{kip}\sum_{l^\prime \in [ \, r_{ki}  ], l^\prime \neq p}\omega_{kil^\prime}\frac{\exp(-\frac{1}{2}\frac{(s-(\delta_{kil^\prime}-\delta_{kip}))^2}{\beta_{kip}^2+\beta_{kil^\prime}^2})}{\sqrt{2\pi(\beta_{kip}^2+\beta_{kil^\prime}^2)}} \\ &\qquad \qquad \qquad\times \Bigg(\frac{\delta_{kil^\prime}-s-\delta_{kip}}{\beta_{kip}^2+\beta_{kil^\prime}^2}(F_{\mathcal{N}}(\frac{t-d^s_{iik,pl^\prime}}{b_{iik,pl^\prime}})-F_{\mathcal{N}}(-\frac{d^s_{iik,pl^\prime}}{b_{iik,pl^\prime}}))\\ &\qquad \qquad \qquad-\frac{\beta_{kil^\prime}}{\beta_{kip}\sqrt{\beta_{kip}^2+\beta_{kil^\prime}^2}}(f_{\mathcal{N}}(\frac{t-d^s_{iik,pl^\prime}}{b_{iik,pl^\prime}})-f_{\mathcal{N}}(-\frac{d^s_{iik,pl^\prime}}{b_{iik,pl^\prime}})) \Bigg) 
\\ &+ \omega_{kip}\sum_{l \in [ \, r_{ki}  ], l \neq p}\omega_{kil}\frac{\exp(-\frac{1}{2}\frac{(s-(\delta_{kip}-\delta_{kil}))^2}{\beta_{kil}^2+\beta_{kip}^2})}{\sqrt{2\pi(\beta_{kil}^2+\beta_{kip}^2)}} \\ &\qquad \qquad \qquad \times\Bigg(\frac{\delta_{kip}+s-\delta_{kil}}{\beta_{kil}^2+\beta_{kip}^2}(F_{\mathcal{N}}(\frac{t-d^s_{iik,lp}}{b_{iik,lp}})-F_{\mathcal{N}}(-\frac{d^s_{iik,lp}}{b_{iik,lp}}))\\ &\qquad \qquad \qquad-\frac{\beta_{kil}}{\beta_{kip}\sqrt{\beta_{kil}^2+\beta_{kip}^2}}(f_{\mathcal{N}}(\frac{t-d^s_{iik,lp}}{b_{iik,lp}})-f_{\mathcal{N}}(-\frac{d^s_{iik,lp}}{b_{iik,lp}})) \Bigg).  
\end{split}
\end{align}
\end{prop}
We can also obtain $\Upsilon$ for mixtures including both Gaussians and Exponentials (but see \cref{rmk:numerical_instability_gaussexp}).
\begin{prop}[Gaussian with Exponential]
We assume that the kernel
\begin{equation*}
\phi_{kj}:=\phi^{\mathcal{E}}_{(\boldsymbol{\omega_{kj}},\boldsymbol{\beta_{kj}})}
\end{equation*}
is an exponential kernel with parameters $\boldsymbol{\theta_{kj}}=\big( \boldsymbol{\omega_{kj}},\boldsymbol{\beta_{kj}} \big)  $. Then for all $t,s \geq 0$
\begin{align}
\Upsilon_{ijk}(t,s) &= \sum_{l=1}^{r_{ki}}\sum_{l^\prime=1}^{r_{kj}}\omega_{kil}\omega_{kjl^\prime}\beta_{kjl^\prime}\exp\bigg(-\beta_{kjl^\prime}\bigg(\delta_{kil}+s-\frac{\beta_{kjl^\prime} \beta_{kil}^2}{2}\bigg) \bigg) \\ 
&\qquad \qquad  \times\bigg( F_{\mathcal{N}}\bigg(\frac{t-(\delta_{kil}-\beta_{kjl^\prime} \beta_{kil}^2)}{\beta_{kil}}\bigg)-F_{\mathcal{N}}\bigg(-\frac{\delta_{kil}-\beta_{kjl^\prime}\beta_{kil}^2}{\beta_{kil}}\bigg) \bigg).
\end{align}
Let $p \in [ \, r_{ki}  ]$ and $q \in [ \, r_{kj}  ]$. The partial derivatives of $\Upsilon_{ijk}$ with respect to the parameters are given by
\begin{align} 
\begin{split}
\frac{\partial \Upsilon_{ijk} }{\partial \omega_{kip}} (t,s) ={}& \sum_{l^\prime=1}^{r_{kj}} \omega_{kjl^\prime}\beta_{kjl^\prime}\exp\bigg(-\beta_{kjl^\prime}\bigg(\delta_{kip}+s-\frac{\beta_{kjl^{\prime}} \beta_{kip}^2}{2}\bigg) \bigg) \\ 
&\qquad  \times\bigg( F_{\mathcal{N}}\bigg(\frac{t-(\delta_{kip}-\beta_{kjl^{\prime}} \beta_{kip}^2)}{\beta_{kip}}\bigg)-F_{\mathcal{N}}\bigg(-\frac{\delta_{kip}-\beta_{kjl^{\prime}}\beta_{kip}^2}{\beta_{kip}}\bigg) \bigg),   
\end{split}\\ 
\begin{split}
\frac{\partial \Upsilon_{ijk} }{\partial \omega_{kjq}} (t,s) ={}& \sum_{l=1}^{r_{ki}} \omega_{kil}\beta_{kjq}\exp\bigg(-\beta_{kjq}\bigg(\delta_{kil}+s-\frac{\beta_{kjq} \beta_{kil}^2}{2}\bigg) \bigg) \\ 
&\qquad\times\bigg( F_{\mathcal{N}}\bigg(\frac{t-(\delta_{kil}-\beta_{kjq} \beta_{kil}^2)}{\beta_{kil}}\bigg)-F_{\mathcal{N}}\bigg(-\frac{\delta_{kil}-\beta_{kjq}\beta_{kil}^2}{\beta_{kil}}\bigg) \bigg),   
\end{split}\\ 
\begin{split}
\frac{\partial \Upsilon_{ijk} }{\partial \beta_{kip}} (t,s) ={}& \omega_{kip} \sum_{l^\prime=1}^{r_{kj}}\omega_{kjl^\prime}\beta_{kjl^\prime}\exp\bigg(-\beta_{kjl^\prime}\bigg(\delta_{kip}+s-\frac{\beta_{kjl^\prime} \beta_{kip}^2}{2}\bigg) \bigg) \\ &\qquad \qquad \times\Bigg( \beta_{kip}\beta_{kjl^\prime}^2\bigg( F_{\mathcal{N}}\Big(\frac{t-(\delta_{kip}-\beta_{kjl^\prime} \beta_{kip}^2)}{\beta_{kip}}\Big)-F_{\mathcal{N}}\Big(-\frac{(\delta_{kip}-\beta_{kjl^\prime} \beta_{kip}^2)}{\beta_{kip}}\Big) \bigg)\\ &  \qquad \qquad+((\beta_{kjl^\prime}-\frac{t-\delta_{kip}}{\beta_{kip}^2})f_{\mathcal{N}}\bigg(\frac{t-(\delta_{kip}-\beta_{kjl^\prime} \beta_{kil}^2)}{\beta_{kip}}\bigg)\\ &\qquad \qquad-(\beta_{kjl^\prime}+\frac{\delta_{kip}}{\beta_{kip}^2})f_{\mathcal{N}}\bigg(-\frac{\delta_{kip}-\beta_{kjl^\prime}\beta_{kip}^2}{\beta_{kip}}\bigg)) \Bigg),
\end{split}\\ 
\begin{split}
\frac{\partial \Upsilon_{ijk} }{\partial \delta_{kip}} (t,s) ={}& \omega_{kip} \sum_{l^\prime=1}^{r_{kj}}\omega_{kjl^\prime}\beta_{kjl^\prime}\exp\bigg(-\beta_{kjl^\prime}\bigg(\delta_{kip}+s-\frac{\beta_{kjl^\prime} \beta_{kip}^2}{2}\bigg) \bigg) \\ &\qquad \qquad \times\Bigg( -\beta_{kjl^\prime}\bigg( F_{\mathcal{N}}\Big(\frac{t-(\delta_{kip}-\beta_{kjl^\prime} \beta_{kip}^2)}{\beta_{kip}}\Big)\\ &\qquad  \qquad-F_{\mathcal{N}}\Big(-\frac{(\delta_{kip}-\beta_{kjl^\prime} \beta_{kip}^2)}{\beta_{kip}}\Big) \bigg)\\ & \qquad  \qquad -\frac{1}{\beta_{kip}} \Bigg(f_{\mathcal{N}}\bigg(\frac{t-(\delta_{kip}-\beta_{kjl^\prime} \beta_{kil}^2)}{\beta_{kip}}\bigg)-f_{\mathcal{N}}\bigg(-\frac{\delta_{kip}-\beta_{kjl^\prime}\beta_{kip}^2}{\beta_{kip}}\bigg) \Bigg) \Bigg),  
\end{split}\\ 
\begin{split}
\frac{\partial \Upsilon_{ijk} }{\partial \beta_{kjq}} (t,s) ={}& \omega_{kjq}\beta_{kjq} \sum_{l=1}^{r_{ki}}\omega_{kil} \exp\bigg(-\beta_{kjq}\bigg(\delta_{kil}+s-\frac{\beta_{kjq} \beta_{kil}^2}{2}\bigg) \bigg)  \\ & \qquad \qquad \qquad \times\Bigg( (\frac{1}{\beta_{kjq}}-\delta_{kil}-s+\frac{\beta_{kjq}\beta_{kil}^2}{2} )\bigg( F_{\mathcal{N}}\Big(\frac{t-(\delta_{kil}-\beta_{kjq} \beta_{kil}^2)}{\beta_{kil}}\Big)\\ &\qquad \qquad \qquad-F_{\mathcal{N}}\Big(-\frac{(\delta_{kil}-\beta_{kjq} \beta_{kil}^2)}{\beta_{kil}}\Big) \bigg)\\ & \qquad \qquad \qquad +\beta_{kil}\Bigg(f_{\mathcal{N}}\bigg(\frac{t-(\delta_{kil}-\beta_{kjq} \beta_{kil}^2)}{\beta_{kil}}\bigg)-f_{\mathcal{N}}\bigg(-\frac{\delta_{kil}-\beta_{kjq}\beta_{kil}^2}{\beta_{kil}}\bigg) \Bigg) \Bigg).
\end{split}
\end{align}
\end{prop}

\begin{prop}[Exponential with Gaussian]
We assume that the kernel
\begin{equation}\label{eq:upsilon_exp_gaussian}
\phi_{kj}:=\phi^{\mathcal{N}}_{(\boldsymbol{\omega_{kj}},\boldsymbol{\beta_{kj}},\boldsymbol{\delta_{kj}})}
\end{equation}
is a Gaussian kernel with parameters $\boldsymbol{\theta_{kj}}=(\boldsymbol{\omega_{kj}},\boldsymbol{\beta_{kj}},\boldsymbol{\delta_{kj}})$. Then for all $t,s\geq 0$

\begin{align}
\Upsilon_{ijk}(t,s) &= \sum_{l=1}^{r_{ki}}\sum_{l^\prime=1}^{r_{kj}}\omega_{kil}\omega_{kjl^\prime}\beta_{kil}\exp\bigg(-\beta_{kil}\bigg(\delta_{kjl^{\prime}}-s-\frac{\beta_{kil} \beta_{kjl^\prime}^2}{2}\bigg) \bigg) \\ & \qquad \quad
\times\bigg[ F_{\mathcal{N}}\bigg(\frac{t-(\delta_{kjl^{\prime}}-s-\beta_{kil} \beta_{kjl^\prime}^2)}{\beta_{kjl^\prime}}\bigg)-F_{\mathcal{N}}\bigg(-\frac{\delta_{kjl^{\prime}}-s-\beta_{kil}\beta_{kjl^\prime}^2}{\beta_{kjl^\prime}}\bigg) \bigg].
\end{align}
Let $p \in [ \, r_{ki}  ]$ and $q \in [ \, r_{kj}  ]$. The partial derivatives of $\Upsilon_{ijk}$ with respect to the parameters are given by
\begin{align} 
\begin{split}
\frac{\partial \Upsilon_{ijk} }{\partial \omega_{kip}} (t,s) ={}& \sum_{l^\prime=1}^{r_{kj}} \omega_{kjl^\prime}\beta_{kip}\exp\bigg(-\beta_{kip}\bigg(\delta_{kjl^{\prime}}-s-\frac{\beta_{kip} \beta_{kjl^\prime}^2}{2}\bigg) \bigg) \\ 
&\qquad \times\bigg( F_{\mathcal{N}}\bigg(\frac{t-(\delta_{kjl^{\prime}}-s-\beta_{kip} \beta_{kjl^\prime}^2)}{\beta_{kjl^\prime}}\bigg)-F_{\mathcal{N}}\bigg(-\frac{\delta_{kjl^{\prime}}-s-\beta_{kip}\beta_{kjl^\prime}^2}{\beta_{kjl^\prime}}\bigg) \bigg),   
\end{split}\\ 
\begin{split}
\frac{\partial \Upsilon_{ijk} }{\partial \omega_{kjq}} (t,s) ={}& \sum_{l=1}^{r_{ki}} \omega_{kil}\beta_{kil}\exp\bigg(-\beta_{kil}\bigg(\delta_{kjq}-s-\frac{\beta_{kil} \beta_{kjq}^2}{2}\bigg) \bigg) \\ 
&\qquad \times\bigg( F_{\mathcal{N}}\bigg(\frac{t-(\delta_{kjq}-s-\beta_{kil} \beta_{kjq}^2)}{\beta_{kjq}}\bigg)-F_{\mathcal{N}}\bigg(-\frac{\delta_{kjq}-s-\beta_{kil}\beta_{kjq}^2}{\beta_{kjq}}\bigg) \bigg),   
\end{split}\\ 
\begin{split}
\frac{\partial \Upsilon_{ijk} }{\partial \beta_{kip}} (t,s) ={}& \omega_{kip}\beta_{kip} \sum_{l^\prime=1}^{r_{kj}}\omega_{kjl^\prime}e^{-\beta_{kip}(\delta_{kjl^\prime}-s-\frac{\beta_{kip}\beta_{kjl^\prime}^2}{2})} \\ &\qquad \qquad \quad \times\Bigg( \left(\frac{1}{\beta_{kip}}-\delta_{kjl^\prime}+s+\frac{\beta_{kip}\beta_{kjl^\prime}^2}{2} \right)\bigg( F_{\mathcal{N}}\left(\frac{t-(\delta_{kjl^{\prime}}-s-\beta_{kil} \beta_{kjl^\prime}^2)}{\beta_{kjl^\prime}}\right)\\ &\qquad \qquad \quad-F_{\mathcal{N}}\left(-\frac{\delta_{kjl^{\prime}}-s-\beta_{kil}\beta_{kjl^\prime}^2}{\beta_{kjl^\prime}}\right)+\beta_{kjl^\prime}\bigg( f_{\mathcal{N}}(\frac{t-(\delta_{kjl^{\prime}}-s-\beta_{kil} \beta_{kjl^\prime}^2)}{\beta_{kjl^\prime}})\\ &\qquad \qquad \quad-f_{\mathcal{N}}(-\frac{(\delta_{kjl^{\prime}}-s-\beta_{kil} \beta_{kjl^\prime}^2)}{\beta_{kjl^\prime}}) \bigg)  \Bigg),   
\end{split}\\ 
\begin{split}
\frac{\partial \Upsilon_{ijk} }{\partial \beta_{kjq}} (t,s) ={}& \omega_{kjl^\prime} \sum_{l=1}^{r_{ki}}\omega_{kil}\beta_{kil} e^{-\beta_{kil}(\delta_{kjq}-s-\frac{\beta_{kil}\beta_{kjq}^2}{2})} \\ 
&\qquad \quad  \times\Bigg( \beta_{kil}^2\beta_{kjq}\bigg( F_{\mathcal{N}}\Big(\frac{t-(\delta_{kjq}-s-\beta_{kil} \beta_{kjq}^2)}{\beta_{kjq}}\Big)\\ 
&\qquad \qquad -F_{\mathcal{N}}\Big(-\frac{\delta_{kjq}-s-\beta_{kil}\beta_{kjq}^2}{\beta_{kjq}}\Big) \bigg)\\ 
&\qquad \qquad +\Bigg( \left(\beta_{kil}+\frac{\delta_{kjq}-s-t}{\beta_{kjq}^2}\right) f_{\mathcal{N}}\left(\frac{t-(\delta_{kjq}-s-\beta_{kil} \beta_{kjq}^2)}{\beta_{kjq}}\right)\\ 
&\qquad \qquad -\left(\beta_{kil}+\frac{\delta_{kjq}-s}{\beta_{kjq}^2}\right)f_{\mathcal{N}}\left(-\frac{(\delta_{kjq}-s-\beta_{kil} \beta_{kjq}^2)}{\beta_{kjq}}\right)  \Bigg),   
\end{split}\\ 
\begin{split}
\frac{\partial \Upsilon_{ijk} }{\partial \delta_{kjq}} (t,s) ={}& -\omega_{kjq}  \sum_{l=1}^{r_{ki}}\omega_{kil}\beta_{kil} e^{-\beta_{kil}(\delta_{kjq}-s-\frac{\beta_{kil}\beta_{kjq}^2}{2})}\\ 
&\qquad \qquad \times \Bigg( \beta_{kil}\bigg( F_{\mathcal{N}}\Big(\frac{t-(\delta_{kjq}-s-\beta_{kil} \beta_{kjq}^2)}{\beta_{kjq}}\Big)-F_{\mathcal{N}}\Big(-\frac{\delta_{kjq}-s-\beta_{kil}\beta_{kjq}^2}{\beta_{kjq}}\Big) \bigg)\\ 
&\qquad \qquad +\frac{1}{\beta_{kjq}} \bigg( f_{\mathcal{N}}(\frac{t-(\delta_{kjq}-s-\beta_{kil} \beta_{kjq}^2)}{\beta_{kjq}})- f_{\mathcal{N}}(-\frac{(\delta_{kjq}-s-\beta_{kil} \beta_{kjq}^2)}{\beta_{kjq}}) \bigg) \Bigg) .    
\end{split}
\end{align}

\end{prop}
\begin{rmk}\label{rmk:numerical_instability_gaussexp}
The term $\exp\bigg(-\beta_{kil}\bigg(\delta_{kjl^{\prime}}-s-\frac{\beta_{kil} \beta_{kjl^\prime}^2}{2}\bigg) \bigg) $ in \eqref{eq:upsilon_exp_gaussian} can lead to overflow. Therefore, implementing directly the formulas in this last proposition leads to numerical instabilities, and is of limited practical use.
\end{rmk}

\subsubsection{Rayleigh}
\begin{definition}[Rayleigh kernel]
Let $r\in \mathbb{N}^*$. For $x\geq 0$, the Rayleigh kernel $\phi^{\textrm{Ry}}_{(\boldsymbol{\omega},\boldsymbol{\beta})}$ is
\begin{equation*}
    \phi^{\textrm{Ry}}_{(\boldsymbol{\omega},\boldsymbol{\beta})}(x)=\sum_{l=1}^r \omega_l \frac{x}{\beta_l^2}\exp\left(-\frac{x^2}{2\beta_l^2}\right),
\end{equation*}
where the parameters are the vector of weights $\boldsymbol{\omega}:=(\omega_l)_{l\in \llbracket 1, r \rrbracket}\in [0,+\infty)^r$, and the vector of scale parameters $\boldsymbol{\beta}:=(\beta_l)_{l\in \llbracket 1, r \rrbracket}\in (0,+\infty)^r$. We have
\begin{equation*}
    \| \phi^{\textrm{Ry}}_{(\boldsymbol{\omega},\boldsymbol{\beta})} \|_1=\sum_{l=1}^r\omega_l .
\end{equation*}
\end{definition}
We assume that the kernel
\begin{equation*}
    \phi_{ki}:=\phi^{\textrm{Ry}}_{(\boldsymbol{\omega_{ki}},\boldsymbol{\beta_{ki}})}
\end{equation*}
is a Rayleigh kernel with parameters $\boldsymbol{\theta_{ki}}=\big(\boldsymbol{\omega_{ki}},\boldsymbol{\beta_{ki}} \big) $, but do not make any assumptions on $\phi_{k^\prime,i^\prime}$ for $(k^\prime,i^\prime)\neq (k,i)$.
\begin{prop}
For all $x\geq 0$, we have
\begin{equation} \label{eq:psi_jk rayleigh}
\psi_{ki}(x) =\sum_{l=1}^r \omega_{kil} \left( 1-\exp{\left( -\frac{x^2}{2\beta^2} \right)} \right)
\end{equation}
Fix $p \in [ \, r_{ki}  ]$. The partial derivatives of $\phi_{ki}$ and $\psi_{ki}$ with respect to model parameters are given by 
\begin{align} 
\begin{split}
\frac{\partial \phi_{ki} }{\partial \omega_{kip}} (x) ={}&  \frac{x}{\beta_{kip}^2}\exp\left(-\frac{x^2}{2\beta_{kip}^2}\right) ,  
\end{split}\\ 
\begin{split}
\frac{\partial \phi_{ki} }{\partial \beta_{kip}} (x) ={}& \frac{\omega_{kip} x}{\beta_{kip}^3} \exp\left(-\frac{x^2}{2\beta_{kip}^2}\right) \left(\frac{x^2}{\beta_{kip}^2}-2 \right) ,  
\end{split}\\ 
\begin{split}
\frac{\partial \psi_{ki} }{\partial \omega_{kip}} (x) ={}&1-\exp{\left( -\frac{x^2}{2\beta_{kip}^2} \right)}  ,   
\end{split}\\ 
\begin{split}
\frac{\partial \psi_{ki} }{\partial \beta_{kip}} (x) ={}& -\frac{\omega_{kip} x^2}{\beta_{kip}^3}\exp{\left( -\frac{x^2}{2\beta_{kip}^2} \right)}.  
\end{split}
\end{align}
\end{prop}
We give closed-form formulas for $\Upsilon_{ijk}$ and its partial derivatives given various parametric classes  for kernel $\phi_{kj}$. 
\begin{prop}[Rayleigh with Rayleigh]
We assume that the kernel
\begin{equation*}
\phi_{kj}:=\phi^{\textrm{Ry}}_{(\boldsymbol{\omega_{kj}},\boldsymbol{\beta_{kj}})}
\end{equation*}
is a Rayleigh kernel with parameters $\boldsymbol{\theta_{kj}}=\big(\boldsymbol{\omega_{kj}},\boldsymbol{\beta_{kj}} \big) $.  Define 
\begin{align} 
b_{ijk,ll^\prime} &=  \frac{\beta_{kil}\beta_{kjl^\prime}}{\sqrt{\beta_{kil}^2+\beta_{kjl^\prime}^2}}, \\ 
b^{(kil)}_{ijk,ll^\prime} &=  \frac{\beta_{kil}^2}{\beta_{kil}^2+\beta_{kjl^\prime}^2},\\ 
b^{(kjl^\prime)}_{ijk,ll^\prime} &=  \frac{\beta_{kjl^\prime}^2}{\beta_{kil}^2+\beta_{kjl^\prime}^2}.
\end{align}

Then for all $t,s \geq 0$

\begin{equation}
\begin{split}
\Upsilon_{ijk}(t,s) & = \sum_{l=1}^{r_{ki}}\sum_{l^\prime=1}^{r_{kj}} \omega_{kil}\omega_{kjl^\prime}\exp{\bigg(-\frac{s^2}{2(\beta_{kil}^2+\beta_{kjl^\prime}^2)}\bigg)}\\
 & \qquad \qquad\times \Bigg[ \frac{sb^{(kjl^\prime)}_{ijk,ll^\prime}}{\beta_{kil}^2+\beta_{kjl^\prime}^2}\exp{\bigg(-\frac{(sb^{(kil)}_{ijk,ll^\prime})^2}{2b^2_{ijk,ll^\prime}} \bigg)} -\frac{t+sb^{(kjl^\prime)}_{ijk,ll^\prime}}{\beta_{kil}^2+\beta_{kjl^\prime}^2}\exp{\bigg(-\frac{(t+sb^{(kil)}_{ijk,ll^\prime})^2}{2b^2_{ijk,ll^\prime}} \bigg)}  \\
 &\qquad \qquad +\sqrt{2\pi}\frac{\beta_{kil} \beta_{kjl^\prime}}{\left(\beta_{kil}^2 +\beta_{kjl^\prime}^2 \right)^{3/2}}\Bigg(1-\frac{s^2}{\beta_{kil}^2 +\beta_{kjl^\prime}^2 }\Bigg)\Bigg(F_{\mathcal{N}}\bigg(\frac{t+sb^{(kil)}_{ijk,ll^\prime}}{b_{ijk,ll^\prime}} \bigg)- F_{\mathcal{N}}\bigg( \frac{sb^{(kil)}_{ijk,ll^\prime}}{b_{ijk,ll^\prime}} \bigg)\Bigg) \Bigg].
\end{split}
\end{equation}
Let $p \in [ \, r_{ki}  ]$ and $q \in [ \, r_{kj}  ]$. If $i \neq j$, the partial derivatives of $\Upsilon_{ijk}$ with respect to the parameters are given by
\begin{align} 
\begin{split}
\frac{\partial \Upsilon_{ijk} }{\partial \omega_{kip}} (t,s) ={}& \sum_{l^\prime=1}^{r_{kj}} \omega_{kjl^\prime}\exp{\bigg(-\frac{s^2}{2(\beta_{kip}^2+\beta_{kjl^\prime}^2)}\bigg)}\\
 & \qquad \times \Bigg[ \frac{sb^{(kjl^\prime)}_{ijk,pl^\prime}}{\beta_{kip}^2+\beta_{kjl^\prime}^2}\exp{\bigg(-\frac{(sb^{(kip)}_{ijk,pl^\prime})^2}{2b^2_{ijk,pl^\prime}} \bigg)} -\frac{t+sb^{(kjl^\prime)}_{ijk,pl^\prime}}{\beta_{kip}^2+\beta_{kjl^\prime}^2}\exp{\bigg(-\frac{(t+sb^{(kip)}_{ijk,pl^\prime})^2}{2b^2_{ijk,pl^\prime}} \bigg)}  \\
 &\qquad +\sqrt{2\pi}\frac{\beta_{kip} \beta_{kjl^\prime}}{\left(\beta_{kip}^2 +\beta_{kjl^\prime}^2 \right)^{3/2}}\Bigg(1-\frac{s^2}{\beta_{kip}^2 +\beta_{kjl^\prime}^2 }\Bigg)\Bigg(F_{\mathcal{N}}\bigg(\frac{t+sb^{(kip)}_{ijk,pl^\prime}}{b_{ijk,pl^\prime}} \bigg)- F_{\mathcal{N}}\bigg( \frac{sb^{(kip)}_{ijk,pl^\prime}}{b_{ijk,pl^\prime}} \bigg)\Bigg) \Bigg], 
\end{split}\\ 
\begin{split}
\frac{\partial \Upsilon_{ijk} }{\partial \omega_{kjq}} (t,s) ={}& \sum_{l=1}^{r_{ki}}  \omega_{kil}\exp{\bigg(-\frac{s^2}{2(\beta_{kil}^2+\beta_{kjq}^2)}\bigg)}\\
 & \qquad \times \Bigg[ \frac{sb^{(kjq)}_{ijk,lq}}{\beta_{kil}^2+\beta_{kjq}^2}\exp{\bigg(-\frac{(sb^{(kil)}_{ijk,lq})^2}{2b^2_{ijk,lq}} \bigg)} -\frac{t+sb^{(kjq)}_{ijk,lq}}{\beta_{kil}^2+\beta_{kjq}^2}\exp{\bigg(-\frac{(t+sb^{(kil)}_{ijk,lq})^2}{2b^2_{ijk,lq}} \bigg)}  \\
 &\qquad  +\sqrt{2\pi}\frac{\beta_{kil} \beta_{kjq}}{\left(\beta_{kil}^2 +\beta_{kjq}^2 \right)^{3/2}}\Bigg(1-\frac{s^2}{\beta_{kil}^2 +\beta_{kjq}^2 }\Bigg)\Bigg(F_{\mathcal{N}}\bigg(\frac{t+sb^{(kil)}_{ijk,lq}}{b_{ijk,lq}} \bigg)- F_{\mathcal{N}}\bigg( \frac{sb^{(kil)}_{ijk,lq}}{b_{ijk,lq}} \bigg)\Bigg) \Bigg],   
\end{split}\\  
\begin{split}
\frac{\partial \Upsilon_{ijk} }{\partial \beta_{kip}} (t,s) ={}& \omega_{kip}\sum_{l^\prime=1}^{r_{kj}}\omega_{kjl^\prime} \exp{\left(-\frac{s^2}{2(\beta_{kip}^2+\beta_{kjl^\prime}^2)}\right)} \\
 & \qquad \qquad \times \Bigg[ -\frac{2sb^{(kjl^\prime)}_{ijk,ll^\prime}(1+2(b^{(kip)}_{ijk,pl^\prime})^2)}{\beta_{kip}^3}\exp{\left(-\frac{(sb^{(kip)}_{ijk,pl^\prime})^2}{2b^2_{ijk,pl^\prime}} \right)}\\
 & \qquad \qquad   -\Bigg(\frac{t+sb^{(kjl^\prime)}_{ijk,ll^\prime}}{\beta_{kip}^3}\bigg(\frac{t^2}{\beta_{kip}^2+\beta_{kjl^\prime}^2}-2(b^{(kip)}_{ijk,pl^\prime})^2\bigg)-\frac{2sb^{(kjl^\prime)}_{ijk,ll^\prime}}{\beta_{kip}^3} \Bigg) \exp{\left(-\frac{(t+sb^{(kip)}_{ijk,pl^\prime})^2}{2b^2_{ijk,pl^\prime}} \right)}\\
 &\qquad \qquad   +\frac{\sqrt{2\pi b^{(kjl^\prime)}_{ijk,ll^\prime}}}{\beta_{kip}^2 +\beta_{kjl^\prime}^2 }\bigg( - \frac{b^{(kip)}_{ijk,pl^\prime}}{(\beta_{kip}^2 +\beta_{kjl^\prime}^2 )^2}s^4+\frac{5b^{(kip)}_{ijk,pl^\prime}-b^{(kjl^\prime)}_{ijk,pl^\prime}}{\beta_{kip}^2 +\beta_{kjl^\prime}^2 }s^2+b^{(kjl^\prime)}_{ijk,pl^\prime}-2b^{(kip)}_{ijk,pl^\prime} \bigg)\\
 &\qquad \qquad \times\Bigg(F_{\mathcal{N}}\bigg(\frac{t+sb^{(kip)}_{ijk,pl^\prime}}{b_{ijk,pl^\prime}} \bigg)- F_{\mathcal{N}}\bigg( \frac{sb^{(kip)}_{ijk,pl^\prime}}{b_{ijk,pl^\prime}} \bigg)\Bigg)  \\
 &\qquad \qquad  +\frac{\sqrt{2\pi}\beta_{kip} \beta_{kjl^\prime}^2}{\left(\beta_{kip}^2 +\beta_{kjl^\prime}^2 \right)^{2}}\left(1-\frac{s^2}{\beta_{kip}^2 +\beta_{kjl^\prime}^2 }\right)\Bigg( \frac{b^{(kip)}_{ijk,pl^\prime}s-t}{\beta_{kip}^2}f_{\mathcal{N}}\left(\frac{t+sb^{(kip)}_{ijk,pl^\prime}}{b_{ijk,pl^\prime}} \right)\\ 
 & \qquad \qquad - \frac{s}{\beta_{kip}^2 +\beta_{kjl^\prime}^2 }f_{\mathcal{N}}\left( \frac{sb^{(kip)}_{ijk,pl^\prime}}{b_{ijk,pl^\prime}} \right)\Bigg) \Bigg],   
\end{split}\\ 
\begin{split}
\frac{\partial \Upsilon_{ijk} }{\partial \beta_{kjq}} (t,s) ={}& \omega_{kjq}\sum_{l=1}^{r_{ki}}\omega_{kil}\exp{\left(-\frac{s^2}{2(\beta_{kil}^2+\beta_{kjq}^2)}\right)} \\
 &\qquad  \qquad \times \Bigg[ \frac{s^3+2\beta_{kjq}^2(b^{(kil)}_{ijk,lq}-b^{(kjq)}_{ijk,lq})s}{\beta_{kjq}(\beta_{kil}^2+\beta_{kjq}^2)^2}\exp{\left(-\frac{(sb^{(kil)}_{ijk,lq})^2}{2b^2_{ijk,lq}} \right)}\\
 & \qquad \qquad -\frac{1}{\beta_{kil}^3}\Bigg(2b^{(kil)}_{ijk,lq}(b^{(kjq)}_{ijk,lq})^2+(b^{(kjq)}_{ijk,lq})^2\bigg(t+sb^{(kjq)}_{ijk,lq}\bigg)\bigg(\frac{s^2}{\beta_{kil}^2+\beta_{kjq}^2}-2\bigg)\\
 & \qquad \qquad+\frac{\bigg(t+b^{(kil)}_{ijk,lq}s\bigg)\bigg(t+b^{(kjq)}_{ijk,lq}s\bigg)\bigg(t+(1+b^{(kjq)}_{ijk,lq})s\bigg)}{\beta_{kil}^2+\beta_{kjq}^2}  \Bigg)\exp{\left(-\frac{(t+sb^{(kil)}_{ijk,lq})^2}{2b^2_{ijk,lq}} \right)}\\
 &\qquad \qquad +\frac{\sqrt{2\pi b^{(kil)}_{ijk,lq}}}{\beta_{kil}^2 +\beta_{kjq}^2 }\bigg( -\frac{b^{(kjq)}_{ijk,lq} }{(\beta_{kjq}^2 +\beta_{kil}^2 )^2}s^4+\frac{5b^{(kjq)}_{ijk,lq}-b^{(kil)}_{ijk,lq}}{\beta_{kjq}^2 +\beta_{kil}^2 }s^2+b^{(kil)}_{ijk,lq}-2b^{(kjq)}_{ijk,lq} \bigg)\\
 &\qquad \qquad\times\left(F_{\mathcal{N}}\left(\frac{t+sb^{(kil)}_{ijk,lq}}{b_{ijk,lq}} \right)- F_{\mathcal{N}}\left( \frac{sb^{(kil)}_{ijk,lq}}{b_{ijk,lq}} \right)\right)    \\
 &\qquad \qquad -\frac{\sqrt{2\pi}\beta_{kil}^2 \beta_{kjq}}{\left(\beta_{kil}^2 +\beta_{kjq}^2 \right)^{2}}\left(1-\frac{s^2}{\beta_{kil}^2 +\beta_{kjq}^2 }\right)\bigg(\Big(\frac{t+s(1+b^{(kjq)}_{ijk,lq}) }{\beta^2_{kjq}}\Big)f_{\mathcal{N}}\left(\frac{t+sb^{(kil)}_{ijk,lq}}{b_{ijk,lq}} \right)\\
 &\qquad \qquad  -\Big(\frac{s(1+b^{(kjq)}_{ijk,lq}) }{\beta^2_{kjq}}\Big)f_{\mathcal{N}}\left( \frac{sb^{(kil)}_{ijk,lq}}{b_{ijk,lq}} \right)\bigg) \Bigg].    
\end{split}
\end{align}
If $i=j$, the partial derivatives of $\Upsilon_{ijk}$ with respect to the parameters are given by
\begin{align} 
\begin{split}
\frac{\partial \Upsilon_{iik} }{\partial \omega_{kip}} (t,s) ={}& 2\omega_{kip}\exp{\left(-\frac{s^2}{4\beta_{kip}^2}\right)}\times \Bigg[ \frac{s}{4\beta_{kip}^2}\exp{\left(-\frac{s^2}{4\beta_{kip}^2} \right)}-\frac{2t+s}{4\beta_{kip}^2}\exp{\left(-\frac{(2t+s)^2}{4\beta_{kip}^2} \right)}  \\
 & +\frac{\sqrt{\pi}}{2}\frac{1}{\beta_{kip}}\left(1-\frac{s^2}{2\beta_{kip}^2 }\right)\left(F_{\mathcal{N}}\left(\frac{2t+s}{\sqrt{2}\beta_{kip}} \right)- F_{\mathcal{N}}\left( \frac{s}{\sqrt{2}\beta_{kip}} \right)\right) \Bigg]\\ 
 &+\sum_{l^\prime \in [ \, r_{ki}  ], l^\prime \neq p}\omega_{kil^\prime} \exp{\left(-\frac{s^2}{2(\beta_{kip}^2+\beta_{kil^\prime}^2)}\right)}\times \Bigg[ \frac{s\beta_{kil^\prime}^2}{(\beta_{kip}^2+\beta_{kil^\prime}^2)^2}\exp{\left(-\frac{(sb^{(kip)}_{iik,pl^\prime})^2}{2b^2_{iik,pl^\prime}} \right)}\\
 & \qquad \qquad  -\frac{(\beta_{kip}^2+\beta_{kil^\prime}^2)t+s\beta_{kil^\prime}^2}{(\beta_{kip}^2+\beta_{kil^\prime}^2)^2}\exp{\left(-\frac{(t+sb^{(kip)}_{iik,pl^\prime})^2}{2b^2_{iik,pl^\prime}} \right)}  \\
 &\qquad \qquad  +\sqrt{2\pi}\frac{\beta_{kip} \beta_{kil^\prime}}{\left(\beta_{kip}^2 +\beta_{kil^\prime}^2 \right)^{3/2}}\left(1-\frac{s^2}{\beta_{kip}^2 +\beta_{kil^\prime}^2 }\right)\left(F_{\mathcal{N}}\left(\frac{t+sb^{(kip)}_{iik,pl^\prime}}{b_{iik,pl^\prime}} \right)- F_{\mathcal{N}}\left( \frac{sb^{(kip)}_{iik,pl^\prime}}{b_{iik,pl^\prime}} \right)\right) \Bigg]\\ 
 &+\sum_{l \in [ \, r_{ki}  ], l \neq p}\omega_{kil}\exp{\left(-\frac{s^2}{2(\beta_{kil}^2+\beta_{kip}^2)}\right)}\times \Bigg[ \frac{s\beta_{kip}^2}{(\beta_{kil}^2+\beta_{kip}^2)^2}\exp{\left(-\frac{(sb^{(kil)}_{iik,lp})^2}{2b^2_{iik,lp}} \right)}\\
 & \qquad \qquad -\frac{(\beta_{kil}^2+\beta_{kip}^2)t+s\beta_{kip}^2}{(\beta_{kil}^2+\beta_{kip}^2)^2}\exp{\left(-\frac{(t+sb^{(kil)}_{iik,lp})^2}{2b^2_{iik,lp}} \right)}  \\
 &\qquad \qquad +\sqrt{2\pi}\frac{\beta_{kil} \beta_{kip}}{\left(\beta_{kil}^2 +\beta_{kip}^2 \right)^{3/2}}\left(1-\frac{s^2}{\beta_{kil}^2 +\beta_{kip}^2 }\right)\left(F_{\mathcal{N}}\left(\frac{t+sb^{(kil)}_{iik,lp}}{b_{iik,lp}} \right)- F_{\mathcal{N}}\left( \frac{sb^{(kil)}_{iik,lp}}{b_{iik,lp}} \right)\right) \Bigg],   
\end{split}\\
\begin{split}
\frac{\partial \Upsilon_{iik} }{\partial \beta_{kip}} (t,s) ={}& \omega_{kip}^2 \exp{\left(-\frac{s^2}{4\beta_{kip}^2}\right)}\\ 
&\times \Bigg[ \frac{1}{2\beta_{kip}^3}\bigg(\frac{s^3}{2\beta_{kip}^2}-s\bigg)\exp{\left(-\frac{s^2}{4\beta_{kip}^2} \right)}  -\frac{1}{2\beta_{kip}^3}\bigg((2t+s)(\frac{s^2}{4\beta_{kip}^2}-1)+\frac{(2t+s)^3}{4\beta_{kip}^2} \bigg)\exp{\left(-\frac{(2t+s)^2}{4\beta_{kip}^2} \right)} \\ 
&+\frac{\sqrt{\pi}}{2\beta^2}(\frac{s^2}{\beta^2}-(1-\frac{s^2}{2\beta^2})^2) \left(F_{\mathcal{N}}\left(\frac{2t+s}{\sqrt{2}\beta_{kip}} \right)- F_{\mathcal{N}}\left( \frac{s}{\sqrt{2}\beta_{kip}} \right)\right) \\ 
&-\sqrt{\frac{\pi}{8}}\frac{1}{\beta_{kip}^3}\left(1-\frac{s^2}{2\beta_{kip}^2 }\right)\left((2t+s)f_{\mathcal{N}}\left(\frac{2t+s}{\sqrt{2}\beta_{kip}} \right)- sf_{\mathcal{N}}\left( \frac{s}{\sqrt{2}\beta_{kip}} \right)\right)    \Bigg]
\\ &+\omega_{kip}\sum_{l^\prime \in [ \, r_{ki}  ], l^\prime \neq p}\omega_{kil^\prime} \exp{\left(-\frac{s^2}{2(\beta_{kip}^2+\beta_{kil^\prime}^2)}\right)} \\
 & \qquad \qquad \qquad \times \Bigg[ -\frac{2sb^{(kil^\prime)}_{iik,ll^\prime}(1+2(b^{(kip)}_{iik,pl^\prime})^2)}{\beta_{kip}^3}\exp{\left(-\frac{(sb^{(kip)}_{iik,pl^\prime})^2}{2b^2_{iik,pl^\prime}} \right)}\\
 & \qquad \qquad  \qquad -\Bigg(\frac{t+sb^{(kil^\prime)}_{iik,ll^\prime}}{\beta_{kip}^3}\bigg(\frac{t^2}{\beta_{kip}^2+\beta_{kil^\prime}^2}-2(b^{(kip)}_{iik,pl^\prime})^2\bigg)-\frac{2sb^{(kil^\prime)}_{iik,ll^\prime}}{\beta_{kip}^3} \Bigg) \exp{\left(-\frac{(t+sb^{(kip)}_{iik,pl^\prime})^2}{2b^2_{iik,pl^\prime}} \right)}\\
 &\qquad \qquad  \qquad +\frac{\sqrt{2\pi b^{(kil^\prime)}_{iik,ll^\prime}}}{\beta_{kip}^2 +\beta_{kil^\prime}^2 }\bigg( - \frac{b^{(kip)}_{iik,pl^\prime}}{(\beta_{kip}^2 +\beta_{kil^\prime}^2 )^2}s^4+\frac{5b^{(kip)}_{iik,pl^\prime}-b^{(kjl^\prime)}_{iik,pl^\prime}}{\beta_{kip}^2 +\beta_{kil^\prime}^2 }s^2+b^{(kil^\prime)}_{iik,pl^\prime}-2b^{(kip)}_{iik,pl^\prime} \bigg)\\
 &\qquad \qquad \qquad \times\Bigg(F_{\mathcal{N}}\bigg(\frac{t+sb^{(kip)}_{iik,pl^\prime}}{b_{iik,pl^\prime}} \bigg)- F_{\mathcal{N}}\bigg( \frac{sb^{(kip)}_{iik,pl^\prime}}{b_{iik,pl^\prime}} \bigg)\Bigg)  \\
 &\qquad \qquad \qquad +\frac{\sqrt{2\pi}\beta_{kip} \beta_{kil^\prime}^2}{\left(\beta_{kip}^2 +\beta_{kil^\prime}^2 \right)^{2}}\left(1-\frac{s^2}{\beta_{kip}^2 +\beta_{kil^\prime}^2 }\right)\Bigg( \frac{b^{(kip)}_{iik,pl^\prime}s-t}{\beta_{kip}^2}f_{\mathcal{N}}\left(\frac{t+sb^{(kip)}_{iik,pl^\prime}}{b_{iik,pl^\prime}} \right)\\ 
 & \qquad \qquad \qquad - \frac{s}{\beta_{kip}^2 +\beta_{kil^\prime}^2 }f_{\mathcal{N}}\left( \frac{sb^{(kip)}_{iik,pl^\prime}}{b_{iik,pl^\prime}} \right)\Bigg) \Bigg] \\ 
 &+ \omega_{kip}\sum_{l \in [ \, r_{ki}  ], l \neq p}\omega_{kil} \exp{\left(-\frac{s^2}{2(\beta_{kil}^2+\beta_{kjq}^2)}\right)} \\
 &\qquad  \qquad \times \Bigg[ \frac{s^3+2\beta_{kip}^2(b^{(kil)}_{iik,lp}-b^{(kip)}_{iik,lp})s}{\beta_{kip}(\beta_{kil}^2+\beta_{kip}^2)^2}\exp{\left(-\frac{(sb^{(kil)}_{iik,lp})^2}{2b^2_{iik,lp}} \right)}\\
 & \qquad \qquad -\frac{1}{\beta_{kil}^3}\Bigg(2b^{(kil)}_{iik,lp}(b^{(kip)}_{iik,lp})^2+(b^{(kip)}_{iik,lp})^2\bigg(t+sb^{(kip)}_{iik,lp}\bigg)\bigg(\frac{s^2}{\beta_{kil}^2+\beta_{kip}^2}-2\bigg)\\
 & \qquad \qquad+\frac{\bigg(t+b^{(kil)}_{iik,lp}s\bigg)\bigg(t+b^{(kip)}_{iik,lp}s\bigg)\bigg(t+(1+b^{(kip)}_{iik,lp})s\bigg)}{\beta_{kil}^2+\beta_{kip}^2}  \Bigg)\exp{\left(-\frac{(t+sb^{(kil)}_{iik,lp})^2}{2b^2_{iik,lp}} \right)}\\
 &\qquad \qquad +\frac{\sqrt{2\pi b^{(kil)}_{iik,lp}}}{\beta_{kil}^2 +\beta_{kip}^2 }\bigg( -\frac{b^{(kip)}_{iik,lp} }{(\beta_{kip}^2 +\beta_{kil}^2 )^2}s^4+\frac{5b^{(kip)}_{iik,lp}-b^{(kil)}_{iik,lp}}{\beta_{kip}^2 +\beta_{kil}^2 }s^2+b^{(kil)}_{iik,lp}-2b^{(kip)}_{iik,lp} \bigg)\\
 &\qquad \qquad\times\left(F_{\mathcal{N}}\left(\frac{t+sb^{(kil)}_{iik,lp}}{b_{iik,lp}} \right)- F_{\mathcal{N}}\left( \frac{sb^{(kil)}_{iik,lp}}{b_{iik,lp}} \right)\right)    \\
 &\qquad \qquad -\frac{\sqrt{2\pi}\beta_{kil}^2 \beta_{kip}}{\left(\beta_{kil}^2 +\beta_{kip}^2 \right)^{2}}\left(1-\frac{s^2}{\beta_{kil}^2 +\beta_{kip}^2 }\right)\bigg(\Big(\frac{t+s(1+b^{(kip)}_{iik,lp}) }{\beta^2_{kip}}\Big)f_{\mathcal{N}}\left(\frac{t+sb^{(kil)}_{iik,lp}}{b_{iik,lp}} \right)\\
 &\qquad \qquad  -\Big(\frac{s(1+b^{(kip)}_{iik,lp}) }{\beta^2_{kip}}\Big)f_{\mathcal{N}}\left( \frac{sb^{(kil)}_{iik,lp}}{b_{iik,lp}} \right)\bigg) \Bigg].  
\end{split}
\end{align}
\end{prop}

\subsubsection{Triangular}
The triangular kernel is a simple example of a finite support non-monotonic kernel.
\begin{definition}[Triangular kernel]
Let $r\in \mathbb{N}^*$. For $x\geq 0$, the Triangular kernel $\phi^{\textrm{Tr}}_{(\boldsymbol{\omega},\boldsymbol{\alpha},\boldsymbol{\beta},\boldsymbol{\delta})}$ is
\begin{equation*}
    \phi^{\textrm{Tr}}_{(\boldsymbol{\omega},\boldsymbol{\alpha},\boldsymbol{\beta},\boldsymbol{\delta})}(x)=\sum_{l=1}^r \omega_l \bigg( \frac{t-\alpha_l}{\beta_l}\mathbbm{1}_{	\{  0 \leq t-\alpha_l \leq \beta_l 	\}}-\frac{t-\alpha_l-\beta_l-\delta_l}{\delta_l}\mathbbm{1}_{	\{  0\leq t-\alpha_l-\beta_l \leq \delta_l	\}} \bigg),
\end{equation*}
where the parameters are the vector of weights $\boldsymbol{\omega}:=(\omega_l)_{l\in \llbracket 1, r \rrbracket}\in [0,+\infty)^r$, the vector of left corners  $\boldsymbol{\alpha}:=(\alpha_l)_{l\in \llbracket 1, r \rrbracket}\in (0,+\infty)^r$, the vector of distances to the altitude feet $\boldsymbol{\beta}:=(\beta_l)_{l\in \llbracket 1, r \rrbracket}\in (0,+\infty)^r$, and the vector of distances between altitude feet and right corners $\boldsymbol{\delta}:=(\delta_l)_{l\in \llbracket 1, r \rrbracket}\in (0,+\infty)^r$. We have
\begin{equation*}
    \| \phi^{\textrm{Tr}}_{(\boldsymbol{\omega},\boldsymbol{\alpha},\boldsymbol{\beta},\boldsymbol{\delta})} \|_1=\sum_{l=1}^r\omega_l \bigg( \beta_l + \delta_l \bigg)/2.
\end{equation*}
\end{definition}

\begin{figure}[h]
    \centering
\resizebox{0.3\textwidth}{!}{
\includegraphics{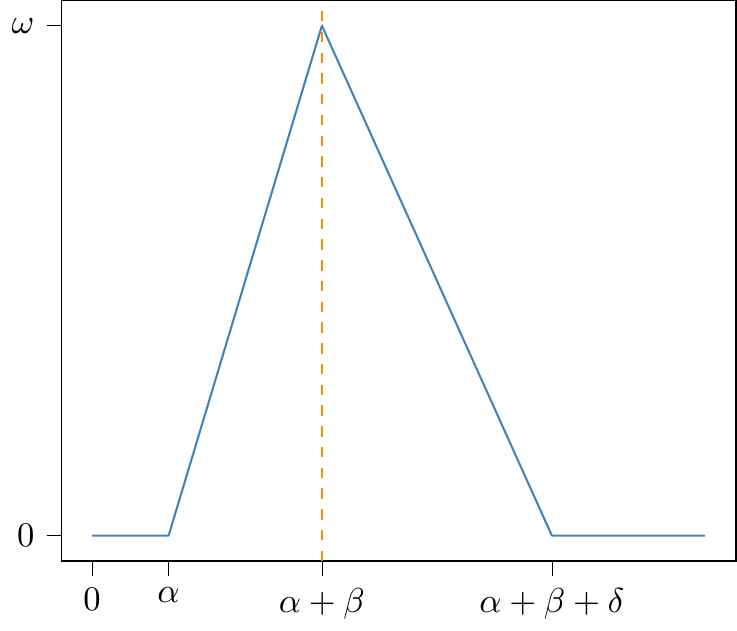}
}
    \caption{Plot of a triangular kernel against time, $r=1$.}
    
\medskip

    \label{fig:phi_triangle}
\end{figure}    
We assume that the kernel
\begin{equation*}
    \phi_{ki}:=\phi^{\textrm{Tr}}_{(\boldsymbol{\omega_{ki}},\boldsymbol{\alpha_{ki}},\boldsymbol{\beta_{ki}},\boldsymbol{\delta_{ki}})}
\end{equation*}
is a triangular kernel with parameters $\boldsymbol{\theta_{ki}}=\big(\boldsymbol{\omega_{ki}},\boldsymbol{\alpha_{ki}},\boldsymbol{\beta_{ki}},\boldsymbol{\delta_{ki}} \big) $, but do not make any assumptions on $\phi_{k^\prime,i^\prime}$ for $(k^\prime,i^\prime)\neq (k,i)$.
\begin{prop}
For all $x\geq 0$, we have
\begin{equation} \label{eq:psi_jk triangle}
\begin{split}
    \psi_{ki}(x) & =\sum_{l=1}^r \omega_{kil} \bigg( \mathbbm{1}_{	\{  0 \leq x-\alpha_{kil} \leq \beta_{kil} 	\}}\frac{(x-\alpha_{kil})^2}{2 \beta_{kil}}\\
 &\qquad +\mathbbm{1}_{	\{  0\leq x-\alpha_{kil}-\beta_{kil} \leq \delta_{kil}	\}} \Big(\frac{\beta_{kil}}{2}+(x-\alpha_{kil}-\beta_{kil})  -\frac{(x-\alpha_{kil}-\beta_{kil})^2}{2\delta_{kil}} \Big) \\
 &\qquad +\frac{\beta_{kil}+\delta_{kil}}{2}\mathbbm{1}_{	\{  x-\alpha_{kil}-\beta_{kil} \geq \delta_{kil}	\}}  \bigg).
\end{split}
\end{equation}
Fix $p \in [ \, r_{ki}  ]$. The partial derivatives of $\phi_{ki}$ and $\psi_{ki}$ with respect to model parameters are given by 
\begin{align} 
\begin{split}
\frac{\partial \phi_{ki} }{\partial \omega_{kip}} (x) ={}& \frac{x-\alpha_{kip}}{\beta_{kip}}\mathbbm{1}_{	\{  0 \leq x-\alpha_{kip} \leq \beta_{kip} 	\}}
\\ & -\frac{x-\alpha_{kip}-\beta_{kip}-\delta_{kip}}{\delta_{kip}}\mathbbm{1}_{	\{  0\leq x-\alpha_{kip}-\beta_{kip} \leq \delta_{kip}	\}}  ,  
\end{split}\\ 
\begin{split}
\frac{\partial \phi_{ki} }{\partial \alpha_{kip}} (x)  ={}& -\frac{\omega_{kip}}{\beta_{kip}}\mathbbm{1}_{	\{  0 < x-\alpha_{kip} < \beta_{kip} 	\}}+\frac{\omega_{kip}}{\delta_{kip}}\mathbbm{1}_{	\{  0< x-\alpha_{kip}-\beta_{kip} < \delta_{kip}	\}} ,
\end{split}\\ 
\begin{split}
\frac{\partial \phi_{ki} }{\partial \beta_{kip}} (x) ={}&  -\omega_{kip}\frac{x-\alpha_{kip}}{\beta_{kip}^2}\mathbbm{1}_{	\{  0 < x-\alpha_{kip} < \beta_{kip} 	\}}+\frac{\omega_{kip}}{\delta_{kip}}\mathbbm{1}_{	\{  0< x-\alpha_{kip}-\beta_{kip} < \delta_{kip}	\}},  
\end{split}\\ 
\begin{split}
\frac{\partial \phi_{ki} }{\partial \delta_{kip}} (x) ={}& \omega_{kip}\frac{x-\alpha_{kip}-\beta_{kip}}{\delta_{kip}^2}\mathbbm{1}_{	\{  0< x-\alpha_{kip}-\beta_{kip} < \delta_{kip}	\}},   
\end{split}\\ 
\begin{split}
\frac{\partial \psi_{ki} }{\partial \omega_{kip}} (x) ={}& \mathbbm{1}_{	\{  0 \leq x-\alpha_{kip} \leq \beta_{kip} 	\}}\frac{(x-\alpha_{kip})^2}{2 \beta_{kip}}\\ &  +\mathbbm{1}_{	\{  0\leq x-\alpha_{kip}-\beta_{kip} \leq \delta_{kip}	\}} \Big(\frac{\beta_{kip}}{2}+(x-\alpha_{kip}-\beta_{kip})
\\ &  -\frac{(x-\alpha_{kip}-\beta_{kip})^2}{2\delta_{kip}} \Big) 
\\ &  +\frac{\beta_{kip}+\delta_{kip}}{2}\mathbbm{1}_{	\{  x-\alpha_{kip}-\beta_{kip} \geq \delta_{kip}	\}} ,   
\end{split}\\ 
\begin{split}
\frac{\partial \psi_{ki} }{\partial \alpha_{kip}} (x) ={}& -\omega_{kip}\frac{x-\alpha_{kip}}{ \beta_{kip}}\mathbbm{1}_{	\{  0 <-\alpha_{kip} < \beta_{kip} 	\}}\\ &  +\omega_{kip}\frac{x-\alpha_{kip}-\beta_{kip}-\delta_{kip}}{\delta_{kip}}\mathbbm{1}_{	\{  0< x-\alpha_{kip}-\beta_{kip} < \delta_{kip}	\}} ,      
\end{split}\\ 
\begin{split}
\frac{\partial \psi_{ki} }{\partial \beta_{kip}} (x) ={}& -\frac{\omega_{kip}}{2}\big(\frac{x-\alpha_{kip}}{ \beta_{kip}}\big)^2\mathbbm{1}_{	\{  0 \leq x-\alpha_{kip} \leq \beta_{kip} 	\}}\\ &  +\omega_{kip}\Big(\frac{1}{2}+\frac{x-\alpha_{kip}-\beta_{kip}-\delta_{kip}}{\delta_{kip}}\Big)\mathbbm{1}_{	\{  0< x-\alpha_{kip}-\beta_{kip} < \delta_{kip}	\}} 
\\ &  +\frac{\omega_{kip}}{2}\mathbbm{1}_{	\{  x> \alpha_{kip}+\beta_{kip}+\delta_{kip}	\}}  ,      
\end{split}\\ 
\begin{split}
\frac{\partial \psi_{ki} }{\partial \delta_{kip}} (x) ={}&  \frac{\omega_{kip}}{2}\Big(\frac{x-\alpha_{kip}-\beta_{kip}}{ \delta_{kip}}\Big)^2\mathbbm{1}_{	\{  0 \leq x-\alpha_{kip}-\beta_{kip} \leq \delta_{kip} 	\}}\\ &  +\frac{\omega_{kip}}{2}\mathbbm{1}_{	\{  x> \alpha_{kip}+\beta_{kip}+\delta_{kip}	\}}  .    
\end{split}
\end{align}
\end{prop}
We give closed-form formulas for $\Upsilon_{ijk}$ and its partial derivatives given various parametric classes  for kernel $\phi_{kj}$. For all $x,y,a,b \in \mathbb{R}$, define
\begin{equation*}
    F_{\tau}(x,y,a,b):= \frac{(y-x)^3}{3}-(a+b)\frac{(y-x)^2}{2}+ab(y-x).
\end{equation*}
The partial derivatives of $F_{\tau}$ are given by
\begin{align}
\frac{\partial F_{\tau}}{\partial x} (x,y,a,b) &= -(y-x)^2+(a+b)(y-x)-ab,   \\
\frac{\partial F_{\tau}}{\partial y} (x,y,a,b) &= (y-x)^2-(a+b)(y-x)+ab,   \\
\frac{\partial F_{\tau}}{\partial a} (x,y,a,b) &=-\frac{(y-x)^2}{2}+b(y-x) , \\
\frac{\partial F_{\tau}}{\partial b} (x,y,a,b) &=-\frac{(y-x)^2}{2}+a(y-x) .
\end{align}
For $z:=(x,y,a,b)$, denote the gradient of $F_{\tau}$ in $z$ by $\nabla F_{\tau}(z)$. We write $\langle\cdot, \cdot \rangle$ for the standard Euclidean inner product.
\begin{prop}[Triangular with triangular]
We assume that the kernel
\begin{equation*}
\phi_{kj}:=\phi^{\textrm{Tr}}_{(\boldsymbol{\omega_{kj}},\boldsymbol{\alpha_{kj}},\boldsymbol{\beta_{kj}},\boldsymbol{\delta_{kj}})}
\end{equation*}
is a triangular kernel with parameters $\boldsymbol{\theta_{kj}}=\big(\boldsymbol{\omega_{kj}},\boldsymbol{\alpha_{kj}},\boldsymbol{\beta_{kj}},\boldsymbol{\delta_{kj}} \big) $. Define the integration lower bounds
\begin{align}
x^{(1)}_{ll^\prime,s} &= \max(0,\alpha_{kil},\alpha_{kjl^\prime}-s),    \\
x^{(2)}_{ll^\prime,s} &= \max(0,\alpha_{kil},\alpha_{kjl^\prime}-s+\beta_{kjl^\prime}) ,   \\
x^{(3)}_{ll^\prime,s} &= \max(0,\alpha_{kil}+\beta_{kil},\alpha_{kjl^\prime}-s) ,  \\
x^{(4)}_{ll^\prime,s} &= \max(0,\alpha_{kil}+\beta_{kil},\alpha_{kjl^\prime}-s+\beta_{kjl^\prime}),
\end{align}
and the integration upper bounds
\begin{align}
 y^{(1)}_{ll^\prime,ts}  &= \min(t,\alpha_{kil}+\beta_{kil},\alpha_{kjl^\prime}-s+\beta_{kjl^\prime}),  \\
  y^{(2)}_{ll^\prime,ts}  &= \min(t,\alpha_{kil}+\beta_{kil},\alpha_{kjl^\prime}-s+\beta_{kjl^\prime}+\delta_{kjl^\prime}),  \\
  y^{(3)}_{ll^\prime,ts}  &= \min(t,\alpha_{kil}+\beta_{kil}+\delta_{kil},\alpha_{kjl^\prime}-s+\beta_{kjl^\prime}),  \\
  y^{(4)}_{ll^\prime,ts} &= \min(t,\alpha_{kil}+\beta_{kil}+\delta_{kil},\alpha_{kjl^\prime}-s+\beta_{kjl^\prime}+\delta_{kjl^\prime}).
\end{align}
Define the coefficients
\begin{align}
a^{(1)}_{l} &= \alpha_{kil},    \\
a^{(3)}_{l} &= \alpha_{kil}+\beta_{kil}+\delta_{kil},   \\
b^{(1)}_{l^\prime,s} &=\alpha_{kjl^\prime}- s,  \\
b^{(2)}_{l^\prime,s} &= \alpha_{kjl^\prime}- s+\beta_{kjl^\prime}+\delta_{kjl^\prime}.
\end{align}
To simplify the expression of $\Upsilon_{ijk}$, denote 
\begin{equation*}
a^{(2)}_{l} :=a^{(1)}_{l}, \quad
a^{(4)}_{l} := a^{(3)}_{l},   \quad
b^{(3)}_{l^\prime,s} :=b^{(1)}_{l^\prime,s},   \quad
b^{(4)}_{l^\prime,s} := b^{(2)}_{l^\prime,s}.     
\end{equation*}
Define the weights
\begin{equation*}
c^{(1)}_{ll^\prime} := \frac{1}{\beta_{kil}\beta_{kjl^\prime}},    \quad
c^{(2)}_{ll^\prime} := - \frac{1}{\beta_{kil} \delta_{kjl^\prime}},  \quad
c^{(3)}_{ll^\prime} :=- \frac{1}{\delta_{kil}\beta_{kjl^\prime}} ,  \quad
c^{(4)}_{ll^\prime} := \frac{1}{\delta_{kil} \delta_{kjl^\prime}}.
\end{equation*}
Then for all $t,s \geq 0$
\begin{equation} 
\Upsilon_{ijk}(t,s)  = \sum_{l=1}^{r_{ki}}\sum_{l^\prime=1}^{r_{kj}}\omega_{kil}\omega_{kjl^\prime}\sum_{m=1}^{4} c^{(m)}_{ll^\prime}\mathbbm{1}_{x^{(m)}_{ll^\prime,s}< y^{(m)}_{ll^\prime,ts}}F_{\tau}\left(x^{(m)}_{ll^\prime,s} , y^{(m)}_{ll^\prime,ts},a^{(m)}_{l} ,b^{(m)}_{l^\prime,s}\right). 
\end{equation}
Define
\begin{equation*}
    \boldsymbol{z^{(m)}_{ll^\prime,ts}}:=\left(x^{(m)}_{ll^\prime,s} , y^{(m)}_{ll^\prime,ts},a^{(m)}_{l} ,b^{(m)}_{l^\prime,s}\right)^\intercal.
\end{equation*}
Let $p \in [ \, r_{ki}  ]$ and $q \in [ \, r_{kj}  ]$. If $i \neq j$, the partial derivatives of $\Upsilon_{ijk}$ with respect to the parameters are given by
\begin{align} 
\frac{\partial \Upsilon_{ijk} }{\partial \omega_{kip}} (t,s) ={}& \sum_{l^\prime=1}^{r_{kj}}\omega_{kjl^\prime} \sum_{m=1}^{4} c^{(m)}_{pl^\prime}\mathbbm{1}_{x^{(m)}_{pl^\prime,s}< y^{(m)}_{pl^\prime,ts}}F_{\tau}\left(\boldsymbol{z^{(m)}_{pl^\prime,ts}}\right),\\ 
\frac{\partial \Upsilon_{ijk} }{\partial \omega_{kjq}} (t,s) ={}& \sum_{l=1}^{r_{ki}}  \omega_{kil}\sum_{m=1}^{4} c^{(m)}_{lq}\mathbbm{1}_{x^{(m)}_{lq,s}< y^{(m)}_{lq,ts}}F_{\tau}\left(\boldsymbol{z^{(m)}_{lq,ts}}\right),\\ 
\frac{\partial \Upsilon_{ijk} }{\partial \alpha_{kip}} (t,s) ={}& \omega_{kip}\sum_{l^\prime=1}^{r_{kj}}\omega_{kjl^\prime}\sum_{m=1}^4 c^{(m)}_{pl^\prime}\mathbbm{1}_{x^{(m)}_{pl^\prime,s}< y^{(m)}_{pl^\prime,ts}}\Big\langle\frac{\partial \boldsymbol{z^{(m)}_{pl^\prime,ts}}}{\partial\alpha_{kip}} , \nabla F_{\tau}\left(\boldsymbol{z^{(m)}_{pl^\prime,ts}}\right)\Big\rangle,\\ 
\frac{\partial \Upsilon_{ijk} }{\partial \alpha_{kjq}} (t,s) ={}& \omega_{kjq}\sum_{l=1}^{r_{ki}}\omega_{kil}\sum_{m=1}^{4} c^{(m)}_{lq}\mathbbm{1}_{x^{(m)}_{lq,s}< y^{(m)}_{lq,ts}} \Big\langle \frac{\partial \boldsymbol{z^{(m)}_{lq,ts}}}{\partial\alpha_{kjq}} , \nabla F_{\tau}\left(\boldsymbol{z^{(m)}_{lq,ts}}\right) \Big\rangle,\\ 
\begin{split}
\frac{\partial \Upsilon_{ijk} }{\partial \beta_{kip}} (t,s) ={}& \omega_{kip}\sum_{l^\prime=1}^{r_{kj}}\omega_{kjl^\prime}\sum_{m=1}^4 c^{(m)}_{pl^\prime}\mathbbm{1}_{x^{(m)}_{pl^\prime,s}< y^{(m)}_{pl^\prime,ts}} \Big\langle \frac{\partial \boldsymbol{z^{(m)}_{pl^\prime,ts}}}{\partial\beta_{kip}} , \nabla F_{\tau}\left(\boldsymbol{z^{(m)}_{pl^\prime,ts}}\right) \Big\rangle\\ &\qquad \qquad \qquad \quad +\mathbbm{1}_{x^{(m)}_{pl^\prime,s}< y^{(m)}_{pl^\prime,ts}}\frac{\partial c^{(m)}_{pl^\prime} }{\partial\beta_{kip}} F_{\tau}\left(\boldsymbol{z^{(m)}_{pl^\prime,ts}}\right),   
\end{split}\\ 
\begin{split}
\frac{\partial \Upsilon_{ijk} }{\partial \beta_{kjq}} (t,s) ={}& \omega_{kjq}\sum_{l=1}^{r_{ki}}\omega_{kil}\sum_{m=1}^{4} c^{(m)}_{lq}\mathbbm{1}_{x^{(m)}_{lq,s}< y^{(m)}_{lq,ts}} \Big\langle \frac{\partial \boldsymbol{z^{(m)}_{lq,ts}}}{\partial\beta_{kjq}} , \nabla F_{\tau}\left(\boldsymbol{z^{(m)}_{lq,ts}}\right) \Big\rangle\\ &\qquad \qquad \qquad +\mathbbm{1}_{x^{(m)}_{lq,s}< y^{(m)}_{lq,ts}} \frac{\partial c^{(m)}_{lq}}{\partial \beta_{kjq}} F_{\tau}\left(\boldsymbol{z^{(m)}_{lq,ts}}\right),   
\end{split}\\ 
\begin{split}
\frac{\partial \Upsilon_{ijk} }{\partial \delta_{kip}} (t,s) ={}& \omega_{kip}\sum_{l^\prime=1}^{r_{kj}}\omega_{kjl^\prime}\sum_{m=1}^4 c^{(m)}_{pl^\prime}\mathbbm{1}_{x^{(m)}_{pl^\prime,s}< y^{(m)}_{pl^\prime,ts}}\Big\langle\frac{\partial \boldsymbol{z^{(m)}_{pl^\prime,ts}}}{\partial\delta_{kip}} ,\nabla F_{\tau}\left(\boldsymbol{z^{(m)}_{pl^\prime,ts}}\right) \Big\rangle\\ &\qquad \qquad \qquad \quad+\mathbbm{1}_{x^{(m)}_{pl^\prime,s}< y^{(m)}_{pl^\prime,ts}}\frac{\partial c^{(m)}_{pl^\prime} }{\partial\delta_{kip}} F_{\tau}\left(\boldsymbol{z^{(m)}_{pl^\prime,ts}}\right),
\end{split}\\ 
\begin{split}
\frac{\partial \Upsilon_{ijk} }{\partial \delta_{kjq}} (t,s) ={}& \omega_{kjq}\sum_{l=1}^{r_{ki}}\omega_{kil}\sum_{m=1}^{4} c^{(m)}_{lq}\mathbbm{1}_{x^{(m)}_{lq,s}< y^{(m)}_{lq,ts}} \Big\langle \frac{\partial \boldsymbol{z^{(m)}_{lq,ts}}}{\partial \delta_{kjq}} ,\nabla F_{\tau}\left(\boldsymbol{z^{(m)}_{lq,ts}}\right) \Big\rangle\\ &\qquad \qquad \qquad \quad+\mathbbm{1}_{x^{(m)}_{lq,s}< y^{(m)}_{lq,ts}} \frac{\partial c^{(m)}_{lq}}{\partial \delta_{kjq}} F_{\tau}\left(\boldsymbol{z^{(m)}_{lq,ts}}\right).    
\end{split}
\end{align}
If $i=j$, the partial derivatives of $\Upsilon_{ijk}$ with respect to the parameters are given by
\begin{align} 
\begin{split}
\frac{\partial \Upsilon_{iik} }{\partial \omega_{kip}} (t,s) ={}& 2\omega_{kip}\sum_{m=1}^{4} c^{(m)}_{pp}\mathbbm{1}_{x^{(m)}_{pp,s}< y^{(m)}_{pp,ts}}F_{\tau}\left(\boldsymbol{z^{(m)}_{pp,ts}}\right)
\\ &+\sum_{l^\prime \in [ \, r_{ki}  ], l^\prime \neq p}\omega_{kil^\prime} \sum_{m=1}^{4} c^{(m)}_{pl^\prime}\mathbbm{1}_{x^{(m)}_{pl^\prime,s}< y^{(m)}_{pl^\prime,ts}}F_{\tau}\left(\boldsymbol{z^{(m)}_{pl^\prime,ts}}\right)
\\ &+\sum_{l \in [ \, r_{ki}  ], l \neq p}\omega_{kil}\sum_{m=1}^{4} c^{(m)}_{lp}\mathbbm{1}_{x^{(m)}_{lp,s}< y^{(m)}_{lp,ts}}F_{\tau}\left(\boldsymbol{z^{(m)}_{lp,ts}}\right),   
\end{split}\\
\begin{split}
\frac{\partial \Upsilon_{iik} }{\partial \alpha_{kip}} (t,s) ={}& \omega_{kip}^2\sum_{m=1}^4 c^{(m)}_{pl^\prime}\mathbbm{1}_{x^{(m)}_{pl^\prime,s}< y^{(m)}_{pp,ts}}\Big\langle\frac{\partial \boldsymbol{z^{(m)}_{pp,ts}}}{\partial\alpha_{kip}} , \nabla F_{\tau}\left(\boldsymbol{z^{(m)}_{pp,ts}}\right) \Big\rangle
\\ &+\omega_{kip}\sum_{l^\prime \in [ \, r_{ki}  ], l^\prime \neq p}\omega_{kil^\prime} \sum_{m=1}^4 c^{(m)}_{pl^\prime}\mathbbm{1}_{x^{(m)}_{pl^\prime,s}< y^{(m)}_{pl^\prime,ts}}\Big\langle\frac{\partial \boldsymbol{z^{(m)}_{pl^\prime,ts}}}{\partial\alpha_{kip}} ,\nabla F_{\tau}\left(\boldsymbol{z^{(m)}_{pl^\prime,ts}}\right) \Big\rangle\\  &+ \omega_{kip}\sum_{l \in [ \, r_{ki}  ], l \neq p}\omega_{kil} \sum_{m=1}^4 c^{(m)}_{lp}\mathbbm{1}_{x^{(m)}_{lp,s}< y^{(m)}_{lp,ts}}\Big\langle\frac{\partial \boldsymbol{z^{(m)}_{lp,ts}}}{\partial\alpha_{kip}} , \nabla F_{\tau}\left(\boldsymbol{z^{(m)}_{lp,ts}}\right) \Big\rangle.  
\end{split}\\
\begin{split}
\frac{\partial \Upsilon_{iik} }{\partial \beta_{kip}} (t,s) ={}& \omega_{kip}^2\Bigg[ \sum_{m=1}^4 c^{(m)}_{pp}\mathbbm{1}_{x^{(m)}_{pp,s}< y^{(m)}_{pp,ts}}\Big\langle\frac{\partial \boldsymbol{z^{(m)}_{pp,ts}}}{\partial\beta_{kip}} ,\nabla F_{\tau}\left(\boldsymbol{z^{(m)}_{pp,ts}}\right) \Big\rangle\\ &\qquad \quad +\mathbbm{1}_{x^{(m)}_{pp,s}< y^{(m)}_{pp,ts}}\frac{\partial c^{(m)}_{pp} }{\partial\beta_{kip}} F_{\tau}\left(\boldsymbol{z^{(m)}_{pp,ts}}\right) \Bigg]
\\ &+\omega_{kip}\sum_{l^\prime \in [ \, r_{ki}  ], l^\prime \neq p}\omega_{kil^\prime} \Bigg[ \sum_{m=1}^4 c^{(m)}_{pl^\prime}\mathbbm{1}_{x^{(m)}_{pl^\prime,s}< y^{(m)}_{pl^\prime,ts}}\Big\langle\frac{\partial \boldsymbol{z^{(m)}_{pl^\prime,ts}}}{\partial\beta_{kip}} ,\nabla F_{\tau}\left(\boldsymbol{z^{(m)}_{pl^\prime,ts}}\right) \Big\rangle\\ &\qquad \qquad \quad +\mathbbm{1}_{x^{(m)}_{pl^\prime,s}< y^{(m)}_{pl^\prime,ts}}\frac{\partial c^{(m)}_{pl^\prime} }{\partial\beta_{kip}} F_{\tau}\left(\boldsymbol{z^{(m)}_{pl^\prime,ts}}\right) \Bigg]\\ &+ \omega_{kip}\sum_{l \in [ \, r_{ki}  ], l \neq p}\omega_{kil}\Bigg[ \sum_{m=1}^4 c^{(m)}_{lp}\mathbbm{1}_{x^{(m)}_{lp,s}< y^{(m)}_{lp^\prime,ts}}\Big\langle\frac{\partial \boldsymbol{z^{(m)}_{lp,ts}}}{\partial\beta_{kip}} ,\nabla F_{\tau}\left(\boldsymbol{z^{(m)}_{lp,ts}}\right) \Big\rangle\\ &\qquad \qquad \quad+\mathbbm{1}_{x^{(m)}_{lp,s}< y^{(m)}_{lp,ts}}\frac{\partial c^{(m)}_{lp} }{\partial\beta_{kip}} F_{\tau}\left(\boldsymbol{z^{(m)}_{lp,ts}}\right) \Bigg].  
\end{split}\\
\begin{split}
\frac{\partial \Upsilon_{iik} }{\partial \delta_{kip}} (t,s) ={}& \omega_{kip}^2 \Bigg[ \sum_{m=1}^{4} c^{(m)}_{pp}\mathbbm{1}_{x^{(m)}_{pp,s}< y^{(m)}_{pp,ts}} \Big\langle\frac{\partial \boldsymbol{z^{(m)}_{pp,ts}}}{\partial \delta_{kip}} ,\nabla F_{\tau}\left(\boldsymbol{z^{(m)}_{pp,ts}}\right) \Big\rangle\\ &\qquad \quad +\mathbbm{1}_{x^{(m)}_{pp,s}< y^{(m)}_{pp,ts}} \frac{\partial c^{(m)}_{pp}}{\partial \delta_{kip}} F_{\tau}\left(\boldsymbol{z^{(m)}_{pp,ts}}\right) \Bigg]
\\ &+\omega_{kip}\sum_{l^\prime \in [ \, r_{ki}  ], l^\prime \neq p}\omega_{kil^\prime} \Bigg[ \sum_{m=1}^{4} c^{(m)}_{pl^\prime}\mathbbm{1}_{x^{(m)}_{pl^\prime,s}< y^{(m)}_{pl^\prime,ts}} \Big\langle\frac{\partial \boldsymbol{z^{(m)}_{pl^\prime,ts}}}{\partial \delta_{kip}} , \nabla F_{\tau}\left(\boldsymbol{z^{(m)}_{pl^\prime,ts}}\right)\Big\rangle\\ &\qquad \qquad \qquad +\mathbbm{1}_{x^{(m)}_{pl^\prime,s}< y^{(m)}_{pl^\prime,ts}} \frac{\partial c^{(m)}_{pl^\prime}}{\partial \delta_{kip}} F_{\tau}\left(\boldsymbol{z^{(m)}_{pl^\prime,ts}}\right) \Bigg]\\ &+ \omega_{kip}\sum_{l \in [ \, r_{ki}  ], l \neq p}\omega_{kil} \Bigg[ \sum_{m=1}^{4} c^{(m)}_{lp}\mathbbm{1}_{x^{(m)}_{lp,s}< y^{(m)}_{lp,ts}} \Big\langle\frac{\partial \boldsymbol{z^{(m)}_{lp,ts}}}{\partial \delta_{kip}} , \nabla F_{\tau}\left(\boldsymbol{z^{(m)}_{lp,ts}}\right)\Big\rangle\\ &\qquad \qquad \qquad+\mathbbm{1}_{x^{(m)}_{lp,s}< y^{(m)}_{lp,ts}} \frac{\partial c^{(m)}_{lp}}{\partial \delta_{kip}} F_{\tau}\left(\boldsymbol{z^{(m)}_{lp,ts}}\right) \Bigg].  
\end{split}
\end{align}
\end{prop}
}

\nomenclature[K]{$\phi^{\mathcal{E}}_{(\boldsymbol{\omega},\boldsymbol{\beta})}$}{Exponential kernel with parameters $(\boldsymbol{\omega},\boldsymbol{\beta})$}
\nomenclature[K]{$\phi^{\mathcal{DE}}_{(\boldsymbol{\omega},\boldsymbol{\beta},\boldsymbol{\delta})}$}{Delayed exponential kernel with parameters $(\boldsymbol{\omega},\boldsymbol{\beta},\boldsymbol{\delta})$}
\nomenclature[K]{$\phi^{\mathcal{N}}_{(\boldsymbol{\omega},\boldsymbol{\beta},\boldsymbol{\delta})}$}{Gaussian mixture kernel with parameters $(\boldsymbol{\omega},\boldsymbol{\beta},\boldsymbol{\delta})$}
\nomenclature[K]{$\phi^{\Gamma}_{(\boldsymbol{\omega},\boldsymbol{\alpha},\boldsymbol{\beta})}$}{Gamma mixture kernel with parameters $(\boldsymbol{\omega},\boldsymbol{\alpha},\boldsymbol{\beta})$}
\nomenclature[K]{$\phi^{\mathcal{DP}}_{(\boldsymbol{\omega},\boldsymbol{\alpha},\boldsymbol{\beta},\boldsymbol{\delta})}$}{Delayed power law kernel with parameters $(\boldsymbol{\omega},\boldsymbol{\alpha},\boldsymbol{\beta},\boldsymbol{\delta})$}

\nomenclature[K]{$\phi^{\mathcal{P}}_{(\boldsymbol{\omega},\boldsymbol{\alpha},\boldsymbol{\beta})}$}{Power law kernel with parameters $(\boldsymbol{\omega},\boldsymbol{\alpha},\boldsymbol{\beta})$}

\nomenclature[F]{$F_{\mathcal{N}}$}{Cumulative distribution function of a standard normal distribution}
\nomenclature[F]{$\Gamma$}{Gamma function}
\nomenclature[F]{$\gamma$}{Lower incomplete gamma function}

\nomenclature[A]{$MHP$}{Multi-variate Hawkes process}

\printnomenclature[1in]

\bibliography{bibliography.bib}

\end{document}